\documentclass[iop]{emulateapj}

\usepackage{natbib,aas_macros}
\usepackage{color}
\usepackage{ulem}
\usepackage{dcolumn}
\usepackage{amsmath}
\usepackage[flushleft]{threeparttable}
\usepackage {threeparttable} 

\newcommand{\del}{\partial}
\renewcommand{\thefootnote}{\fnsymbol{footnote}}

\renewcommand{\apjl}{ApJL}

\shorttitle{Mass Ejection from NS-NS Merger remnant}
\shortauthors{S. Fujibayashi, K. Kiuchi, N. Nishimura, Y. Sekiguchi, and M. Shibata}

\begin{document}

\title{Mass Ejection from the Remnant of Binary Neutron Star Merger:
  Viscous-Radiation Hydrodynamics Study}

\author{Sho Fujibayashi\altaffilmark{1}, Kenta Kiuchi\altaffilmark{1}, Nobuya Nishimura\altaffilmark{1}, Yuichiro Sekiguchi\altaffilmark{1,2}, and Masaru Shibata\altaffilmark{1,3}
}

\email{sho.fujibayashi@yukawa.kyoto-u.ac.jp}
\altaffiltext{1}{Center for Gravitational Physics, Yukawa Institute for Theoretical Physics, Kyoto University, Kyoto 606-8502, Japan}
\altaffiltext{2}{Department of Physics, Toho University, Funabashi, Chiba 274-8510, Japan}
\altaffiltext{3}{Max Planck Institute for Gravitational Physics (Albert Einstein Institute), Am M\"uhlenberg 1, Potsdam-Golm, D-14476, Germany}


\begin{abstract}
We perform long-term general relativistic neutrino-radiation hydrodynamics simulations (in axisymmetry) for a massive neutron star (MNS) surrounded by a torus, which is a canonical remnant formed after the binary neutron star merger.
We take into account effects of viscosity which is likely to arise in the merger remnant due to magnetohydrodynamical turbulence.
The viscous effect plays key roles for the mass ejection from the remnant in two phases of the evolution.
In the first $t\lesssim10$ ms, a differential rotation state of the MNS is changed to a rigidly rotating state.
A shock wave caused by the variation of its quasi-equilibrium state induces significant mass ejection of mass $\sim$ (0.5--2.0) $\times 10^{-2}M_\odot$ for the alpha viscosity parameter of 0.01--0.04.
For the longer-term evolution with $\sim$ 0.1--10\,s, a significant fraction of the torus material is ejected.
We find that the total mass of the viscosity-driven ejecta ($\gtrsim10^{-2}M_\odot$) could dominate over that of the dynamical ejecta ($\lesssim 10^{-2}M_\odot$).
The electron fraction, $Y_e$, of the ejecta is always high enough ($Y_e\gtrsim0.25$) that this post-merger ejecta is lanthanide-poor, and hence, the opacity of the ejecta is likely to be $\sim 10-100$ times lower than that of the dynamical ejecta.
This indicates that the electromagnetic signal from the ejecta would be rapidly evolving, bright, and blue if it is observed from a small viewing angle ($\lesssim 45^\circ$) for which the effect of the dynamical ejecta is minor.

\end{abstract}

\section{Introduction}

The merger of binary neutron stars is one of the most promising
gravitational-wave sources for ground-based detectors such as
advanced LIGO~\citep{2010NIMPA.624..223A},
advanced VIRGO~\citep{2011CQGra..28k4002A}, and
KAGRA~\citep{2010CQGra..27h4004K}.  Gravitational waves from binary
neutron stars carry the information on the mass and the tidal
deformability of neutron stars.  The total mass and mass ratio will be
used for exploring the formation process of binary neutron
stars~\citep{2017ApJ...846..170T}.  The tidal deformability together
with the mass will give us invaluable information for exploring the
properties of the high-density nuclear matter~\citep{2008PhRvD..77b1502F}.  The latest detection
of gravitational waves from a system of binary neutron stars,
GW170817, indeed shows that the detection will give us such information on neutron stars~\citep{2017PhRvL.119p1101A}.

Associated with gravitational waves, a variety of electromagnetic
signals are likely to be emitted, because an appreciable amount of
mass is likely to be ejected from the system with high velocity
(10--30\% of the speed of light) during the merger and post-merger
phases~\citep[e.g.,][]{2013PhRvD..87b4001H,2013RSPTA.37120272R,2013ApJ...773...78B,2015PhRvD..91f4059S,2016MNRAS.460.3255R,2016PhRvD..93d4019F, 2013MNRAS.435..502F,2009ApJ...690.1681D,2014MNRAS.441.3444M,2014MNRAS.443.3134P,2015MNRAS.448..541J,2015PhRvD..91l4021F}.
One promising scenario for the electromagnetic emission is described
by the so-called macronova/kilonova model~\citep{1998ApJ...507L..59L,2010MNRAS.406.2650M}.  In this
model, a fraction of neutron-rich matter of mass $\sim0.01-0.1M_\odot$
is ejected from the merger event, then the $r$-process
nucleosynthesis proceeds in the ejecta for synthesizing a variety of
neutron-rich heavy nuclei
~\citep{1974ApJ...192L.145L,1982ApL....22..143S,1989Natur.340..126E,1999ApJ...525L.121F,2011ApJ...738L..32G,2014ApJ...789L..39W, 2015ApJ...813....2M, 2015MNRAS.448..541J, 2017MNRAS.472..904L}.
By the continuous radioactive decay of the $r$-process elements, the ejecta is heated up.
For the early phase of its expansion, the ejecta is so dense that it is optically thick to photons.
Thus, the generated heat is consumed by the adiabatic expansion in this phase.
However, when the density of the matter becomes low enough for photons to
diffuse out in a timescale shorter than the expansion timescale, the
ejecta starts shining brightly.  According to the macronova/kilonova model,
the time to reach the peak emission is $\sim 1$--10\,days after the merger, and the
luminosity at the peak is $\sim10^{41}$--$10^{42}$\, ${\rm
erg\ s^{-1}}$ in the optical to infrared bands, depending on the mass, velocity, and opacity of the ejecta~\citep{2010MNRAS.406.2650M}.
Such electromagnetic signals consistent with macronova/kilonova were indeed observed simultaneously with
GW170817~\citep[e.g.,][]{2017ApJ...848L..12A, 2017PASJ...69..102T, 2017Natur.551...64A, 2017ApJ...848L..16S, 2017Sci...358.1556C, 2017Natur.551...67P, 2017Natur.551...75S, 2017Sci...358.1570D, 2017Sci...358.1574S, 2017Sci...358.1559K, 2017ApJ...848L..17C, 2017ApJ...848L..18N}.

For extracting the information of the merger process from the observed
electromagnetic signals, we need reliable theoretical modeling for the mass ejection and resulting electromagnetic
emission.  For this purpose, numerical-relativity simulation taking
into account a variety of physical ingredients, such as neutrino emission
and absorption, angular momentum transport, and equation of state
(EOS) for the neutron star matter is the unique approach.

In this paper, we focus on the evolution of a remnant of a binary
neutron star merger.  As found in our previous
works~\citep[e.g.,][]{2013PhRvD..88d4026H, 2015PhRvD..91f4059S,2016PhRvD..93l4046S}, after
the merger of binary neutron stars with a typical total mass
2.6--$2.8M_\odot$, a massive neutron star (MNS) surrounded by a dense
torus of mass 0.2--0.3 $M_\odot$ is formed as a typical remnant.  Such
remnants are always differentially and rapidly rotating and,
furthermore, are likely to be highly magnetized, possibly in a magnetohydrodynamic (MHD) turbulence state~\citep{2015PhRvD..92f4034K, 2017arXiv171001311K}.
As a result, the turbulent viscosity is likely to arise effectively.
In addition, the remnant is so hot that it emits a huge amount of neutrinos.
Consequently, the neutrino irradiation process would change the electron fraction of the matter in the envelope and ejecta and influence the feature of the observational signals significantly because the electron fraction is closely related to the efficiency for synthesizing lanthanide elements, which are the major players for enhancing the opacity of the ejecta~\citep{2013ApJ...774...25K,2013ApJ...775..113T,2015MNRAS.450.1777K,2018ApJ...852..109T}.

Motivated by this consideration, we perform
radiation-viscous hydrodynamics simulations for a remnant of a binary neutron star merger.
Following \cite{2017ApJ...846..114F}, we first perform a radiation
hydrodynamics simulation in numerical relativity for a merger of
binary neutron stars of equal
mass~\citep{2015PhRvD..91f4059S}.  We then evolve
the remnant MNS surrounded by a torus by radiation-viscous
hydrodynamics simulations in axisymmetric numerical relativity.
We use a general relativistic viscous hydrodynamics code based on a formalism developed by Israel and Stuart~\citep{1979AnPhy.118..341I}.
The assumption of the axisymmetry is justified because the central part of the remnant relaxes to
an approximately axisymmetric state in tens of ms after the onset of the merger for the equal-mass system.
Assuming the axisymmetry, we can save computational costs and follow the long-term evolution of the remnant for more than a few seconds.
We note that we still need about 70,000 CPU hours for simulating one model for 2 s in our axisymmetric simulations (in our computational resources, it takes about 2 months to finish each simulation).

The paper is organized as follows.
In Sec.~2, we briefly describe our method of the simulation.
Then, the results of the simulation are shown in Sec.~3.
In Sec.~4, we discuss the properties of the ejecta and resulting electromagnetic counterparts.
Finally, we summarize this paper in Sec.~5.
Throughout this paper, we employ the geometrical units in which $c=1=G$, where $c$ and $G$ are the speed of light and gravitational constant, respectively.
In some cases, we explicitly use $c$ and $G$ to clarify the physical units.

\section{Method}

We developed a fully general relativistic, neutrino-radiation-viscous hydrodynamics code.
In this code, the neutrino radiation transfer equation and Einstein's equation are solved with the same methods as those in our previous work \citep{2017ApJ...846..114F}, and the viscous effect is incorporated using a simplified version of the Israel--Stewart formalism~\citep{1979AnPhy.118..341I} following \cite{2017PhRvD..95h3005S}.  In this section, we briefly describe the basic equations, particularly, focusing on viscous hydrodynamics.

\subsection{Einstein's Equation}\label{sec2.1}

In our code, Einstein's equation is solved using a version of the Baumgarte--Shapiro--Shibata--Nakamura puncture formalism~\citep{1995PhRvD..52.5428S,1999PhRvD..59b4007B,2006PhRvL..96k1101C,2006PhRvL..96k1102B}.
The quantities which we evolve in our formalism (in the Cartesian coordinate basis) are listed in Table~\ref{tab:var}.
Here, from the spacetime metric $g_{\alpha\beta}$ and the timelike unit vector normal to spatial hypersurfaces $n_\alpha$, we define the induced spatial metric as $\gamma_{\alpha\beta} \equiv g_{\alpha\beta} + n_\alpha n_\beta$.
In addition, we define the determinant of the induced metric as $\gamma = \det(\gamma_{ij})$ and the extrinsic curvature as $K_{\alpha\beta} = - (1/2) {\cal L}_n \gamma_{\alpha\beta}$, where ${\cal L}_n$ is the Lie derivative with respect to $n^\alpha$.
For the gauge conditions, we employ dynamical lapse and shift gauge conditions that are the same as Equations~(1) and (2) in \cite{2017ApJ...846..114F}.
We adopt the so-called cartoon method~\citep{2001IJMPD..10..273A, 2003ApJ...595..992S} to impose axisymmetric conditions for the geometric quantities.

\begin{table}[t]
\caption{ List of the Quantities that We Evolve in Our Baumgarte--Shapiro--Shibata--Nakamura Puncture Formulation.}
\begin{center}
\begin{tabular}{ll}
\hline \hline
Notation & Definition\\
\hline
$\tilde{\gamma}_{ij} = \gamma^{-1/3} \gamma_{ij}$ & Conformal three-metric\\
$W=\gamma^{-1/6}$ & Conformal factor\\
$K=\gamma^{ij}K_{ij}$ & Trace of the extrinsic curvature $K_{ij}$\\
$\tilde{A}_{ij}=\gamma^{-1/3} (K_{ij}-\gamma_{ij} K/3)$ 
& Trace-free part of $K_{ij}$\\
$F_{i} = \delta^{jk} \del_j \tilde{\gamma}_{ki}$& Auxiliary variable\\
\hline
\end{tabular}
\end{center}
\label{tab:var}
\end{table}

\subsection{Radiation-viscous Hydrodynamics Equations}\label{sec2.2}

As in \cite{2017ApJ...846..114F}, we decompose neutrinos into ``streaming" and ``trapped" components and write the total energy-momentum tensor of the matter (fluid and neutrinos) as
\begin{align}
T^{\alpha\beta}_{\rm (tot)} = T^{\alpha\beta} + \sum_i T_{(\nu_i, {\rm S})}^{\alpha\beta},
\end{align}
where $T^{\alpha\beta} = T_{\rm (fluid)}^{\alpha\beta}+\sum_i T_{(\nu_i, {\rm T})}^{\alpha\beta}$ is the energy-momentum tensor composed of the sum of the fluid and trapped neutrinos, and  $T_{(\nu_i, {\rm S})}^{\alpha\beta}$ denotes the energy-momentum  tensor for free-streaming neutrinos. 
These energy-momentum tensors obey the following equations:
\begin{align}
\nabla _\beta T^{\alpha\beta} &= -Q_{\rm (leak)}^\alpha = - \sum_i
Q^\alpha_{{\rm (leak)} \nu_i},\label{eq:fluid}\\ \nabla _\beta
T_{(\nu_i, {\rm S})}^{\alpha\beta} &= Q^\alpha_{{\rm (leak)}\nu_i}, \label{eq:rad}
\end{align}
where $Q^\alpha_{{\rm (leak)}\nu_i}$ denotes the leakage rate of $i$th species of neutrinos. For solving the evolution equations for free-streaming neutrinos, we employ the so-called M1 scheme with a closure relation \citep{2011PThPh.125.1255S}, and, for trapped neutrinos, we employ a leakage-based scheme developed by \cite{2010PThPh.124..331S}.
The detailed description of these schemes is found in \cite{2010PThPh.124..331S} and \cite{2017ApJ...846..114F}.

Then, we describe the basic equations for viscous hydrodynamics.
The formulation of our general relativistic viscous hydrodynamics is described in~\cite{2017PhRvD..95h3005S} and \cite{2017PhRvD..95l3003S}.
In the following, we briefly outline our formulation.

The energy-momentum tensor of the viscous fluid with trapped neutrinos is given as
\begin{align}
T_{\alpha\beta} = \rho h u_\alpha u_\beta + Pg_{\alpha\beta} - \rho h
\nu \tau^0{}_{\alpha\beta},
\end{align}
where $\rho$ is the baryon rest-mass density, $h=1+\epsilon + P/\rho$ is the specific enthalpy, $\epsilon$ is the specific internal energy, $u^\alpha$ is the fluid four-velocity, $P$ is the pressure, $\nu$ is the viscosity coefficient, and $\tau^0{}_{\alpha\beta}$ is the viscous stress tensor.
Here $\tau^0{}_{\alpha\beta}$ is a symmetric tensor that satisfies the relation $\tau^0{}_{\alpha\beta}u^\alpha=0$, and, in the
formalism of \cite{1979AnPhy.118..341I}, it obeys the following evolution equation
\begin{align}
{\cal L}_u \tau^0{}_{\alpha\beta} = -\zeta(\tau^0{}_{\alpha\beta} -
\sigma_{\alpha\beta}), \label{eq:tau0}
\end{align}
where $\sigma_{\alpha\beta}$ is the shear tenor defined by
\begin{align}
\sigma_{\alpha\beta} = h_\alpha{}^\mu h_\beta{}^\nu (\nabla_\mu u_\nu + \nabla_\nu u_\mu) = {\cal L}_u h_{\alpha\beta},
\end{align}
and $h_{\alpha\beta} = g_{\alpha\beta} + u_\alpha u_\beta$.
Here $\zeta$ is a nonzero constant of (time)$^{-1}$ dimension, and it has to be chosen in an appropriate manner for $\tau^0{}_{\alpha\beta}$ to approach $\sigma_{\alpha\beta}$ in a short timescale because it is reasonable to suppose that $\tau^0{}_{\alpha\beta}$ should approach $\sigma_{\alpha\beta}$ in a microscopic timescale.

We can rewrite Eq.~\eqref{eq:tau0} as
\begin{align}
{\cal L}_u \tau_{\alpha\beta} = -\zeta \tau^0{}_{\alpha\beta}, \label{eq:tau}
\end{align}
where $\tau_{\alpha\beta} \equiv \tau^0{}_{\alpha\beta} -\zeta h_{\alpha\beta}$.
Thus, in addition to hydrodynamics equations, we solve Eq.~\eqref{eq:tau} as a basic equation that describes the evolution of $\tau_{\alpha\beta}$.

The energy-momentum tensor of the fluid is rewritten as
\begin{align}
T_{\alpha\beta} &= \rho h (1- \nu\zeta) u_\alpha u_\beta + (P - \rho h
\nu\zeta )g_{\alpha\beta} - \rho h \nu \tau_{\alpha\beta}\notag \\ &=
\rho w \hat{e} n_\alpha n_\beta + J_\alpha n_\beta + J_\beta n_\alpha
+ S_{\alpha\beta}, \label{eq:tmunu}
\end{align}
where we defined
\begin{align}
&\hat{e} \equiv T_{\alpha\beta} n^\alpha n^\beta/\rho w \notag \\ 
&~~= hw (1-\nu\zeta) - \frac{P-\rho h \nu\zeta}{\rho w} - h \nu
w^{-3}\tau_{ij} V^i V^j,
\end{align}
\begin{align}
&J_i \equiv -T_{\alpha\beta} n^\alpha \gamma_i{}^\beta \notag\\ 
&~~~=\rho h w u_i (1-\nu\zeta) - \rho h \nu
w^{-1}\tau_{ij}V^j,\\
&S_{ij} \equiv T_{\alpha\beta}\gamma^\alpha{}_i \gamma^\beta{}_j\notag\\ 
&~~~~= \rho h (1-\nu\zeta)u_i u_j + (P-\rho h \nu\zeta)\gamma_{ij} -
\rho h \nu \tau_{ij}. 
\end{align}
Here $w\equiv-n_\alpha u^\alpha$ (the Lorentz factor of the fluid), and $V^i \equiv
\gamma ^{ij}u_j$.  The general relativistic Navier--Stokes and energy
equations are derived, respectively, from the space-like and time-like
components of Eq.~\eqref{eq:fluid}, i.e., $\gamma_{i \alpha}
\nabla_\beta T^{\alpha \beta} = -\gamma_{i \alpha} Q_{\rm
  (leak)}^\alpha$ and $n_\alpha \nabla_\beta T^{\alpha \beta} =
-n_\alpha Q_{\rm (leak)}^\alpha$, as
\begin{widetext}
\begin{align}
\del_t(\sqrt{\gamma} J_i) + \del_k [\sqrt{\gamma}(\alpha S^k{}_i
  -\beta^kJ_i)] &= \sqrt{\gamma}\biggl(-\rho_*\hat{e}\del_i \alpha +
J_k \del_i \beta^k -\frac{\alpha}{2}S_{kl}\del_i \gamma^{kl}
-\gamma_{i \alpha} Q_{\rm (leak)}^\alpha \biggr), \label{eq12}\\ 
\del_t(\rho_* \hat{e}) + \del_k [\sqrt{\gamma}(\alpha J^k
  -\beta^k\rho_*\hat{e})] &= \sqrt{\gamma}\biggl(S_{kl} K^{kl} -
\gamma^{kl} J_k \del_l \alpha + n_\alpha Q_{\rm (leak)}^\alpha
\biggr),\label{eq13}
\end{align}
where $\rho_*\equiv\rho w \sqrt{\gamma}$, and $\alpha$ and $\beta^k$ denote
the lapse function and shift vector, respectively.  On the other hand,
the evolution equation of $\tau_{ij}$ is derived from
Eq.~\eqref{eq:tau} as
\begin{align}
\del_t(\rho_* \tau_{ij}) + \del_k(\rho_* \tau_{ij} v^k) &=
-\rho_*\bigl(\tau_{ik} \del_j v^k + \tau_{jk} \del_i v^k\bigr) -
\rho_* \alpha w^{-1} \zeta \tau^0{}_{ij},
\end{align}
\end{widetext}
where we used the equation of continuity $\del_t\rho_* + \del_k
(\rho_* v^k)=0$ and $v^k\equiv u^k/u^t$.  We solve these equations in
cylindrical coordinates as in \cite{2017PhRvD..95h3005S}.

\subsection{Microphysics}

For the neutron star EOS, we employ a tabulated EOS referred to as DD2
\citep{2014ApJS..214...22B}.  We extend this EOS to low-temperature
and low-density ranges down to $10^{-3}$\,MeV and $\approx 0.1\,{\rm g\,cm^{-3}}$ using an EOS by \cite{2000ApJS..126..501T} as
in our previous work \citep{2017ApJ...846..114F}.  This extended DD2
EOS includes contributions of nucleons, heavy nuclei, electrons,
positrons, and photons to the pressure and internal energy.

In our formalism \citep{2010PThPh.124..331S, 2017ApJ...846..114F}, we
solve the equation for the energy-momentum density of streaming neutrinos and number density of electrons and trapped neutrinos.
For the source terms of them, we take into account the relevant processes due to the weak interaction.  For (anti)neutrino production processes, we adopt the electron and positron capture of nucleons and heavy nuclei, electron--positron pair annihilation, nucleon--nucleon bremsstrahlung, and plasmon decay based on \cite{1985ApJ...293....1F}, \cite{1986ApJ...309..653C}, \cite{2006NuPhA.777..356B}, and \cite{1996A&A...311..532R}, respectively.  
On the other hand, for (anti)neutrino absorption processes, we adopt the absorption of
neutrinos and antineutrinos on nucleons and heavy nuclei and
neutrino--antineutrino pair annihilation to electron--positron pairs.
The detailed description of calculating the rates is also found in
\cite{2017ApJ...846..114F}.

Since we use the energy-integrated formalism for neutrino radiation transfer, we have to assume the average energy of neutrinos for calculating the rates of neutrino absorption and neutrino pair-annihilation processes, which depend strongly on the energy of neutrinos.
In \cite{2017ApJ...846..114F}, we found that the results of the simulation, such as the ejecta mass and kinetic energy, depend weakly on the assumption for the average neutrino energy.  In this work, we
adopt the average energy estimated by Eq.~(41) in
\cite{2017ApJ...846..114F} because it can be derived directly from the
energy integration of the energy-dependent neutrino absorption heating
rate.

\subsection{Prescription of the Viscosity Coefficient}\label{sec:vis}

In this work, we model the dynamical shear viscosity coefficient following \cite{1973A&A....24..337S} as
\begin{align}
\nu = \alpha_{\rm vis} c_s H_{\rm tur}, \label{eq:visc}
\end{align}
where $\alpha_{\rm vis}$ is the so-called $\alpha$-viscosity parameter, $c_s$ is the sound speed, and $H_{\rm tur}$ is the possible largest scale of eddies in a turbulence state. In this work, we set $H_{\rm tur} =10$\,km, because it should be approximately equal to the size of the MNS (for the central region of the system).

As already mentioned in the previous subsection, $\zeta^{-1}$ has to be a timescale short enough that $\tau^0{}_{\alpha\beta}$ quickly approaches $\sigma_{\alpha\beta}$.
In this work, we set $\zeta=3\times 10^4\,{\rm s^{-1}}$.
This value is about 4 times larger than the maximum angular velocity of the system ($\Omega \sim 8000\,{\rm rad\,s^{-1}}$); thus, the requirement for $\zeta$ is reasonably satisfied. 

We note that for the outer part of the torus, the viscosity coefficient would be underestimated because the scale height of the torus increases with the radius in general.
Thus, in our prescription, we conservatively take into account the effect of the viscosity for large radii.

\begin{table}[t]
\caption{
List of the Simulation Models.
}
\begin{center}
\begin{tabular*}{\hsize}{@{\extracolsep{\fill}}llcccc}
\hline \hline
Model && $\alpha_{\rm vis}$ & $\Delta x_0$& $N$ & $L$\\
&&& (m)& & (km) \\
\hline
DD2-135135-0.00-H&                            & 0.00  & 150 & 951 & 5440\\
DD2-135135-0.01-H &                          & 0.01  & 150 & 951 & 5440\\
DD2-135135-0.02-H &(fiducial,high)     & 0.02  & 150  & 951 & 5440\\
DD2-135135-0.04-H &                           & 0.04  & 150 & 951 & 5440\\
\hline
DD2-135135-0.01-M &(medium)           & 0.01 & 200  & 871 & 5790 \\
DD2-135135-0.01-L & (low)                   & 0.01 & 250 & 809 & 5690 \\
\hline
\end{tabular*}
{\bf Note.} Here $\Delta x_0$, $N$, and $L$ are the grid size of inner region, grid number, and size of the computational domain, respectively (see the text in Sec.~\ref{sec2.4}).
For all the models, DD2 EOS is employed, and the gravitational mass of each neutron star in a binary is $1.35M_\odot$.
\end{center}
\label{tab:models}
\end{table}

\subsection{Initial Condition and Grid Setting}\label{sec2.4}

The initial condition in this work is given by the same method as that in \cite{2017ApJ...846..114F}: we first perform a three-dimensional, numerical-relativity simulation for the merger of equal-mass binary neutron stars with the total mass $2.7M_\odot$~\citep{2015PhRvD..91f4059S} using DD2 EOS~\citep{2014ApJS..214...22B}.
For these choices of the total mass and the EOS, the remnant formed after the merger is a long-lived MNS surrounded by a torus. At a few tens of ms after the formation of this system, the MNS and torus relax approximately to an axisymmetric configuration; hence, we employ such a system as the initial condition.
The total baryon and Arnowitt--Deser--Misner (ADM) mass of the initial condition are $\approx 2.95M_\odot$ and $2.65M_\odot$, respectively.
We note that $\approx 0.05M_\odot$ is carried by gravitational radiation during the inspiral and merger.
The baryon mass of the torus of the initial condition is $\approx 0.2 M_\odot$.
Here we define the torus mass by
\begin{align}
M_{\rm b,torus} &= \int_{\rho<10^{13}{\rm g\,cm^{-3}}} d^3x\ \rho_*.
\label{eq:bmasstorus}
\end{align}
We note that the mass of the merger-remnant torus also depends on the total mass, mass ratio of the binary, and the neutron star EOS.
Therefore, the result in this work would be quantitatively different for different parameters and EOSs.

In addition to the dynamical variables employed in
\cite{2017ApJ...846..114F}, we need to prepare an initial condition
for the spatial components of the viscous tensor $\tau_{ij}$.  For
simplicity, we assume that the viscous tensor $\tau^0{}_{\alpha\beta}$
vanishes in the initial state.  That is, as the initial condition for
$\tau_{ij}$, we set $\tau_{ij}=-\zeta h_{ij}$.

For evolving radiation-viscous hydrodynamics equations in cylindrical coordinates $(x, z)$, we employ the same nonuniform grid as that in \cite{2017ApJ...846..114F}, in which the grid structure is determined
by the uniform grid spacing of the inner region, $\Delta x_0$; the range of
the inner region, $R_{\rm star}$; the increase rate of the
grid spacing in the outer region, $1+\delta$; and the grid number, $N$.
For all the models in this paper, we set $R_{\rm star}=30$ km and
$\delta=0.0075$.  
We list $\Delta x_0$ and $N$, together with the size of the computational domain, $L$, are listed in Table~\ref{tab:models}.
We performed simulations with three different viscosity parameters, $\alpha_{\rm vis}=$ 0.01, 0.02, and 0.04, among which we refer to the model with $\alpha_{\rm vis} = $ 0.02 as our fiducial model.
Our choice of the viscosity parameter $\alpha_{\rm vis}=O(0.01)$ is justified, at least for the outer part of the MNS ($\rho \lesssim 10^{14}\,{\rm g\,cm^{-3}}$), by the recent high-resolution global MHD simulations for a binary neutron star merger~\citep{2017arXiv171001311K}.
For a higher-density region, such as an inner region of the MNS, the value of the viscosity parameter is still uncertain, but for simplicity, we use the same value of the viscosity parameter for the entire region, supposing that a turbulence state would be achieved entirely in the MNS\footnote{
We note that for our choice of the parameters in Israel--Stewart formalism, $\zeta$ and $\nu=\alpha \rho h$, the causality is satisfied at least locally.
Indeed, we do not find any pathological behavior of the fluid in our simulations, so that we conclude that the causality is satisfied globally in our simulation.
}.

To investigate the long-term evolution of the system, the simulations are performed for $\sim 3$\,s for our fiducial and $\alpha_{\rm vis}=0.04$ models, while the simulations of the other models are performed for a shorter time ($\alt 1$\,s).

In addition to these models, to confirm that numerical results with different grid resolutions reasonably agree, we performed simulations for $\alpha_{\rm vis}=0.01$ with two lower resolutions as $\Delta x_0=200$\,m (M) and 250\,m (L).
The grid structure for these cases is also listed in Table~\ref{tab:models}.
The result of this check is described in Appendix~A.

\subsection{Rates of Viscous Heating and Angular Momentum Transport in Our Formalism}\label{sec2.5}
From the component of Eq.~\eqref{eq:fluid} along $u^\alpha$, we obtain
\begin{align}
-Q_{\rm (leak)}^\alpha u_\alpha &= - \rho u^\alpha \nabla_\alpha h +
u^\alpha \nabla_\alpha P - u^\alpha \nabla^\beta (\rho h \nu
\tau^0{}_{\alpha\beta} ) \notag \\
&= - \rho T u^\alpha \nabla_\alpha s + \rho h \nu \tau^0{}_{\alpha\beta}
\nabla^\alpha u^\beta,
\end{align}
where we used the relation $\tau^0{}_{\alpha\beta} u^\alpha = 0$ and
the first law of thermodynamics $dh = dP/\rho + Tds$ with $T$ and $s$
being the temperature and specific entropy.  From the above relation,
the viscous heating rate in the fluid rest frame in our formalism can
be written approximately as
\begin{align}
Q_{\rm vis}^{(+)} = \rho h \nu \tau^0{}^{\alpha\beta} \nabla_\alpha
u_\beta = \frac{1}{2} \rho h \nu
\tau^0{}^{\alpha\beta}\tau^0{}_{\alpha\beta},
\end{align}
where we assumed that $\tau^0{}_{\alpha\beta}$ quickly approaches 
$\sigma_{\alpha\beta}$.

Using the $y$ component of Eq.~(\ref{eq12}), the angular
momentum transport flux due to the viscous effect is described by
\begin{align}
- \rho h \nu \bar{\tau}^0{}^{i}{}_y = - \rho h \nu \gamma^{ik}
\tau^0{}_{ky}.
\end{align}
From the surface integral of this, the local angular momentum
transport rate is derived.

\subsection{Timescales for Viscous Evolution}\label{sec2.6}

Before describing the numerical results, we estimate the timescales for the
evolution of the MNS-torus system in the presence of viscosity.
In this problem, the viscous effect plays two crucial roles. 

In the first $\lesssim 20$\,ms, the viscous effect plays an important role in the MNS.
The MNS formed after the merger initially has a differentially rotating velocity profile. By the viscous effect, the angular momentum in a high angular velocity region is transported to a low angular velocity region; thus, the MNS approaches a rigidly rotating state.
The timescale for the angular momentum transport in the MNS is estimated by
\begin{align}
\frac{R_{\rm eq}{}^2}{\nu} \approx& ~10\,{\rm ms}
\biggl(\frac{\alpha_{\rm vis}}{0.02}\biggr)^{-1} 
\biggl(\frac{c_s}{c/3}\biggr)^{-1} \nonumber \\
&~~~~~~\times \biggl(\frac{R_{\rm eq}}{15\,{\rm km}}\biggr)^2 
\biggl(\frac{H_{\rm tur}}{10\,{\rm km}}\biggr)^{-1}, \label{eq:mnsvis}
\end{align}
where $R_{\rm eq}$ denotes the equatorial radius of the MNS.  Thus, the
differential rotation profile of the MNS is expected to be erased
within $\sim 10 (\alpha_{\rm vis}/0.02)^{-1}$\,ms by the viscous 
effect. 

For longer timescales, the viscous effect plays a primary role in determining the evolution of the torus surrounding the MNS.
Assuming that the torus evolution could be described by a standard accretion disk theory \citep{1973A&A....24..337S}, the viscous accretion timescale is estimated by
\begin{align}
t_{\rm vis} \sim&\ \alpha_{\rm vis}{}^{-1} \biggl(\frac{R_{\rm
    torus}{}^3}{GM_{\rm MNS}}\biggr)^{1/2}
\biggl(\frac{H}{R_{\rm torus}}\biggr)^{-2}\notag \\ 
\approx &\ 0.55\,{\rm
  s}\ \biggl(\frac{\alpha_{\rm vis}}{0.01}\biggr)^{-1} \biggl(\frac{M_{\rm
    MNS}}{2.5\ M_\odot}\biggr)^{-1/2}\notag \\ 
&~~~~~~\times \biggl(\frac{R_{\rm torus}}{50\,{\rm km}}\biggr)^{3/2}
\biggl(\frac{H/R_{\rm torus}}{1/3} \biggr)^{-2}, \label{eq:tvis}
\end{align}
where $M_{\rm MNS}$ is the gravitational mass of the MNS, $R_{\rm
  torus}$ is the typical radius of the torus, and $H/R_{\rm torus}$ is
the aspect ratio of the torus.  Then, the mass accretion rate is
estimated as
\begin{align}
\dot{M}_{\rm MNS}\sim& \frac{M_{\rm torus}}{t_{\rm
    vis}}\notag\\ 
\approx &\ 0.36\ M_\odot\,{\rm
  s^{-1}}\ \biggl(\frac{\alpha_{\rm vis}}{0.01}\biggr) \biggl(\frac{M_{\rm
    MNS}}{2.5\ M_\odot}\biggr)^{1/2}\notag \\ &\times
\biggl(\frac{R_{\rm torus}}{50\,{\rm km}}\biggr)^{-3/2}
\biggl(\frac{H/R_{\rm torus}}{1/3} \biggr)^2 \biggl(\frac{M_{\rm
    torus}}{0.2\ M_\odot} \biggr). \label{eq:mdestimate}
\end{align}
We note that in these estimates, we used the standard accretion disk in a stationary state as a model.
In reality, the MNS-torus system is evolved by the viscous timescale (i.e., the values of $R_{\rm torus}$ and  $M_{\rm torus}$ change with time); hence, we should check whether the stationary disk model is valid or not by performing the simulation of this system.

\section{Result}
\subsection{Dynamics of the System}\label{sec3.1}

First, we briefly describe the evolution process of the system for the fiducial model DD2-135135-0.02-H.

\begin{figure}[t]
\begin{center}
\includegraphics[width=\hsize]{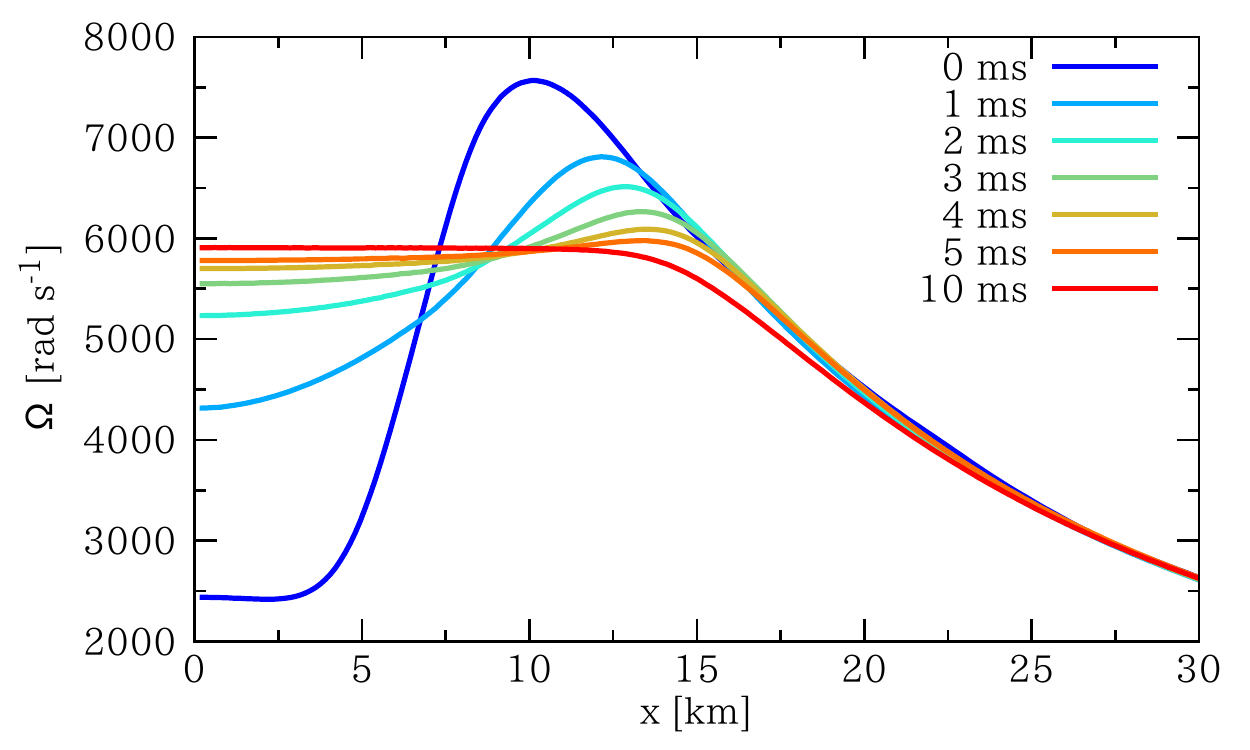}
\caption{ Angular velocity on the equatorial plane at 
  $t=$ 0, 1, 2, 3, 4, 5, and 10 ms for the fiducial model
  DD2-135135-0.02-H ($\alpha_{\rm vis}=0.02$).  }
\label{fig:eqomega}
\end{center}
\end{figure}

In the early stage of the evolution, the angular momentum transport occurs in the MNS.
Figure \ref{fig:eqomega} shows the angular velocity profiles in the equatorial plane at different time slices for the fiducial model.
This shows that initial differential rotation of the MNS within the radius of $\sim 12$\,km disappears and the MNS settles into a rigid rotation state in $\sim 10$\,ms as estimated by Eq.~\eqref{eq:mnsvis}.

Then, a sound wave is formed in the vicinity of the MNS.  This is due
to the variation of the MNS density profile caused by the angular momentum redistribution.

\begin{figure*}[t]
\begin{minipage}{0.5\hsize}
\includegraphics[bb=0 50 1000 500,width=\hsize]{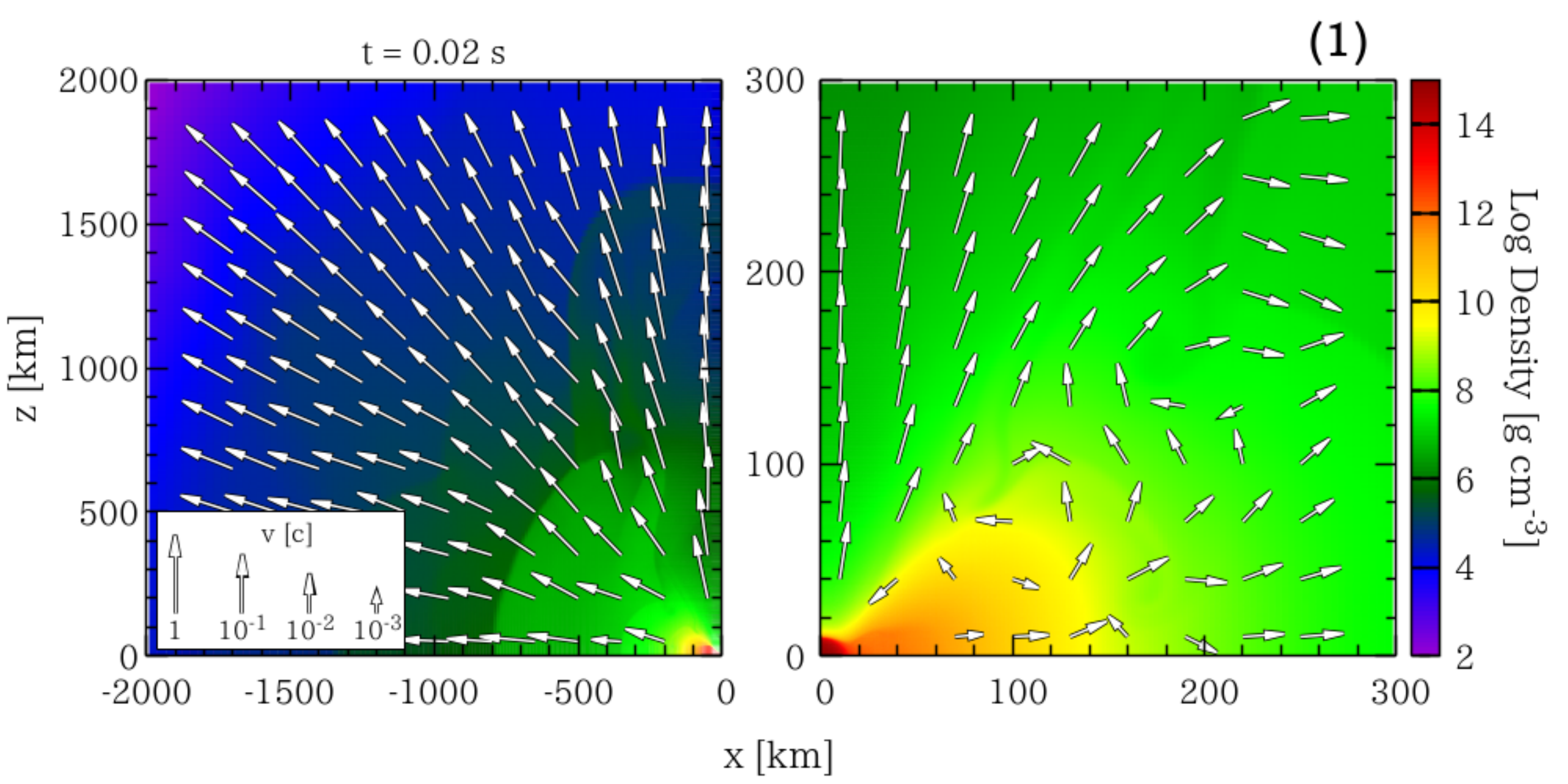}
\includegraphics[bb=0 50 1000 500,width=\hsize]{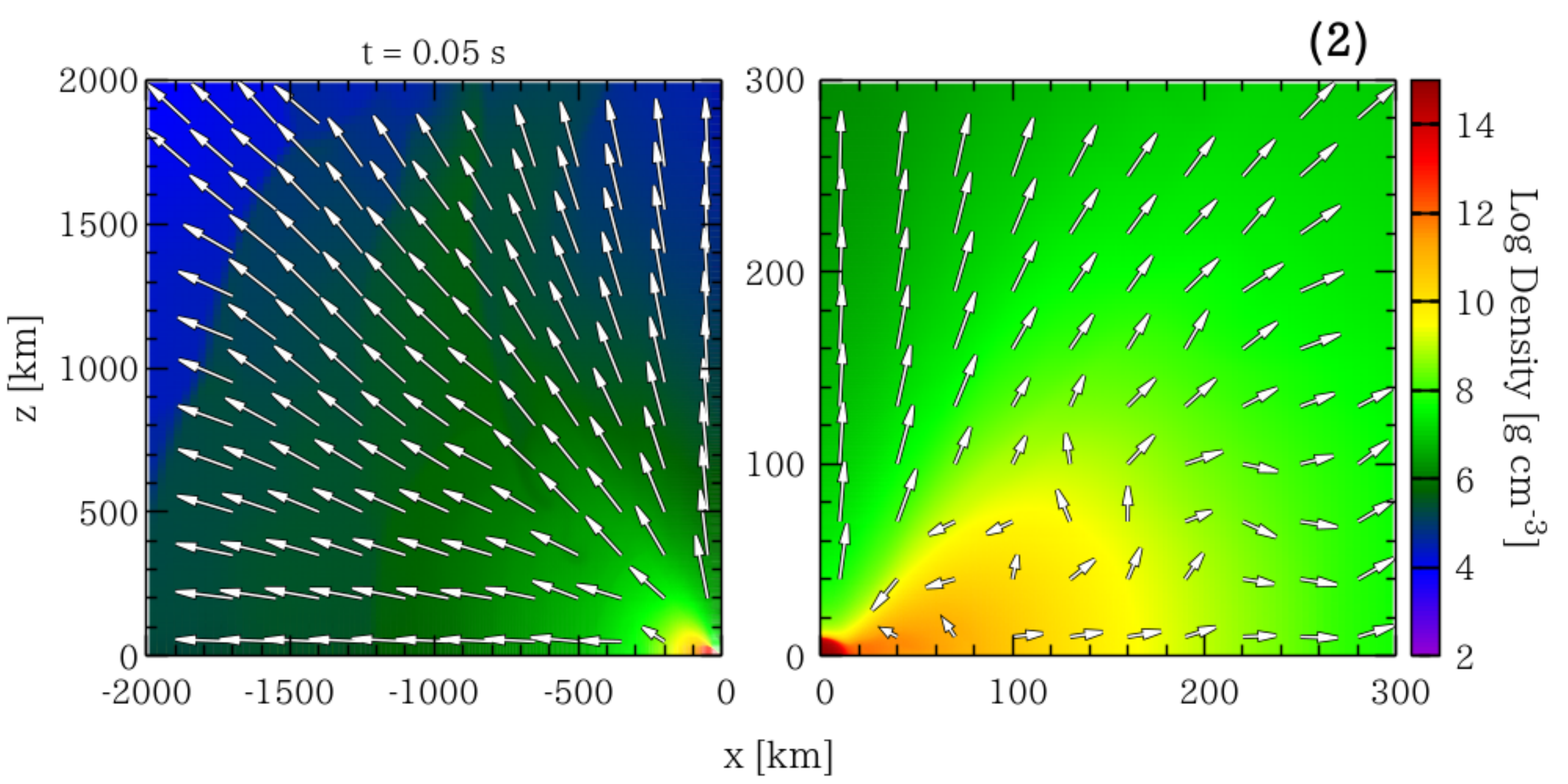}
\includegraphics[bb=0 50 1000 500,width=\hsize]{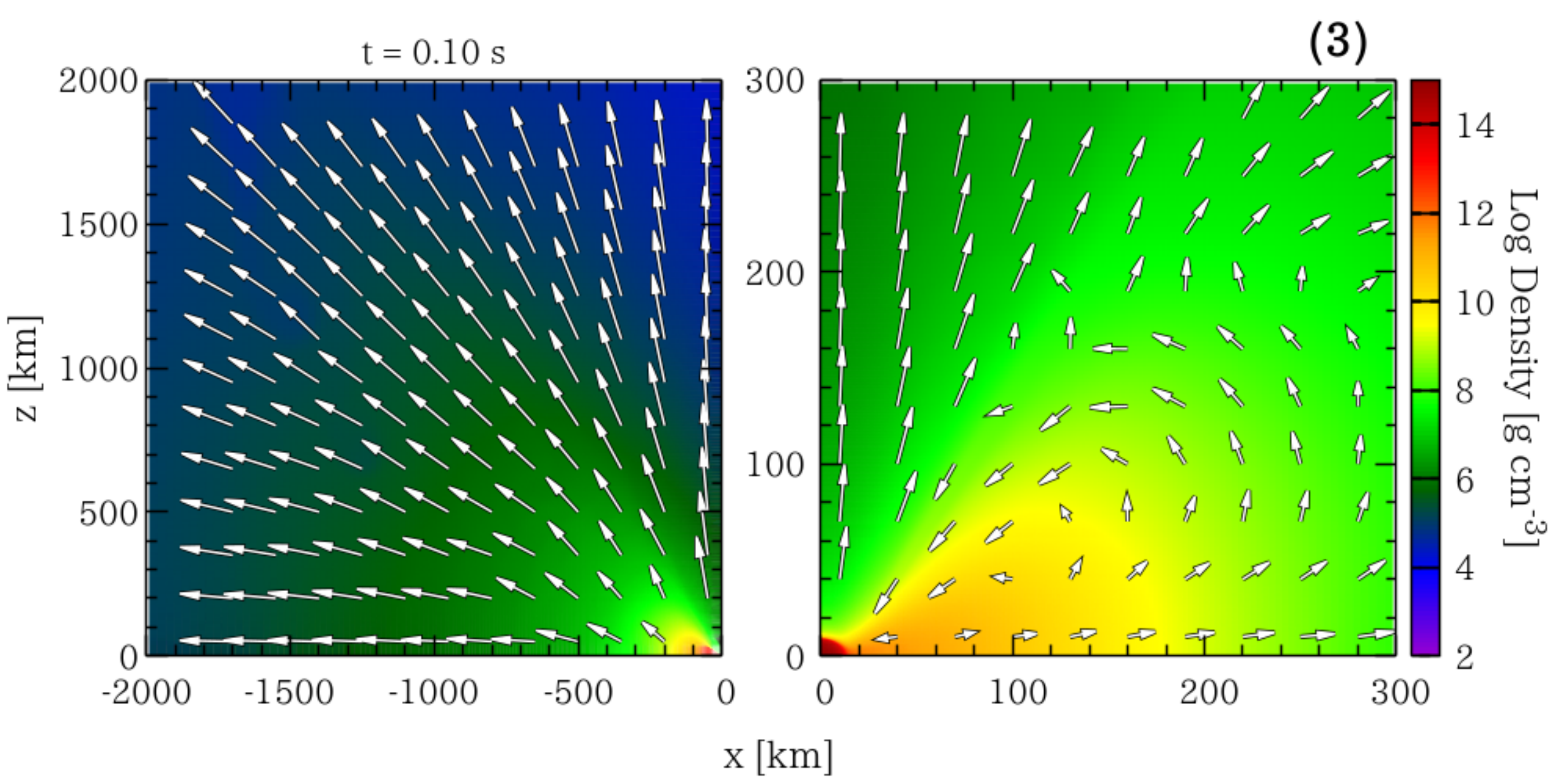}
\includegraphics[bb=0   0 1000 500,width=\hsize]{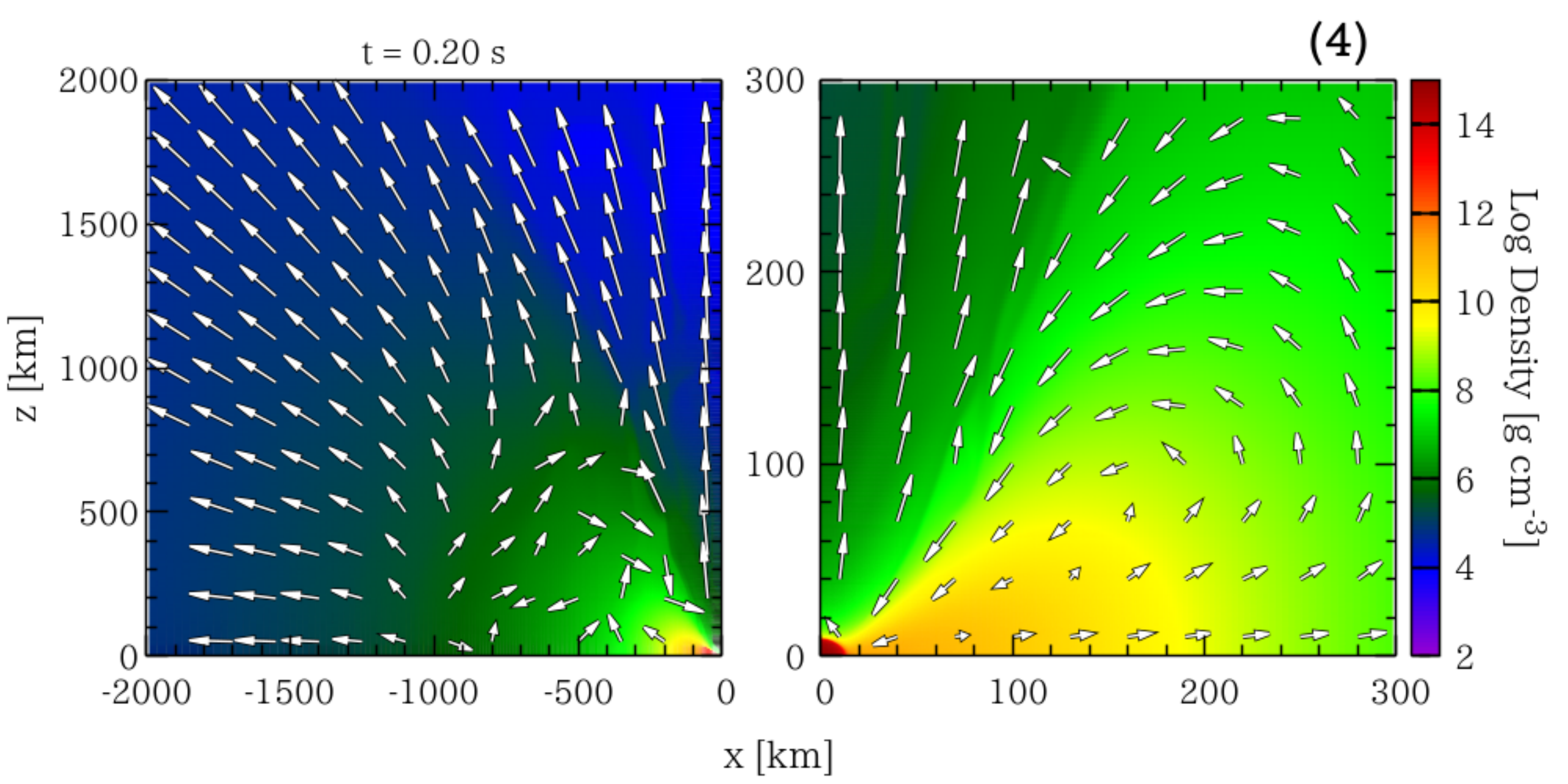}
\end{minipage}
\begin{minipage}{0.5\hsize}
\includegraphics[bb=0 50 1000 500,width=\hsize]{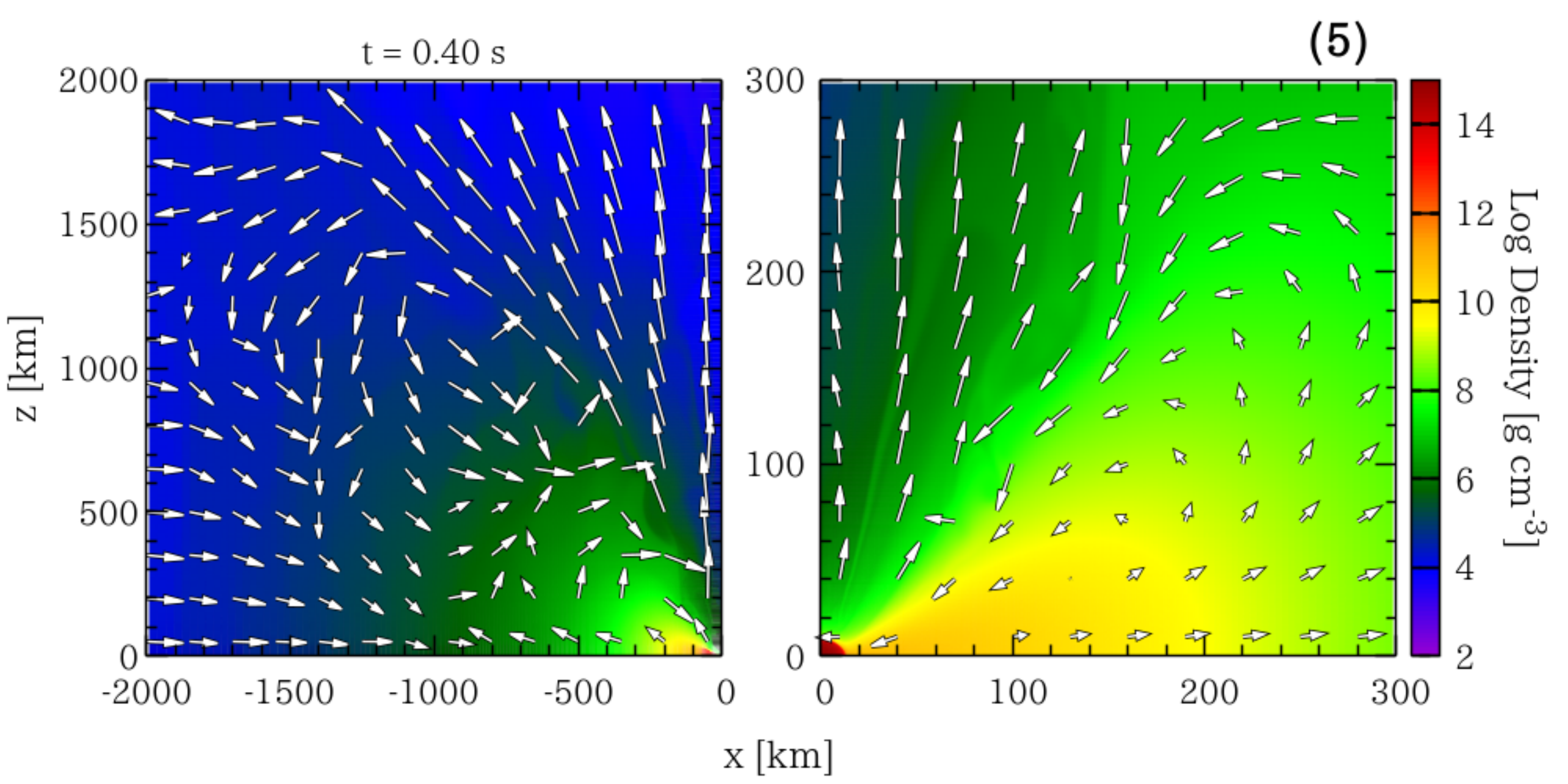}
\includegraphics[bb=0 50 1000 500,width=\hsize]{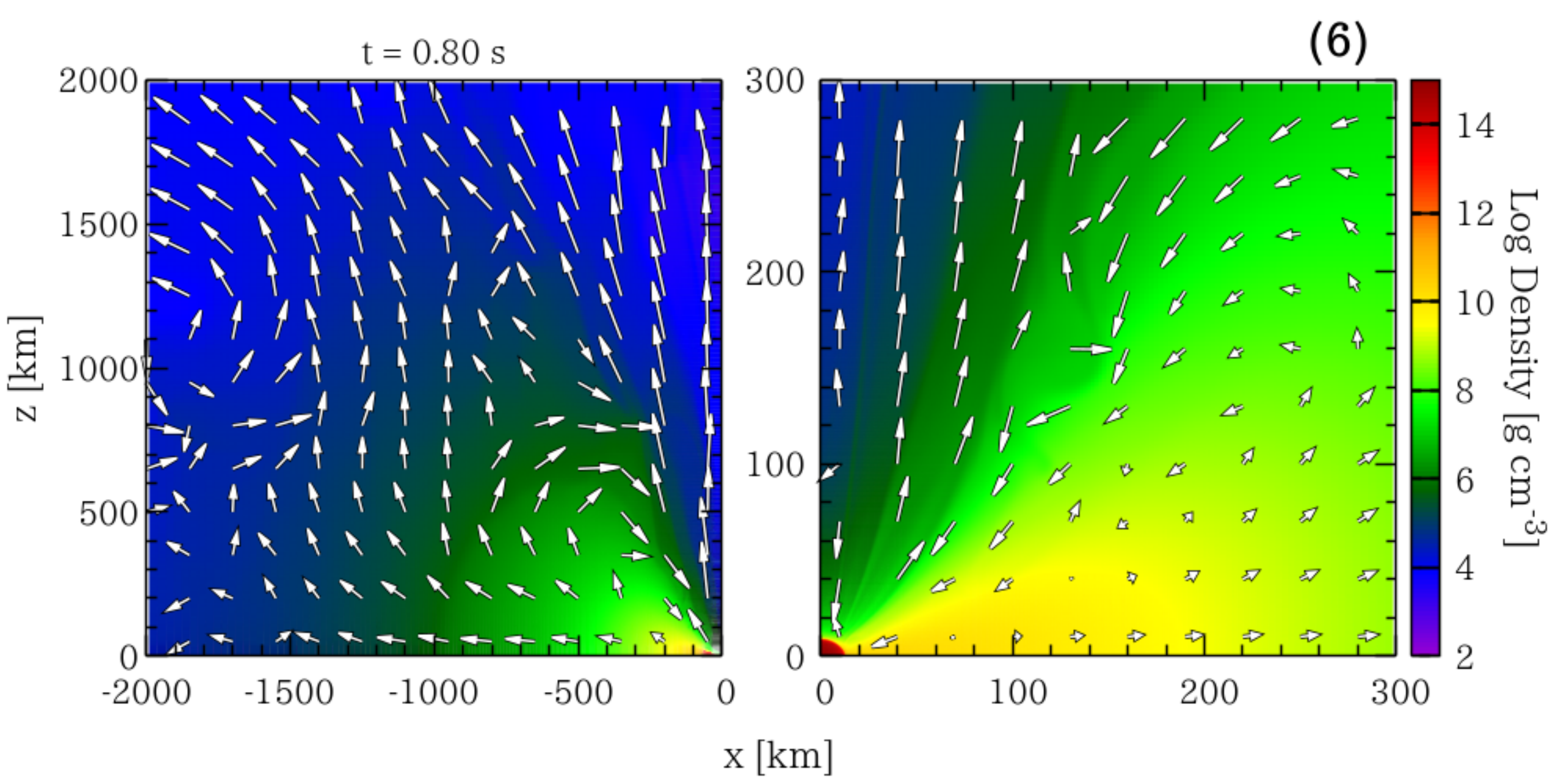}
\includegraphics[bb=0 50 1000 500,width=\hsize]{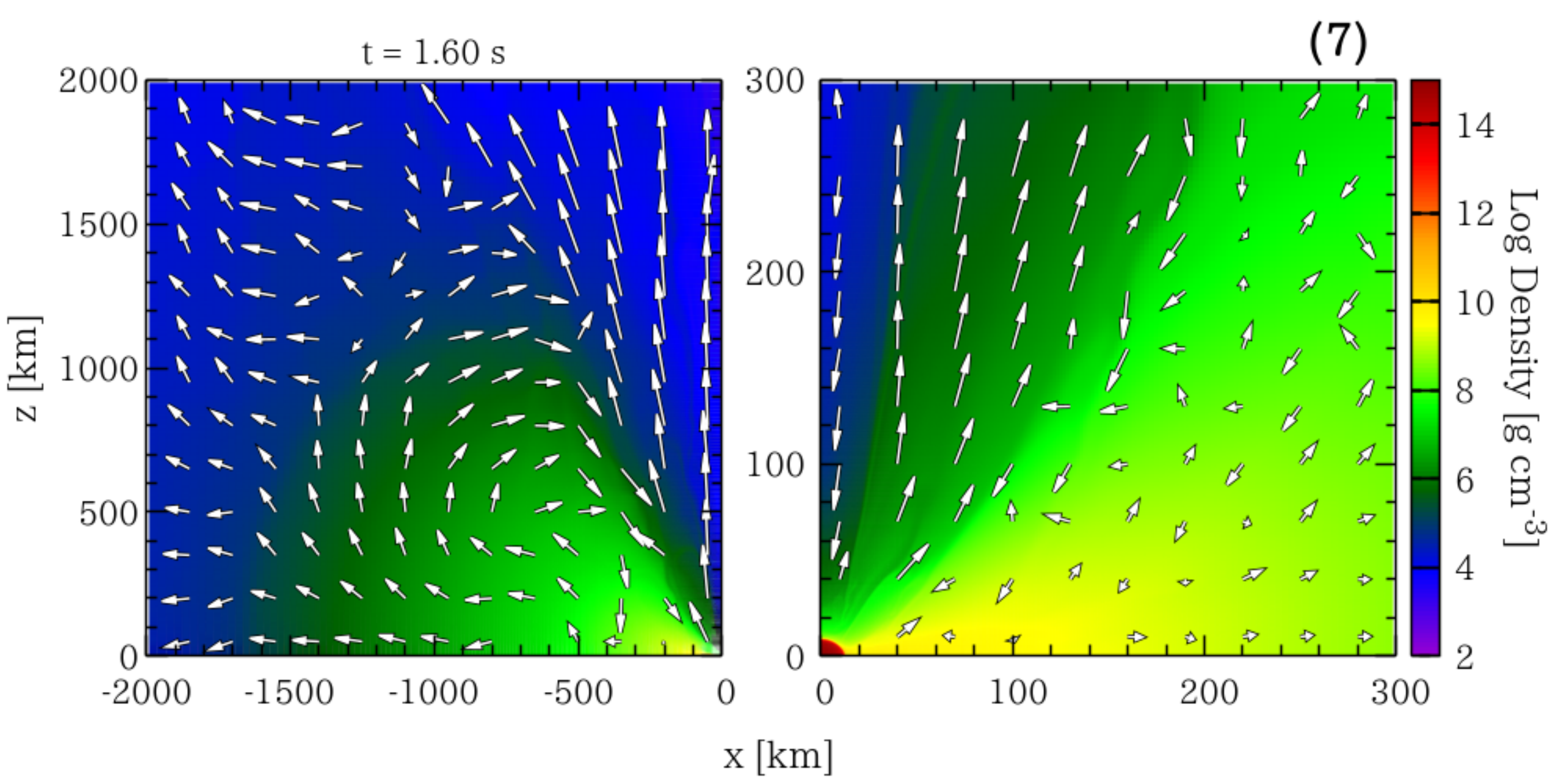}
\includegraphics[bb=0   0 1000 500,width=\hsize]{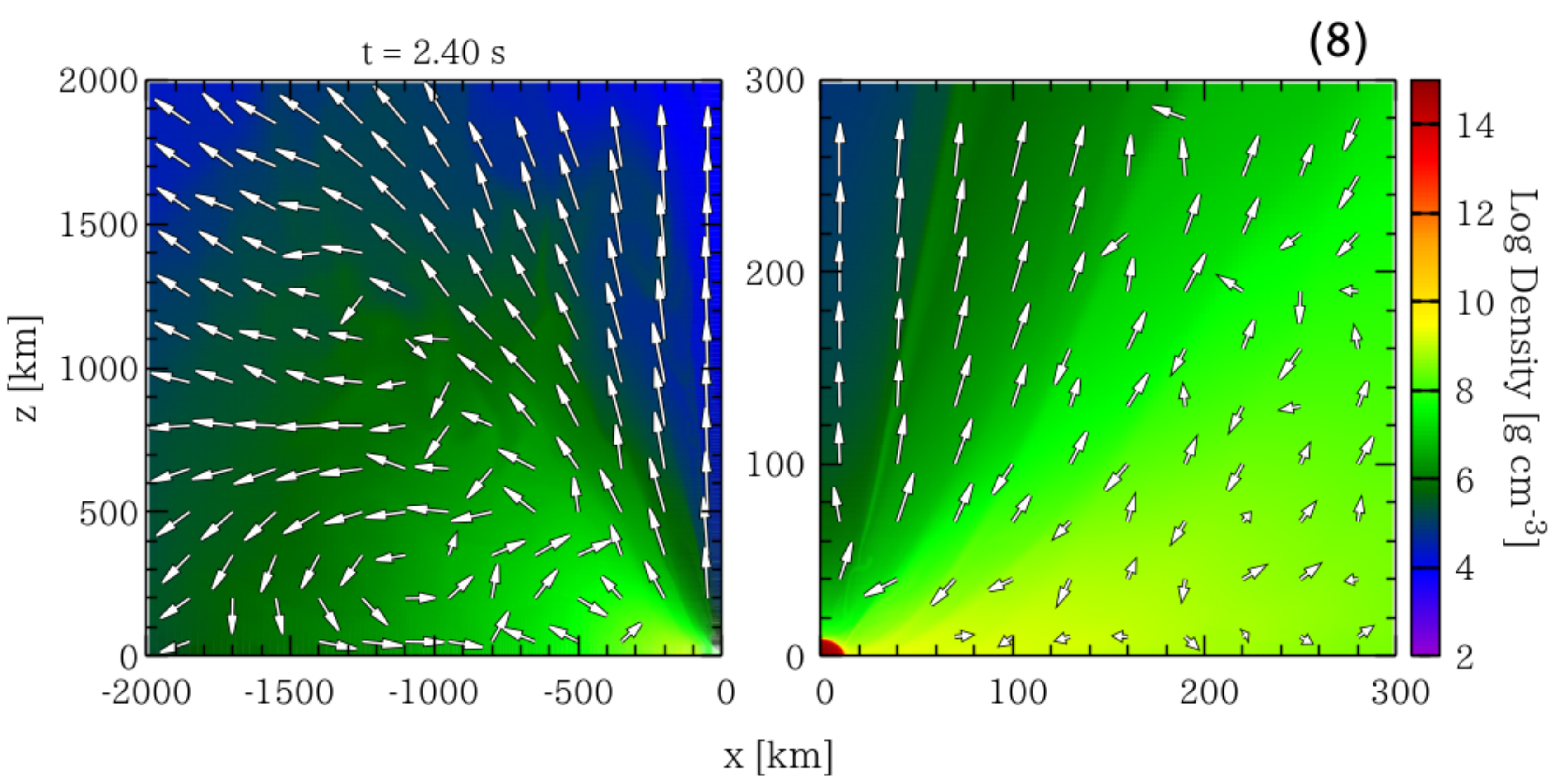}
\end{minipage}
\caption{ Snapshots of the density and poloidal velocity field for the fiducial
  model DD2-135135-0.02-H at $t=0.02$, 0.04, 0.1, 0.2, 0.4, 0.8, 1.6, and
  2.4\,s.  The length of the velocity vector corresponds to the
  logarithm of the poloidal velocity. The scale is shown in the panel (1).
For each panel, the left and right subpanels show the wide region ($r\lesssim 2000$\,km) and narrow region ($r\lesssim300$\,km), respectively.
  }
\label{fig:snap}
\end{figure*}

The variation of the rotational kinetic energy in the redistribution of the rotational profile is approximately estimated by
\begin{align}
\frac{1}{2} \Delta (v^2) M &\sim \frac{1}{2}[\Delta (\Omega^2) R^2] M 
\notag \\
&\approx 2.3\times 10^{52}\,{\rm erg}\ \biggl(\frac{\Delta (\Omega^2)}
{\rm 9\times10^6\,rad^2\,s^{-2}}\biggr)\notag \\
&\times \biggl(\frac{R}{\rm 10\,km}\biggr)^2 \biggl(\frac{M}{2.5M_\odot}
\biggr),
\end{align}
where $\Delta (v^2)$ and $\Delta (\Omega^2)$ are the variations of the
rotational velocity squared and angular velocity squared, and $M$ is
the mass suffered from this variation (approximately equal to $M_{\rm
  MNS}$).  This shows that the energy of $\sim 10^{52}$\,erg could be
redistributed in the inner region of the MNS in its viscous timescale, which is proportional to
$\alpha_{\rm vis}{}^{-1}$.  This implies that the sound wave becomes
stronger for the models with higher viscosity parameters.

During its outward propagation, the sound wave becomes a shock wave (see the panel (1) of Fig.~\ref{fig:snap}).
It sweeps the material surrounding the torus and then induces mass ejection of the material for the first $\sim 50$ ms (see the panels (1) and (2) of Fig~\ref{fig:snap}).
We refer to this mass ejection as `` early viscosity-driven mass ejection".

The material swept up by the shock wave for small radii ($r\lesssim 500$ km) is still gravitationally bound while it continues to expand for $\sim 0.1$\,s as shown in the panel (3) of Fig~\ref{fig:snap}.
After $\sim 0.1$\,s, the material begins to turn over and fall again (see the panel (4) of Fig~\ref{fig:snap}).
We note that for the model with $\alpha_{\rm vis}=0.04$, this turnover occurs only weakly and more mass is ejected from the system (see the discussion of Sec.~3.4.2).

After the early viscosity-driven mass ejection, the neutrino- and viscosity-driven mass ejection is developed in the polar region.
It is difficult to separate the contributions of the heating due to the neutrino irradiation and viscosity, but for $t\gtrsim 0.2$\,s, the viscous heating becomes important rather than the neutrino heating.
This is because the neutrino pair-annihilation heating, which is the primary neutrino heating source, plays a minor role due to the decrease of the neutrino luminosity in this phase for the inviscid model~\citep{2017ApJ...846..114F}.

\begin{figure}[t]
\begin{center}
\includegraphics[width=\hsize]{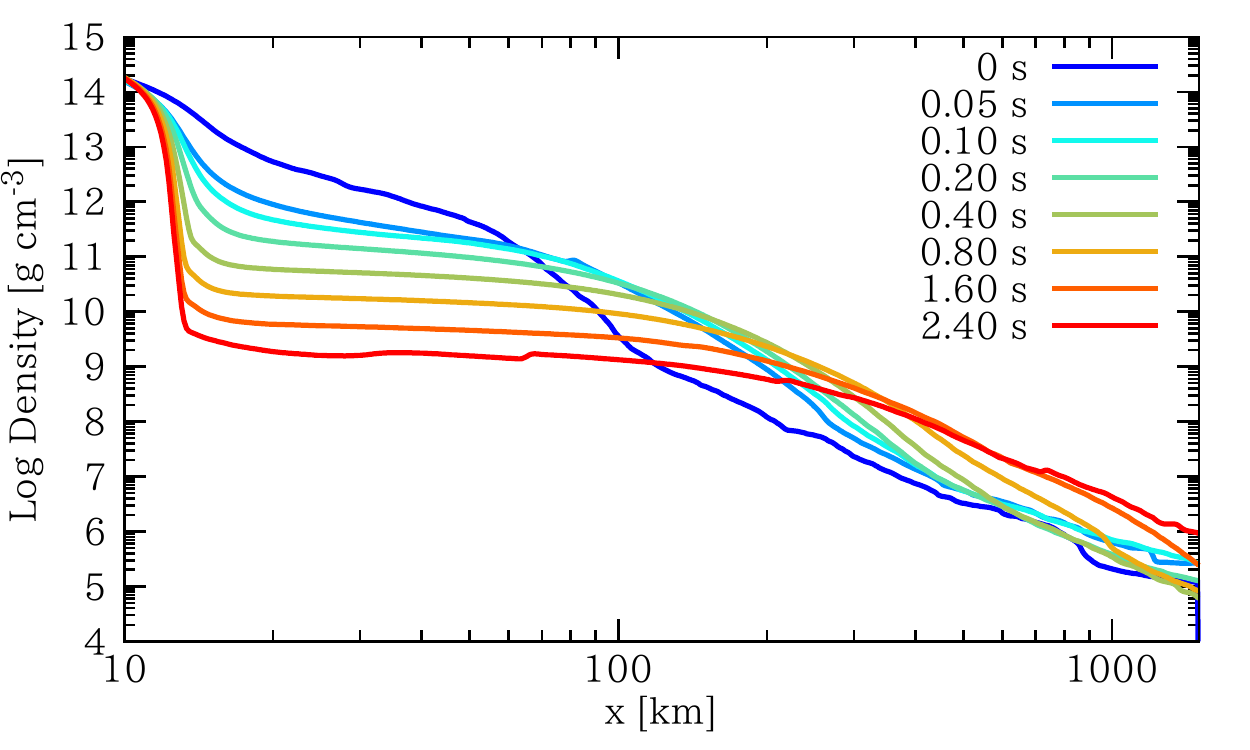}
\caption{ Density profiles on the equatorial plane for the fiducial model at 0, 0.05, 0.1,
  0.2, 0.4, 0.8, 1.6, and 2.4\,s.}
\label{fig:dentime}
\end{center}
\end{figure}

The material in the torus moves in various directions for the first tens of ms (see the panel (1) of Fig.~\ref{fig:snap}), but the
flow structure gradually relaxes to a laminal state due to the
viscosity-driven redistribution (see the panels (3) and (4) of Fig.~\ref{fig:snap}).
After the relaxation, the MNS-torus system evolves in a quasi-stationary manner.
The torus material around the equatorial plane expands outward, while the torus material near the torus surface accretes onto the central MNS (see also Fig.~\ref{fig:flow} in Sec.~3.2).
Figure~\ref{fig:dentime} shows the density profiles on the equatorial plane at different time slices.
As found in this figure, the torus gradually expands with time; hence, the torus density decreases.
This behavior of the torus is determined by the viscous angular momentum transport (see Sec.~3.2 for the details).

\begin{figure}[t]
\begin{center}
\includegraphics[width=\hsize]{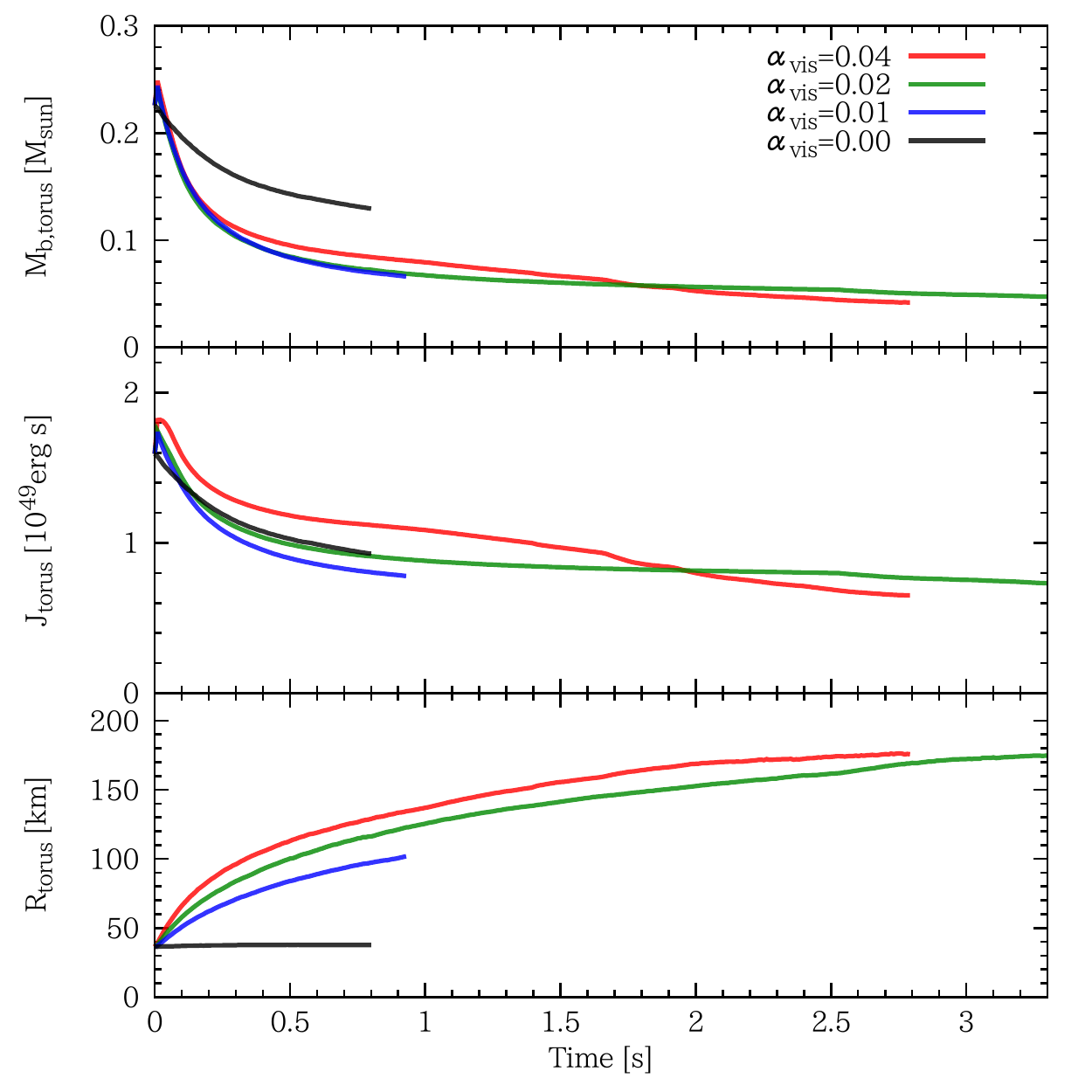}
\caption{ Time evolution of the baryon mass (top), angular momentum
  (middle), and typical radius (bottom) of the torus.  The baryon
  mass, angular momentum, and typical radius of the torus are defined
  by Eqs.~\eqref{eq:bmasstorus}, \eqref{eq:jtorus}, and \eqref{eq:radtorus}, respectively. 
   For all panels, the different colors of the curves indicate the results for
  different models DD2-135135-0.00-H, DD2-135135-0.01-H,
  DD2-135135-0.02-H, and DD2-135135-0.04-H.  }
\label{fig:accretion1}
\end{center}
\end{figure}
The top and middle panels of Fig.~\ref{fig:accretion1} show the time
evolution of the baryon mass and the angular momentum of the torus
defined by
\begin{align}
J_{\rm torus} &= \int_{\rho<10^{13}{\rm g\,cm^{-3}}} d^3x\ \rho_* 
h u_y x.\label{eq:jtorus}
\end{align}
For all of the models, they decrease with time due to the mass accretion onto the central MNS and outflow\footnote[2]{Here $M_{\rm b,torus}$ and $J_{\rm torus}$ slowly decrease even in the inviscid model.
This inflow is due to the cooling of the torus by the neutrino emission; the loss of the pressure support causes the torus accretion.}.

In the phase of the early viscosity-driven mass ejection, in which the density profiles of the MNS and torus vary in a short timescale, the decrease timescale is slightly shorter than that estimated by Eq.~\eqref{eq:tvis}. However, after the torus relaxes to a quasi-stationary state, the timescale agrees approximately with that by Eq.~\eqref{eq:tvis}.

The decrease rates of $M_{\rm b,torus}$ and $J_{\rm torus}$ approach zero for $t \agt 1$\,s. This implies that the accretion onto the MNS becomes inefficient. This point will be revisited in Sec.~3.3.

Because the angular momentum profile of the torus is approximately described by the Keplerian profile around the MNS, we define a typical torus radius by
\begin{align}
R_{\rm torus} &\equiv \frac{J_{\rm torus}^2}{GM_{\rm b,torus}^2}
\frac{1}{M_{\rm MNS}}\notag\\
&\approx 70 \,{\rm km} \biggl(\frac{J_{\rm torus}}{10^{49}\,{\rm erg\,s}}
\biggr)^2 \biggl(\frac{M_{\rm b,torus}}{0.1\ M_\odot}\biggr)^{-2} 
\biggl(\frac{M_{\rm MNS}}{2.6\ M_\odot}\biggr)^{-1}. \label{eq:radtorus}
\end{align}
Here we assumed that $J_{\rm torus}$ would be approximately given by $M_{\rm b,torus} \sqrt{GM_{\rm MNS} R_{\rm torus}}$.
The bottom panel of Fig.~\ref{fig:accretion1} shows the radii for all the models.
Here we used $M_{\rm MNS}=2.6\ M_\odot$ for all of the models, which is approximately equal to the ADM mass of the system.
The typical radius is 30--40 km at the beginning of the simulation.
For the viscous models, it increases with time, while the radius is approximately constant in time for the inviscid model.
This shows that, by viscous angular momentum transport, the torus expands along the equatorial direction. This effect eventually induces viscosity-driven mass ejection from the torus.
We will describe this process in Sec.~3.4.

\subsection{Structure of the Quasi-stationary Flow for $0.1\,{\rm s} \lesssim t\lesssim 2\,{\rm s}$}

\begin{figure*}[t]

\begin{center}
\includegraphics[bb=0 -70 2000 1200,width=0.7\hsize]{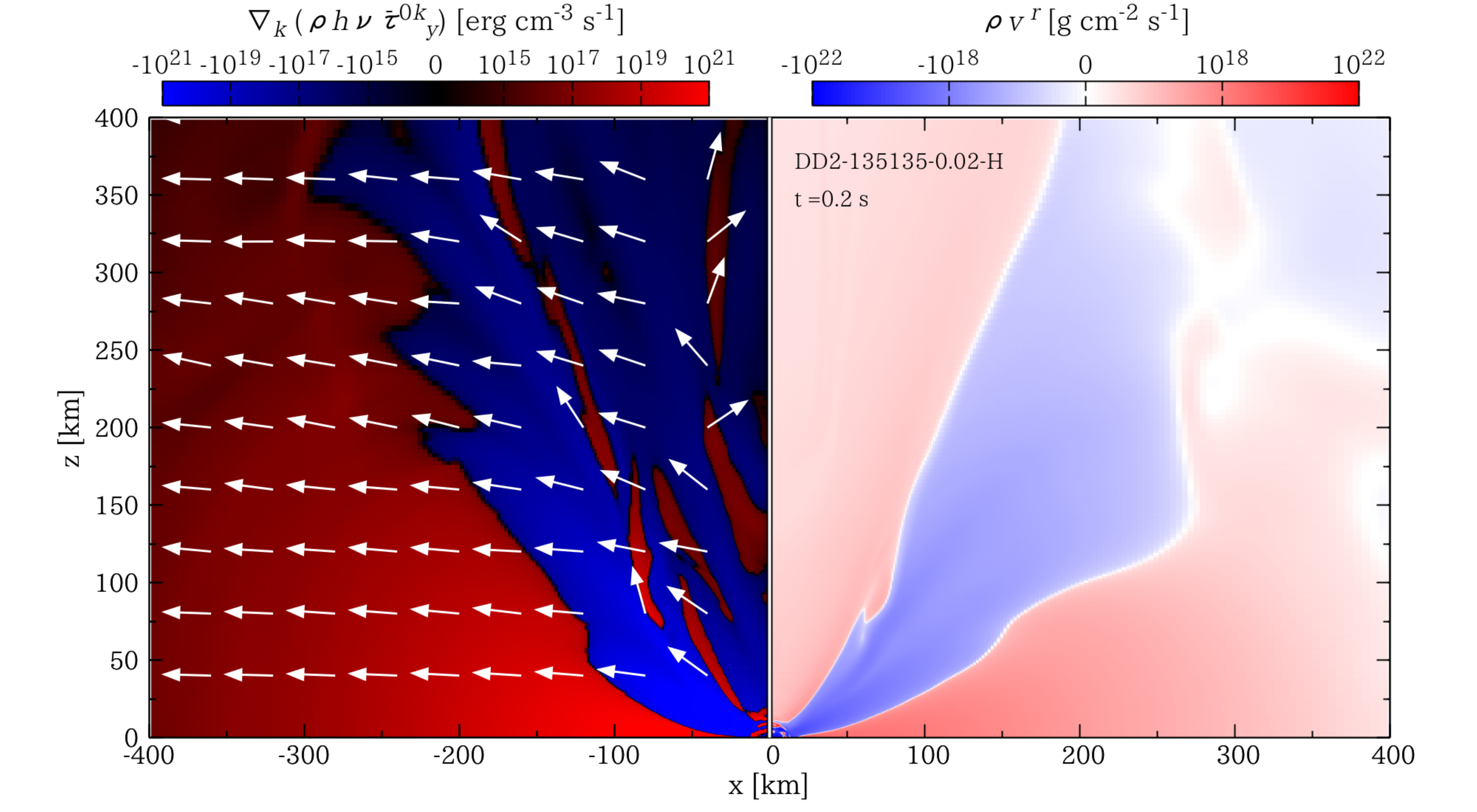}
\includegraphics[bb=0 -70 2000 1200,width=0.7\hsize]{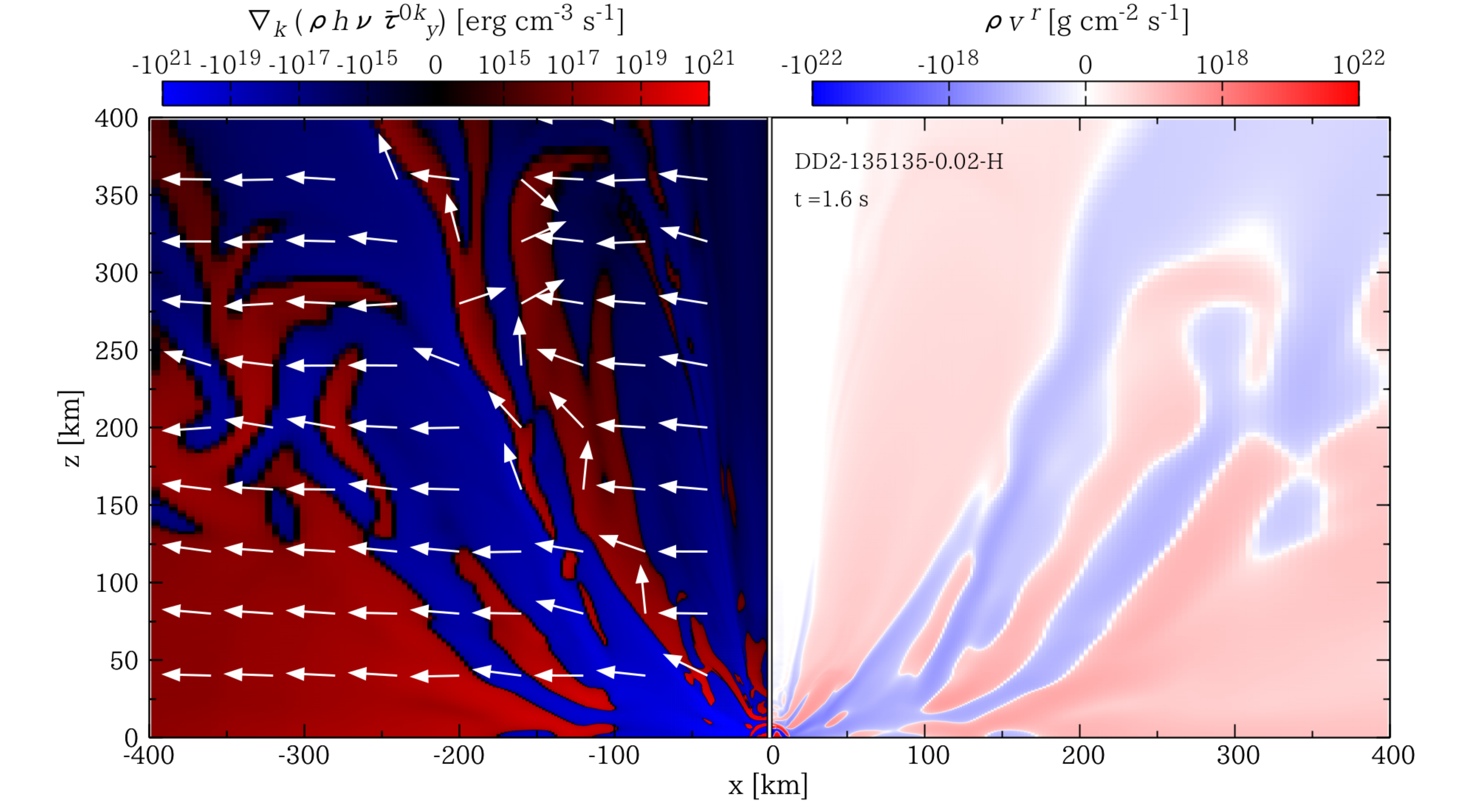}
\caption{Snapshot of the viscous angular momentum flux
  (left panels) and the radial mass flux (right panels)  for the fiducial model at $t=0.2$\,s (top) and 1.6\,s (bottom).  The
  vector field represents only the direction of the poloidal angular
  momentum flux $-\rho h \nu(\bar{\tau}^{0x}{}_y,
  \bar{\tau}^{0z}{}_y)$.  }
\label{fig:flow}
\end{center}
\end{figure*}

\begin{figure*}[t]
\begin{center}
\includegraphics[width=0.45\hsize]{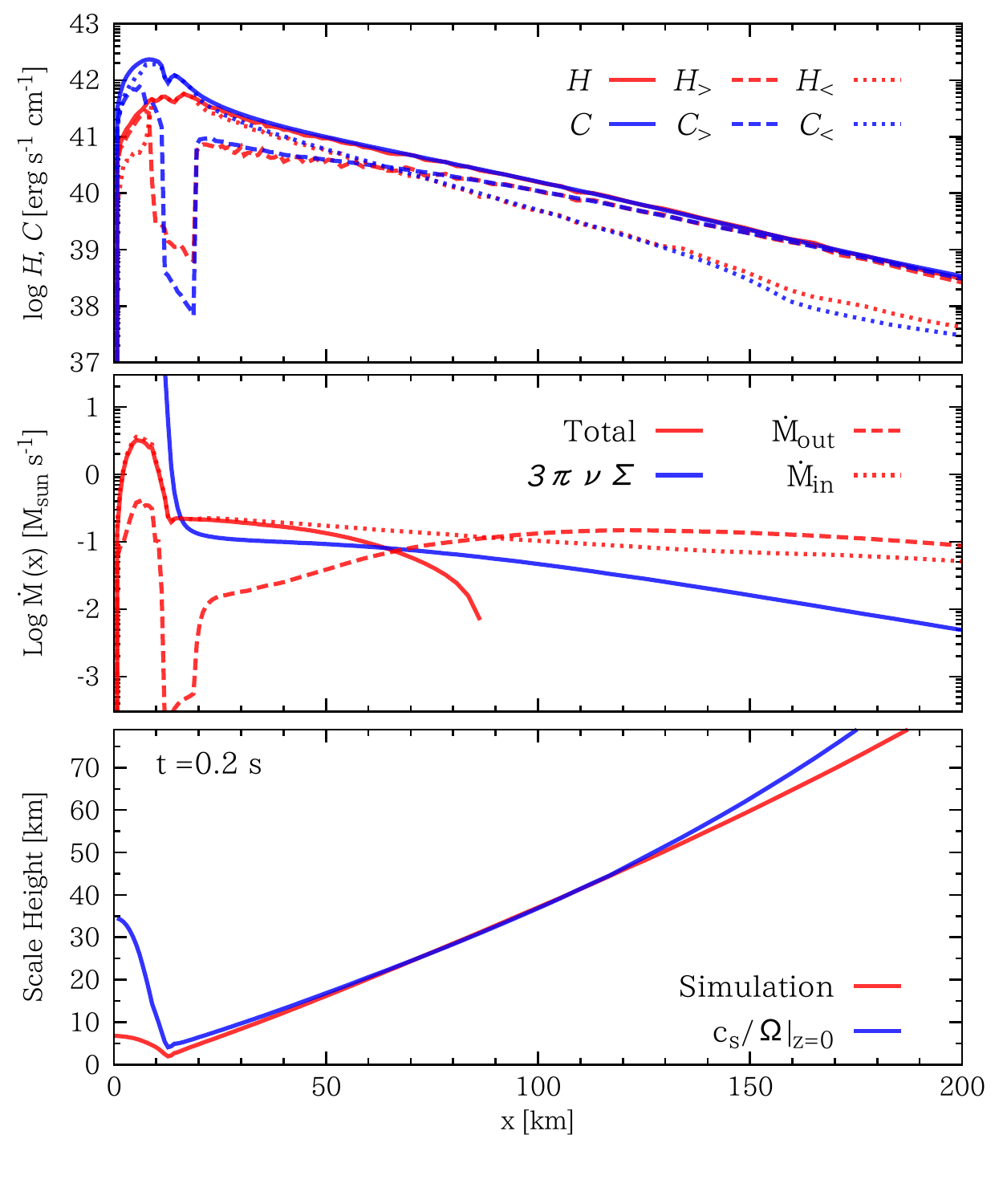}
\includegraphics[width=0.45\hsize]{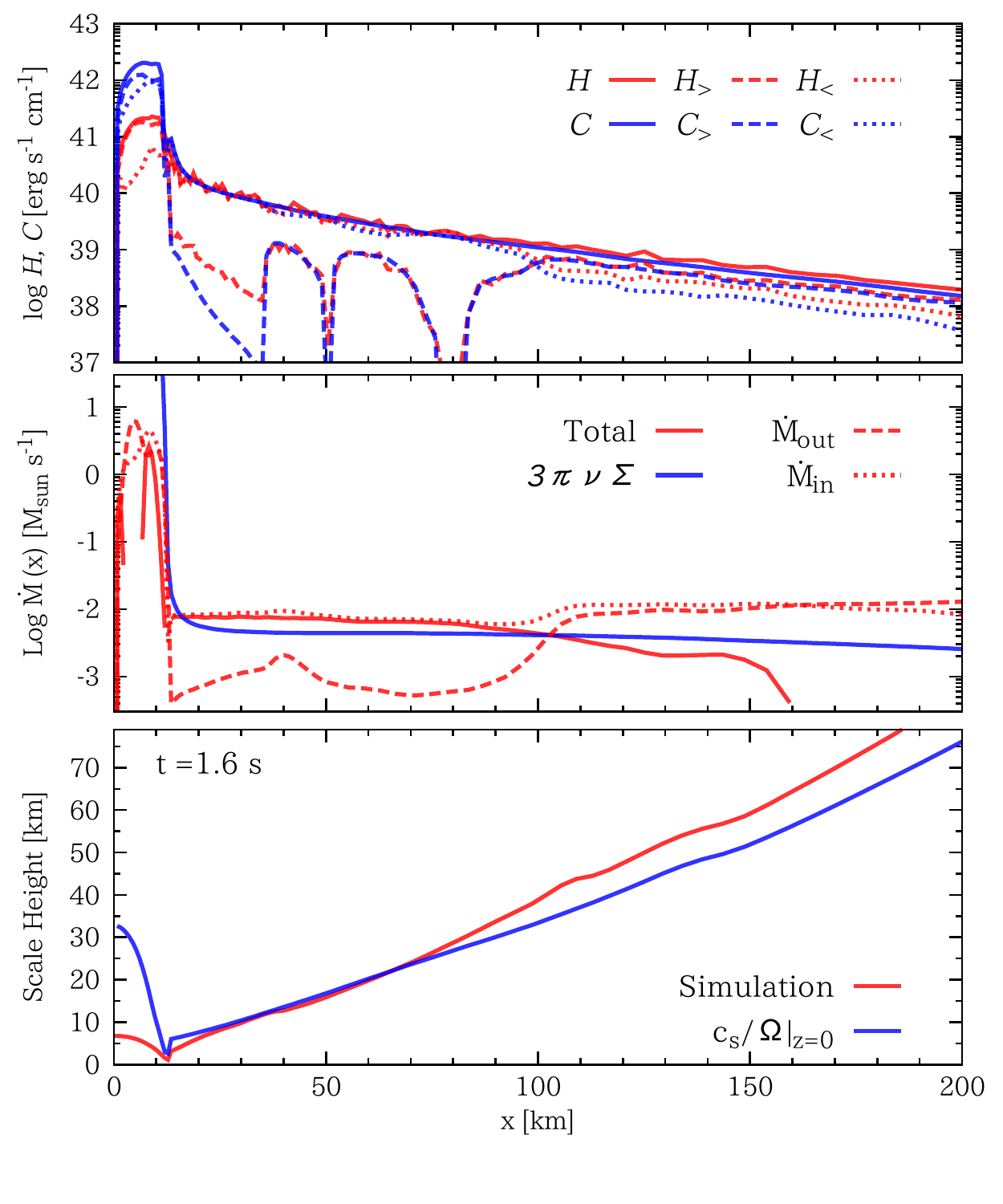}
\caption{ Flow structure for the fiducial model at
  $t=0.2$\,s (left panels) and 1.6\,s (right panels).  Top:
  vertically integrated heating and cooling rates.  In addition to the
  total rate (i.e., Equations~\eqref{eq:htot} and \eqref{eq:ctot}) shown by
  the solid curves, the heating and cooling rates integrated over only
  for $v^x<0$ or $v^x>0$ regions (i.e., Equations~\eqref{eq:hreg} and
  \eqref{eq:creg}) are shown in the dotted and dashed curves,
  respectively.  Middle: mass accretion and outflow rates.  The solid
  red curve indicates the net accretion rate defined by
  Eq.~\eqref{eq:mdot}, and the dotted and dashed curves indicate the
  mass inflow and outflow rates defined by Equations~\eqref{eq:mdin} and
  \eqref{eq:mdout}, respectively.  The solid blue curve indicates the
  mass accretion rate defined by Eq.~\eqref{eq:mdsd}.  Bottom: scale
  height of the torus.  The red curve indicates the scale height
  defined by Eq.~\eqref{eq:hscdata}, and the blue curve indicates the
  scale height obtained by assuming the hydrostatic structure in the
  vertical direction, i.e., $c_s/\Omega|_{z=0}$.  }
\label{fig:qstructure}
\end{center}
\end{figure*}

When the central remnant composed of a rigidly rotating MNS and torus relaxes to a quasi-stationary state, the flow structure also reaches a quasi-stationary state approximately.
This state is preserved until the matter temperature of the torus decreases sufficiently.
The top right panel of Fig.~\ref{fig:flow} shows the radial mass flux in the meridian plane at $t=0.2$\,s for the fiducial model.
Broadly speaking, the structure of the flow is divided into three parts: the outflow near the polar region (Region I; red), the expanding flow along the equatorial plane (Region II; red), and inflow between the two regions (Region III; blue).

The top left panel of Fig.~\ref{fig:flow} displays the profile of the source of the angular momentum transport due to the viscosity (i.e., $x^{-2} \del_j (x^2\rho h \nu \bar{\tau}^{0j}{}_y)$).
This panel also shows the direction of the angular momentum flux due to the viscosity (i.e., $-\rho h \nu(\bar{\tau}^x{}_y, \bar{\tau}^z{}_y)$).
This clearly illustrates that the angular momentum is transported from Region III to Region II.
Therefore, the expanding part in Region II is driven by the angular momentum transport from the inflow in Region III.

The top left panel of Fig.~\ref{fig:qstructure} shows the vertically and azimuthally integrated heating rate (solid red curve)
\begin{align}
{\cal H} = 2 \int_0^L dz\ 2\pi x\ \bigl( Q_{\rm vis}^{(+)} + Q_{\nu}^{(+)} 
\bigr) \label{eq:htot}
\end{align}
and cooling rate (solid blue curve)
\begin{align}
{\cal C}= 2 \int_0^L dz\ 2\pi x\ Q_{\rm (leak)}^{(-)}. \label{eq:ctot}
\end{align}
Here $Q_{\nu}^{(+)}$ is the sum of the matter-heating source terms due to the neutrino absorption and pair-annihilation processes (see \cite{2017ApJ...846..114F} for the description of the individual heating rates) and $Q_{\rm (leak)}^{(-)}$ is the sum of the leakage source terms for all flavors of neutrinos.
This figure shows that the viscous heating rate balances approximately with the
net neutrino cooling rate.

In the same panel, the heating and cooling rates in the outgoing and ingoing regions,
\begin{align}
{\cal H}_{\tiny
\{
\begin{array}{c}
<\\
>
\end{array}
\}
}
&= 2 \int^L_{0,{\tiny
\{
\begin{array}{c}
v^x<0\\
v^x>0
\end{array}
\}
}}
 dz\ 2\pi x\ \bigl( Q_{\rm vis}^{(+)} + Q_{\nu}^{(+)} \bigr), \label{eq:hreg}\\
{\cal C}_{\tiny
\{
\begin{array}{c}
<\\
>
\end{array}
\}
}
&= 2 \int^L_{0,{\tiny
\{
\begin{array}{c}
v^x<0\\
v^x>0
\end{array}
\}
}}
dz\ 2\pi x\ Q_{\rm (leak)}^{(-)} \label{eq:creg}
\end{align}
are plotted in the dashed and dotted curves, respectively.  These
clearly show that the balances between the viscous heating and the net
neutrino cooling are approximately achieved both in Region II
(expanding region) and Region III (inflowing region).  This indicates that
the neutrino-dominated accretion flow (NDAF) is achieved in Region
III.

The middle left panel of Fig.~\ref{fig:qstructure} shows the mass inflow and outflow rates ($\dot{M}_{\rm in}$ and $\dot{M}_{\rm out}$) along the equatorial plane, together with the net rates ($\dot{M}$), defined, respectively, by 
\begin{align}
\dot{M}_{\rm in} &=  - 2\int^L_{0, v^x<0} dz\ 2\pi x \rho_* v^x,\label{eq:mdin}\\
\dot{M}_{\rm out} &= 2\int^L_{0, v^x>0} dz\ 2\pi x \rho_* v^x, \label{eq:mdout}
\end{align}
and
\begin{align}
\dot{M} &= \dot{M}_{\rm in} -\dot{M}_{\rm out}. \label{eq:mdot}
\end{align}
In the same panel, we plot the mass accretion rate in the ``standard" disk
picture, i.e., a stationary thin accretion disk,
\begin{align}
\dot{M}_{\rm sd} &\approx 3\pi \nu \Sigma. \label{eq:mdsd}
\end{align}
Here we neglected the terms from the inner boundary condition and defined the column density of the torus by $\Sigma = 2 \rho c_s/\Omega|_{z=0}$, which is evaluated using the
values on the equatorial plane.  The net mass accretion rate $\dot{M}$
agrees approximately with the accretion rate in the standard disk
picture $\dot{M}_{\rm sd}$ for the innermost region of the torus
$x\lesssim 50$ km.

In the bottom left panel of Fig.~\ref{fig:qstructure}, we compare the
scale height of the torus assuming the vertically hydrostatic
structure $H_{\rm sd}=c_{s}/\Omega|_{z=0}$ and the scale height
calculated from the simulation data $H_{\rm num}$ to check the validity of the
approximation of the flow as the standard disk.  Here $H_{\rm num}$ is defined
as
\begin{align}
\rho(z=H_{\rm num}) = \rho(z=0)/e, \label{eq:hscdata}
\end{align}
where $e$ is the base of natural logarithms.  In the innermost region
($x\lesssim 100$ km) of the torus, the two scale heights agree well.

The flow structure, heating and cooling rates, mass accretion rate,
and scale height at the late time $t=1.6$\,s are shown in the bottom
panels of Fig.~\ref{fig:flow} and the right panels of
Fig.~\ref{fig:qstructure}, respectively.  We find in
Fig.~\ref{fig:flow} that the flow structure at $t=1.6$\,s is
qualitatively the same as that at $t=0.2$\,s, while the mass flux
becomes smaller than that in the early phase (see the middle panel of Fig. 6).  At $t=1.6$\,s, the
region for which the net mass accretion rate agrees with the standard
picture accretion rate is wider than that at $t=0.2$\,s.

To conclude, the global structure of the flow can be described
approximately by the standard NDAF for a long timescale 0.1\,s
$\lesssim t \lesssim 2$ s.

\subsection{Mass Accretion onto the MNS}

\begin{figure}[t]
\begin{center}
\includegraphics[width=\hsize]{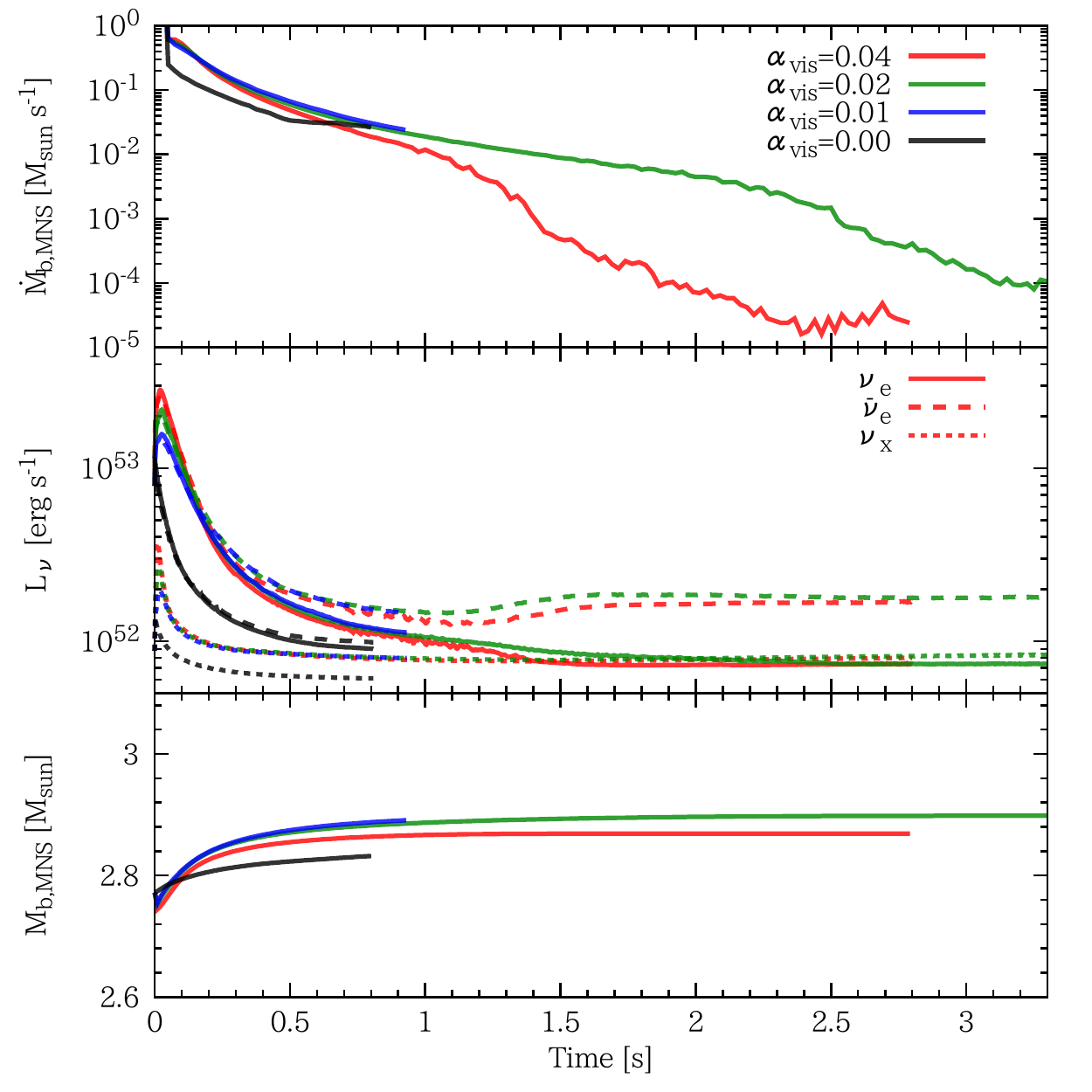}
\caption{ Time evolution of the mass accretion rate (top), neutrino
  emission rate (middle), and baryon mass of the MNS (bottom).  The baryon
  mass of the MNS is defined by Eq.~\eqref{eq:bmassmns}.  In the
  middle panel, the emission rates of electron neutrinos, electron
  antineutrinos, and the other neutrino species are shown by the
  solid, dashed, and dotted curves, respectively.  For all the panels,
  the different colors of the curves indicate the results for different
  models DD2-135135-0.00-H, DD2-135135-0.01-H, DD2-135135-0.02-H, and
  DD2-135135-0.04-H.}
\label{fig:accretion2}
\end{center}
\end{figure}

The top panel of Fig.~\ref{fig:accretion2} shows the accretion rate of
the torus material onto the MNS, which is defined by the radial mass inflow rate (i.e.,
Eq.~\eqref{eq:mdot}) at $x=20$ km.
The mass accretion rate is $\sim 0.5 M_\odot\,{\rm s^{-1}}$ at $\sim 0.1$\,s, which is
consistent with the estimation in Eq.~\eqref{eq:mdestimate}.  The mass
accretion rate monotonically decreases, and it becomes $\sim
0.004M_\odot\,{\rm s^{-1}}$ at $t=2$\,s for the fiducial model.  This decrease
stems partly from the decrease of the torus mass, but the major reason is the
radial expansion of the torus due to the outward angular momentum transport, as described in the
previous subsection.  Unlike the estimation of Eq.~\eqref{eq:mdestimate},
the mass accretion rate for $t\lesssim 1$\,s depends only weakly on $\alpha_{\rm vis}$
(compare the results for $\alpha_{\rm vis}=0.01$, 0.02, and 0.04).
This is because the typical radius of the torus becomes large more
rapidly for the model with higher viscosity parameters (see the bottom
panel of Fig.~\ref{fig:accretion1}).
The mass accretion rate for $t\gtrsim 1$\,s depends on $\alpha_{\rm vis}$; the rate for the model with $\alpha_{\rm vis}=0.04$ decreases more rapidly than that for the fiducial model.
This is because the neutrino cooling in the torus becomes inefficient at the earlier time; hence, the torus material starts being ejected rather than accreted onto the MNS (see Sec.~\ref{sec:ej} for details).

The middle panel of Fig.~\ref{fig:accretion2} shows the time evolution
of the neutrino emission rates.  Because of the presence of the viscous
heating, the neutrino emission rate for $\alpha_{\rm vis}\neq 0$ is
higher than that of the inviscid model.  For the
viscous models, the high emission rate of $\sim 10^{53}\,{\rm
  erg\,s^{-1}}$ is sustained for $\sim 0.1$\,s.  However, for the late
time $t\gtrsim 1$\,s, the emission rate decreases to $\sim 10^{52}\,{\rm
  erg\,s^{-1}}$ as the accretion rate, $\dot{M}_{\rm b,MNS}$,
decreases.  Since the mass accretion rate depends weakly on
$\alpha_{\rm vis}$ for $t\lesssim 1$\,s, the neutrino emission rate also depends weakly on the
viscosity parameter.

The bottom panel of Fig.~\ref{fig:accretion2} shows the baryon mass of
the MNS, defined by
\begin{align}
M_{\rm b,MNS} &= \int_{x<20\,{\rm km}} d^3x\ \rho_*. \label{eq:bmassmns}
\end{align}
As found from the top panel of Fig.~\ref{fig:accretion2}, the MNS mass
increases with time, but eventually, it is saturated because the mass
accretion from the torus decreases significantly.  The final relaxed
value, $\approx2.9M_\odot$, depends only weakly on $\alpha_{\rm vis}$.

For the DD2 EOS, the MNS with $M_{\rm b,MNS}=2.9M_\odot$ does not collapse to a black hole, at least in a few seconds after the onset of merger, because the maximum gravitational mass for cold spherical neutron stars for this EOS is quite high, as $2.42M_\odot$.
However, for softer EOSs, in which the maximum gravitational mass is smaller, say $2.1M_\odot$, the remnant MNS of the baryon mass of $\sim 2.9M_\odot$ may collapse to a black hole due to the mass accretion.

\subsection{Mass Ejection} \label{sec:ej}

\subsubsection{Early Viscosity-driven Mass Ejection}

\begin{figure}[t]
\begin{center}
\includegraphics[width=\hsize]{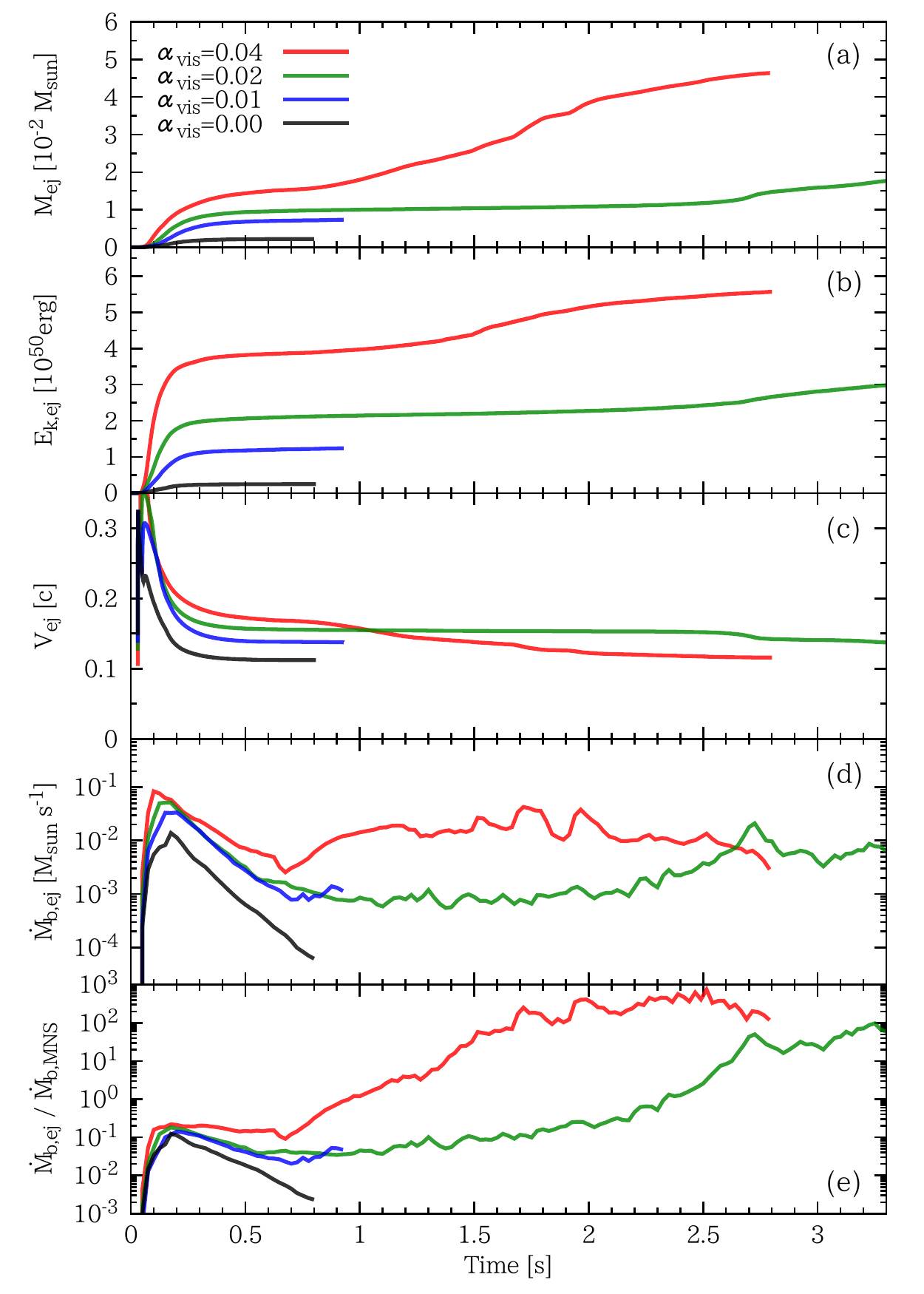}
\caption{ Time evolution of the baryon mass (panel (a)), kinetic
  energy (panel (b)), and average velocity (panel (c)) of the ejecta,
  mass ejection rate (panel (d)), and ratio of $\dot{M}_{\rm b,ej}$ to $\dot{M}_{\rm b,MNS}$ (panel
  (e)).  For all the panels, the different colors of the curves indicate
  the results for different models DD2-135135-0.00-H,
  DD2-135135-0.01-H, DD2-135135-0.02-H, and DD2-135135-0.04-H.
}
\label{fig:ejecta}
\end{center}
\end{figure}

Panel (a) of Fig.~\ref{fig:ejecta} shows the evolution of the baryon mass of the
ejecta, which is calculated by integrating the flux of the unbound
material at an outer surface as
\begin{align}
M_{\rm b,ej}(t) &= \int_0^t dt' \,2\int dS_k \rho_* v^k\Theta (|hu_t|-1)\notag\\ 
&= \int_0^t dt' \,2\int_{0}^{L_{\rm bnd}} dx\ 2\pi x \rho_* v^z\Theta (|hu_t|-1) \notag\\
&+ \int_0^t dt' \,2\int_{0}^{L_{\rm bnd}} dz\ 2\pi x \rho_* v^x\Theta (|hu_t|-1), \label{eq:mej}
\end{align}
where $dS_k$ is the two-dimensional area element on a cylinder of both radius and height $L_{\rm bnd}$.
Here we set $L_{\rm bnd}= 4000$\,km and suppose that fluid elements with $|hu_t|>1$ are gravitationally unbound\footnote{We note that this criterion may overestimate the amount of the ejecta, since some amount of the thermal energy is radiated by neutrinos before the energy is transformed into the kinetic energy. However, the temperature of the material at $L_{\rm bnd}= 4000$\,km is lower than 0.1 MeV, for which the neutrino cooling timescale is much longer than the expansion timescale, and thus the estimation of the ejecta mass works reasonably.}.

This figure shows that the mass ejection rate in the early phase $t\lesssim 0.2$\,s is rather high, 0.03--$0.1M_\odot\,{\rm s^{-1}}$.
This is achieved by the early viscosity-driven ejecta (see Sec.~3.1).

We also calculate the evolution of the total ejecta energy by
\begin{align}
E_{\rm tot,ej}(t) &= \int_0^t dt' \int dS_k \rho_*\hat{e}_0 v^k\Theta (|hu_t|-1).
\end{align}
Using $ E_{\rm tot,ej} $ and $ M_{\rm b,ej}$, the kinetic energy of the ejecta is defined by
\begin{align}
E_{\rm kin,ej} = E_{\rm tot,ej} - M_{\rm b,ej},
\end{align}
which is shown in the panel (b) of Fig.~\ref{fig:ejecta}.
Here we assumed that the internal energy of the ejecta would be totally transformed into the kinetic energy during the subsequent expansion.
Unlike the mass accretion rate and the neutrino emission rate shown in Sec. 3.3, the mass and kinetic energy of the ejecta increase monotonically with $\alpha_{\rm vis}$ for $t \alt 0.2$\,s.

Panel (c) shows the evolution of the average velocity of the ejecta, which is defined by $V_{\rm ej} = \sqrt{2E_{\rm k,ej}/M_{\rm b,ej}}$.
For the early viscosity-driven ejecta, which is ejected for $t\lesssim 0.2$\,s, the average velocity is in the range $0.15$--$0.2\,c$, and monotonically increases with $\alpha_{\rm vis}$ as the mass and the kinetic energy of the ejecta.
This is because the shock driven by the early viscous effect on the MNS is stronger for the higher viscosity parameter model.

In our simulations, we suppose that the viscosity suddenly arises after the merger remnant settles to a quasi-stationary state.
We should note that the variation of the quasi-equilibrium configuration occurs only if the viscous effect is enhanced after the dynamical phase of the merger ends, e.g., by MHD turbulence induced by the significantly amplified magnetic fields at the onset of the merger~\citep{2014PhRvD..90d1502K, 2015PhRvD..92f4034K, 2017arXiv171001311K}.
If the viscosity in the MNS arises slowly enough, the variation of the quasi-stationary states occurs smoothly, so that the density wave would not appear strongly at the MNS surface and the mass ejection due to the shock would not occur significantly.

\subsubsection{Late-time Viscosity-driven Mass Ejection}

The mass ejection for $t\gtrsim 0.2$\,s is driven primarily by the viscous effects in the torus.  The ejecta is launched mainly toward the polar region for the relatively early time with $t \alt 1$\,s, as found in panels (4)--(6) of Fig.~\ref{fig:snap}.
However, the major mass ejection mechanism is subsequently changed as discussed below.
In the following, we focus only on the models for $\alpha_{\rm vis}=0.02$ and 0.04 because long-term simulations are performed only for these models.

\begin{figure}[t]
\begin{center}
\includegraphics[width=\hsize]{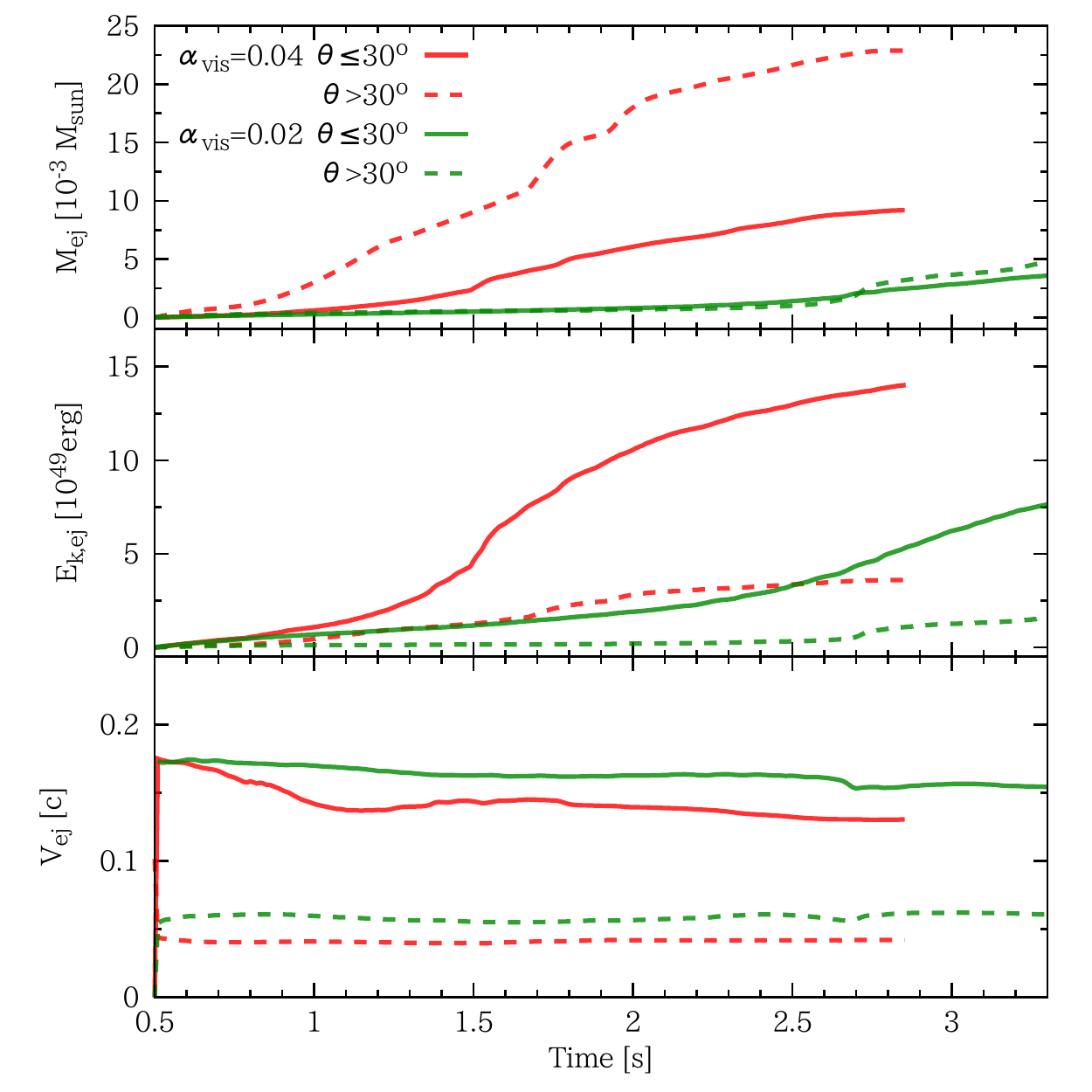}
\caption{
Same figures as panels (a)--(c) of Fig.~\ref{fig:ejecta} but for the
material ejected toward the polar angle $0^\circ\leq \theta \leq
30^\circ$ (solid curves) and $30^\circ< \theta \leq 90^\circ$ (dotted
curves) by the late-time viscosity-driven mass ejection for $t \geq 0.5$\,s for the models DD2-135135-0.02-H (red curves) and DD2-135135-0.04-H (green curves), respectively.}
\label{fig:ejectalate}
\end{center}
\end{figure}

\begin{figure*}[t]
\begin{center}
\begin{minipage}{0.49\hsize}
\includegraphics[width=\hsize]{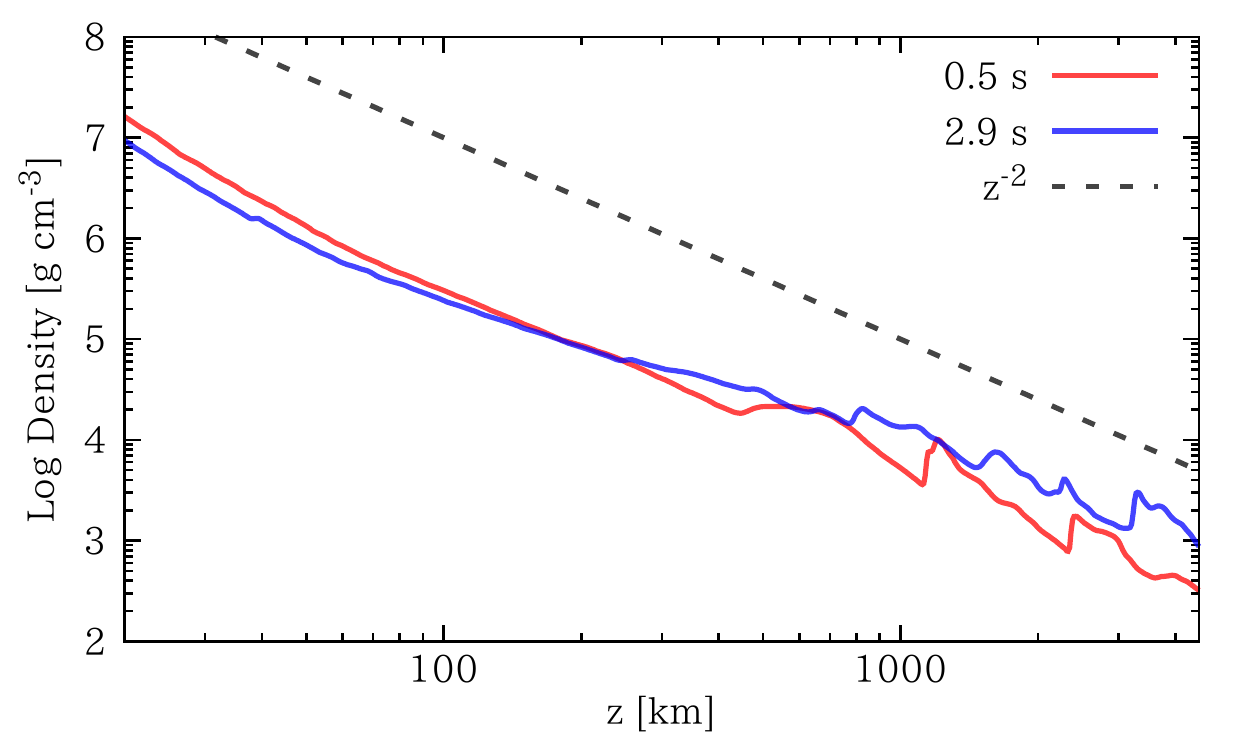}
\end{minipage}
\begin{minipage}{0.49\hsize}
\includegraphics[width=\hsize]{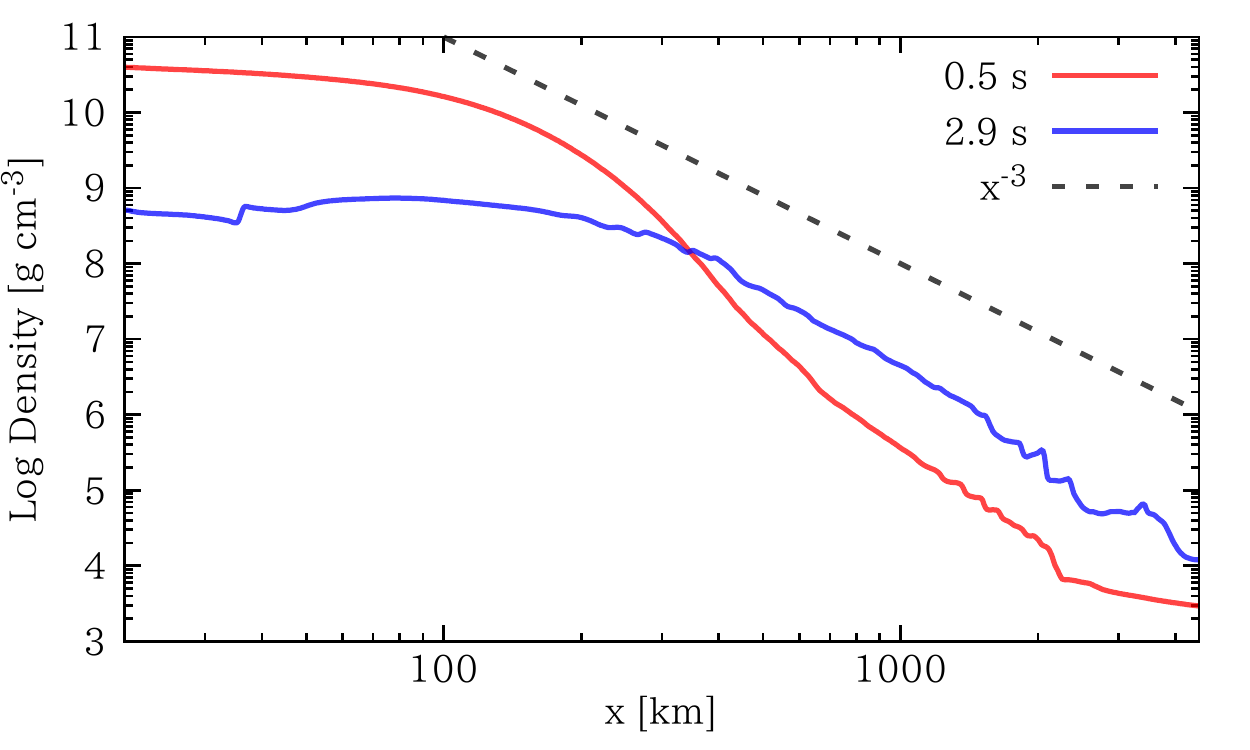}
\end{minipage}

\caption{
Density profiles on the rotational axis (left) and equatorial plane (right) at 0.5\,s (red curves) and 2.9\,s (blue curves) for the fiducial model.
}  
\label{fig:ejden}
\end{center}
\end{figure*}

Figure~\ref{fig:ejectalate} shows the evolution of the mass, kinetic energy, and average velocity of the ejecta that becomes unbound after $t=0.5$\,s for the components toward the polar direction of $0^\circ\leq \theta \leq 30^\circ$ and the other (equatorial) direction of $30^\circ< \theta \leq 90^\circ$, respectively.
From this figure, it is found that the material ejected toward the polar direction is smaller than or approximately as large as that ejected toward the equatorial direction.
On the other hand, the polar ejecta has larger kinetic energy than that of the equatorial ejecta because the average velocity of the polar ejecta is larger ($\sim 0.15\,c$) than that of the equatorial ejecta ($\sim 0.05\,c$).
This clearly shows that there are two components for the ejecta in the late time.
We note that the two components are distinct because they are launched from the completely different regions in the system.
The polar ejecta is launched from the vicinity of the MNS, while the equatorial ejecta is launched from the outer region of the torus.

As found in the top panel of Fig.\,\ref{fig:ejectalate}, for $\alpha_{\rm vis}=0.04$, the mass ejection in this late phase (for $t\gtrsim 0.5$\,s) is primarily toward the equatorial direction, while for the fiducial model ($\alpha_{\rm vis}=0.02$), the equatorial mass ejection is as weak as the polar one up to $t\sim 2.7$\,s.
For $\alpha_{\rm vis}=0.04$, the early viscosity-driven mass ejection continues for a later time than that for the fiducial ($\alpha_{\rm vis}=0.02$) model because the shock driven by the early viscous effect on the MNS is stronger for the larger viscosity parameter models.
This is the reason for the larger mass ejection rate toward the equatorial direction for $\alpha_{\rm vis}=0.04$ than that for the fiducial model for $t\lesssim 0.8$\,s.
On the other hand, the reason for the rapid increase of the mass ejection rate for $t\gtrsim 0.8$\,s is the viscous effect on the torus.
The torus expands due to the viscous angular momentum transport; hence, its density and temperature decrease.
As a result, the neutrino cooling becomes inefficient in the outer region of the torus.
Then, the materials in the outer region can be ejected toward the equatorial region by the viscous heating and angular momentum transport without suffering from the neutrino cooling.

This late-time viscosity-driven mass ejection from the torus is first suggested in \cite{2013MNRAS.435..502F} and \cite{2014MNRAS.441.3444M} and is also found in \cite{2015MNRAS.448..541J}.
The velocity of the ejecta in our work, $\sim 0.05\,c$, as well as the mass ejection mechanism, is totally consistent with their results.
We confirm their results in a more realistic setup with an initial condition derived from three-dimensional simulation by a fully general relativistic simulation that self-consistently solves the evolution of both the MNS and the torus surrounding the MNS.

After the neutrino cooling becomes subdominant in the torus, the torus expands in the viscous timescale defined by Eq.~\eqref{eq:tvis}; hence, this late-time mass ejection rate can be estimated by dividing the torus mass by the viscous timescale as
\begin{align}
\dot{M} &\sim \frac{M_{\rm b,torus}}{t_{\rm vis}} \notag\\
&\sim 0.012M_\odot\,{\rm s^{-1}} \biggl(\frac{\alpha_{\rm vis}}{0.04}\biggr)\biggl(\frac{H_{\rm tur}}{\rm 10\,km}\biggr)^2 \notag\\
&\times \biggl(\frac{M_{\rm b,torus}}{0.05M_\odot}\biggr)\biggl(\frac{R_{\rm torus}}{100\,{\rm km}}\biggr)^{-7/2}\biggl(\frac{M_{\rm MNS}}{2.6M_\odot}\biggr)^{1/2},
\end{align}
where we used the torus mass for $t\agt 1$\,s (see the top panel of Fig.~\ref{fig:accretion1}).
This mass ejection rate agrees reasonably well with our results (see the panel (d) of Fig.~\ref{fig:ejecta}).

For the fiducial model, this late-time equatorial viscosity-driven mass ejection also sets in at $t\sim2.7$\,s.
As found in the last panel of Fig.~\ref{fig:snap}, the velocity in the dense region ($\rho \sim 10^7\,{\rm g\,cm^{-3}}$) turns outward.
In addition, the top panel of Fig.~\ref{fig:ejectalate} indicates that the equatorial mass ejection rate increases for $t\gtrsim 2.7$\,s.
The time delay for the onset of the late-time viscosity-driven mass ejection toward the equatorial plane for the smaller value of $\alpha_{\rm vis}$ is simply due to the less viscous power.

The mass ejection efficiency, defined by $\dot{M}_{\rm b,ej}/\dot{M}_{\rm b,MNS}$, is shown in the panel (e) of Fig.~\ref{fig:ejecta}.
We find that this ratio becomes larger than the unity for $t\agt 1$\,s and $t \agt 2.3$\,s for $\alpha_{\rm vis}=0.04$ and 0.02, respectively, and it eventually exceeds 10 because the mass accretion onto the MNS approaches zero.
Therefore, a large fraction of the torus material, which is $\sim 0.05\ M_\odot$ at $t\agt 1$\,s (see the top panel of Fig.~\ref{fig:accretion1}), is likely to be ejected from the system eventually.
This speculation is supported by the result for the model with $\alpha_{\rm vis}=0.04$, for which the mass of the late-time equatorial viscosity-driven ejecta is already $\sim 0.05M_\odot$ at $t \sim 2.7$\,s.

For the $\alpha=0.02$ model, a large fraction of the material that would eventually be ejected is still in the region $0\le x<L_{\rm bnd}$ and $0\le z<L_{\rm bnd}$, and hence, they are still not considered as an ejecta at the end of the simulations.
The main neutrino emitter is the MNS for $t\gtrsim 0.5$\,s, and the neutrino emission rate is $\sim 10^{52}\,{\rm erg\,s^{-1}}$.
For this emission rate, the timescale for the neutrino absorption
\begin{align}
&\frac{4\pi r^2 \left<\omega\right>}{L_\nu} \frac{1}{G_{\rm F}^2 \left<\omega\right>^2} \notag\\
&\sim 0.4\,{\rm s}\,  \biggl(\frac{\left<\omega \right>}{10\ {\rm MeV}}\biggr)^{-1}\biggl(\frac{L_\nu}{10^{52}\ {\rm erg/s}}\biggr)^{-1}\biggl(\frac{r}{100\ {\rm km}}\biggr)^2
\end{align}
is much longer than the time for the expansion timescale
\begin{align}
r/v \sim 7\times 10^{-3}\,{\rm s}\,  \biggl(\frac{r}{100\,{\rm km}}\biggr)\biggl(\frac{v}{0.05\,c}\biggr)^{-1}. \label{eq:extime}
\end{align}
Thus, the electron fraction of the material in the outer part of the torus, which is in the range 0.3--0.4, is frozen out; hence, the electron fraction of the material that would eventually be ejected is also in the range 0.3--0.4.

Figure~\ref{fig:ejden} shows the density profiles on the rotational axis (left panel) and equatorial plane (right panel) for the fiducial model.
We find that the density decreases approximately in proportion to $z^{-2}$ along the rotational axis.
This behavior does not depend strongly on the time, since the polar mass ejection develops from the early phase of the evolution of the system.
On the other hand, along the equatorial plane, the density for the large radius ($x\gtrsim 500$\,km) decreases in proportion approximately to $x^{-3}$ after the equatorial viscosity-driven mass ejection sets in (see the blue curve in the right panel).
This radius dependence does not change significantly in time for the late phase ($t\gtrsim2.7$\,s); hence, we expect that the density structure of the equatorial ejecta is also proportional approximately to $x^{-3}$.
This radius dependence of the equatorial ejecta is similar to that of the dynamical ejecta with the velocity $\lesssim0.4\,c$~\citep{2014ApJ...784L..28N}.

To summarize this subsection, we find two mass ejection mechanisms for the system
  composed of the long-lived MNS and torus. One is the early
  viscosity-driven mass ejection, in which the differential rotation of the
  MNS is the engine, and the other is the late-time viscosity-driven mass
  ejection from the torus, in which the differential rotation of the
  torus is the engine. Specifically, there are further two different
  components for the late-time viscosity-driven ejecta from the torus, one of which is the
  viscosity-driven ejecta toward the polar direction in
  assistance with the neutrino irradiation, and the other is launched
  primarily toward the equatorial direction in the later phase
  for which the neutrino cooling becomes inefficient because the temperature of the torus decreases sufficiently (see Table~\ref{tab:ejecta} for a summary).


\section{Discussion}

\subsection{Overall Mass Ejection Processes of the Merger and Post-merger of Binary Neutron Stars}\label{sec4.1}

\begin{figure*}[t]
\begin{center}
\includegraphics[width=0.8\hsize]{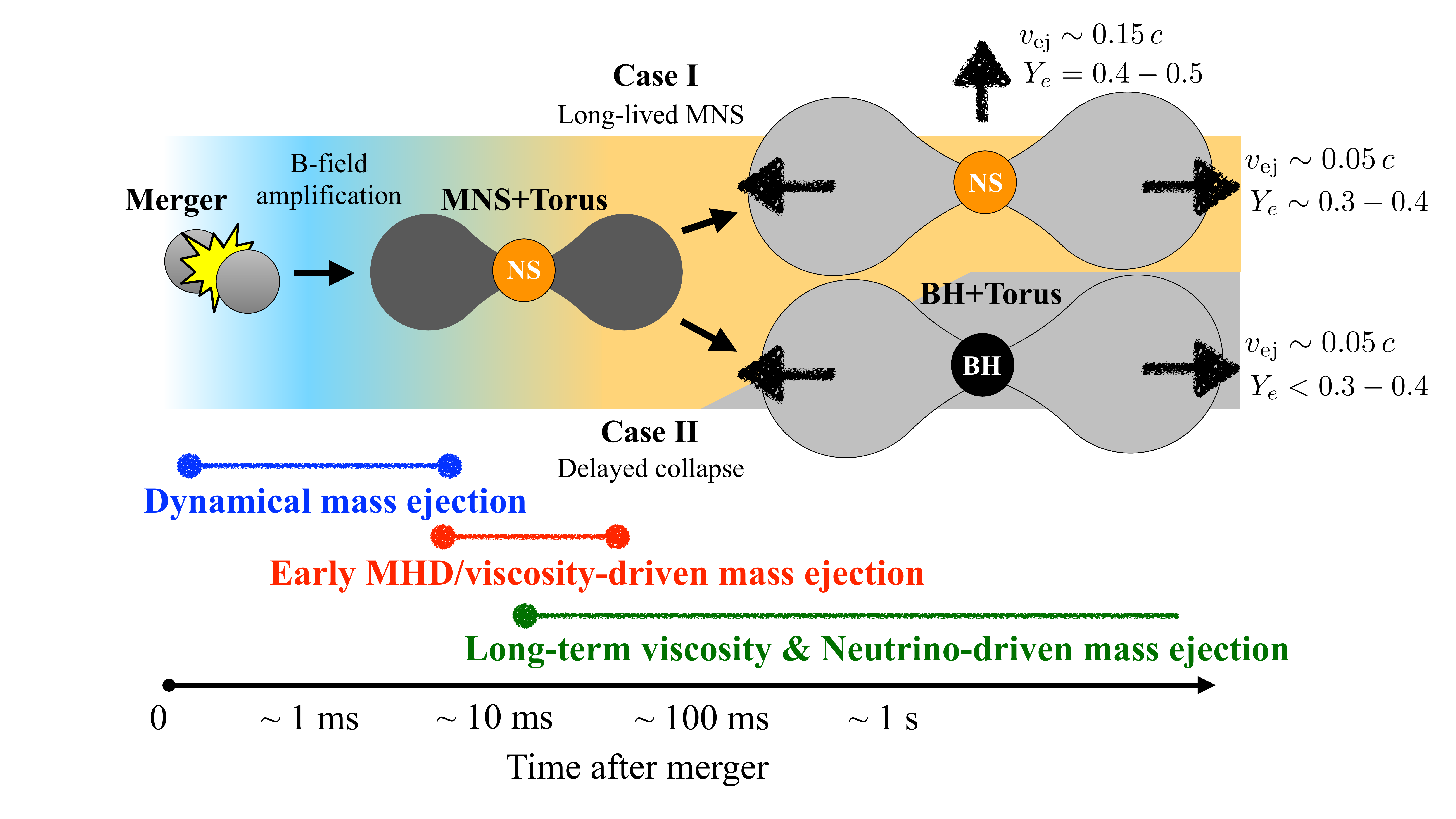}
\caption{ Schematic picture of the overall mass ejection processes in the
  merger and post-merger phases of binary neutron stars.  The time of 
  the delayed collapse of the MNS depends on the neutron star 
  EOS and the total mass of the binary.  If the neutron star EOS is
  stiff enough or the total mass of the binary is small enough, the
  MNS survives for a long timescale (Case I), at least the timescale of
  neutrino cooling $\sim 10$\,s.  Otherwise, the MNS collapses to a
  black hole in the thermal timescale, which is determined by the
  thermal energy of the MNS and neutrino emission rate, or the timescale
  for angular momentum transport in the MNS (Case II).  }
\label{fig:ponchi}
\end{center}
\end{figure*}

Here we discuss the properties of the ejecta for individual mass ejection processes based on the results of our numerical-relativity simulations.

We summarize possible mass ejection processes in the merger and
  post-merger phases in Fig.~\ref{fig:ponchi}. First, during the
  merger, the dynamical mass ejection
  occurs~\citep[e.g.,][]{2013PhRvD..87b4001H,2015PhRvD..91f4059S,2016PhRvD..93l4046S}.
  This mass ejection proceeds for $\sim 10$\,ms primarily toward the
  equatorial direction.  The ejecta mass depends on the stiffness of
  the neutron star EOS, total mass, and mass ratio of the binary, and
  it is in the range 0.001--$0.02M_\odot$.  For the DD2 EOS, which is
  used in this work, it is 0.002--0.005 $M_\odot$ for the total mass
  of $2.7\ M_\odot$ with the mass ratio between 0.85 and 1
  \citep{2016PhRvD..93l4046S}.  In this work, we consider the
  equal-mass merger of $1.35\ M_\odot$ neutron stars, in which the dynamically
  ejected mass is $\sim 0.002\ M_\odot$.  In contrast to the ejecta
  mass, the typical velocity and the profile of the electron fraction
  of the dynamical ejecta depend only weakly on the neutron star EOS,
  and they are in the ranges 0.15--$0.25c$ and 0.05--0.5,
  respectively \citep{2015PhRvD..91f4059S}.

In the post-merger phase, an MNS surrounded by a torus
  is typically formed. However, the long-term evolution process of this system depends
  on the neutron star EOS and total mass of the binary.  If the EOS is
  stiff enough (i.e., the maximum mass for cold spherical neutron stars, $M_{\rm max}$, is high enough) or the total mass of the binary is small enough, the
  MNS survives for seconds or longer after the onset of the merger as shown
  in this paper (Case I of Fig.~\ref{fig:ponchi}).  Otherwise, the MNS
  collapses to a black hole after it loses its thermal energy and/or
  angular momentum (Case II of Fig.~\ref{fig:ponchi}).
The time at which the collapse takes place would be determined by the value of $M_{\rm max}$.

The magnetic field strength inside the remnant MNS is likely to
  be significantly enhanced due to the Kelvin-Helmholtz instability in
  the shear layer at the onset of the merger~\citep{2015PhRvD..92l4034K, 2017arXiv171001311K}.
  Then, it is natural to suppose that MHD turbulence would be induced, and, consequently, the MHD-driven viscosity is likely to arise.
If the lifetime of the MNS is sufficiently long, i.e., $\agt$ several tens of ms, the early viscosity-driven mass ejection could occur due to an induced sound wave and resulting shock wave generated associated with the variation of the density profile of the MNS caused by the angular momentum transport driven by the MHD turbulence.
Our present study indicates that the early viscosity-driven mass ejection proceeds for $\alt 0.1$\,s in a moderately isotropic manner with $\theta \agt 30^\circ$.
The mass of this ejecta depends on the magnitude
  of the turbulent viscosity, but the mass of $\sim 0.01\ M_\odot$ could be ejected for a reasonable value of the viscosity
  parameter of $\alpha_{\rm vis}=0.01$--0.04~\citep{2017arXiv171001311K}.
  For the fiducial model ($\alpha_{\rm vis}=0.02$), the mass of the early viscosity-driven ejecta is $\sim 0.01\ M_\odot$.
We note that this mass can be larger if the initial torus mass is larger. The typical velocity is in the range 0.15--$0.2\,c$.
  
\begin{figure*}[t]
\begin{center}
\begin{minipage}{0.68\hsize}
\includegraphics[bb=0 0 1000 550,width=\hsize]{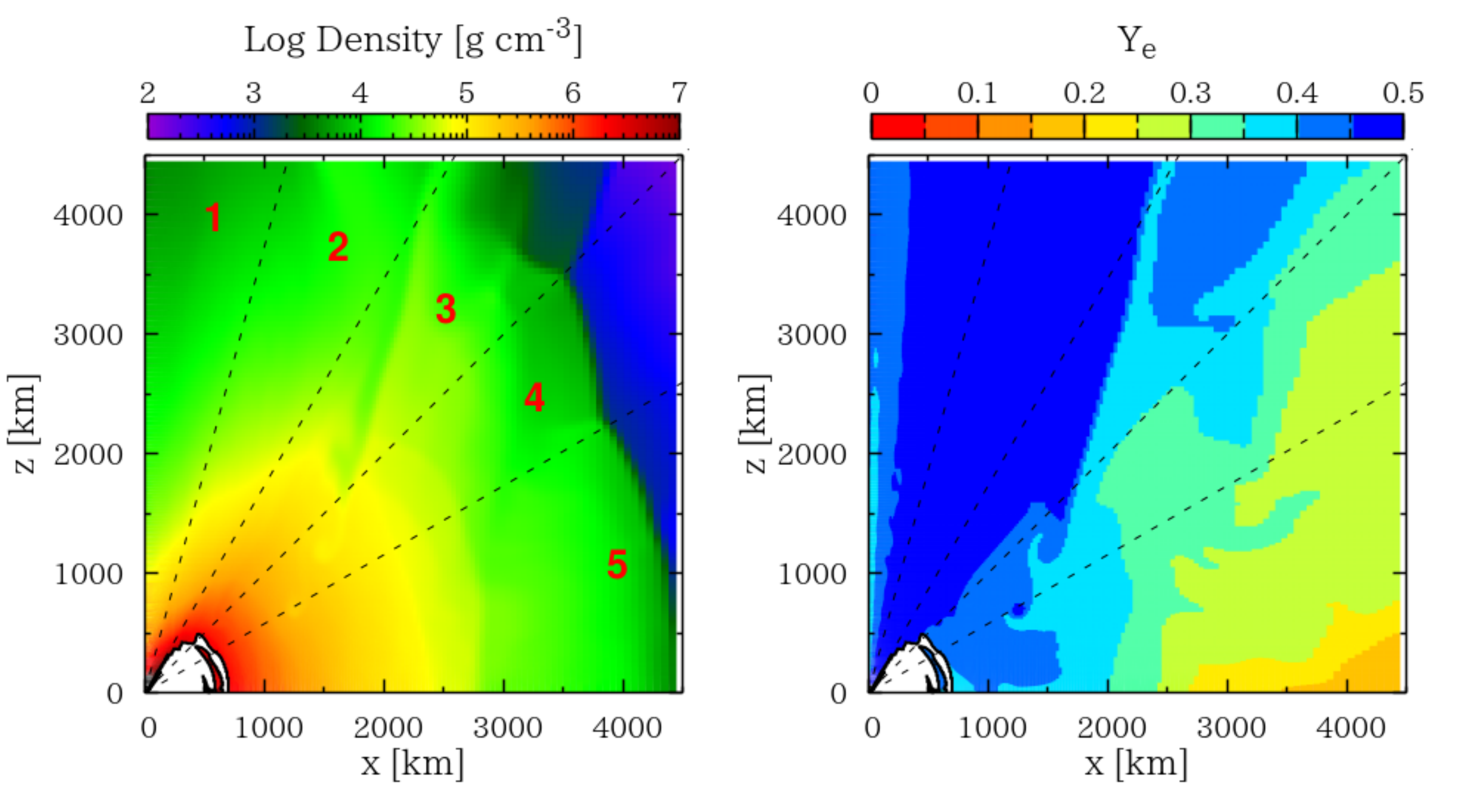}
\end{minipage}
\begin{minipage}{0.3\hsize}
\vspace{8mm}
\includegraphics[width=\hsize]{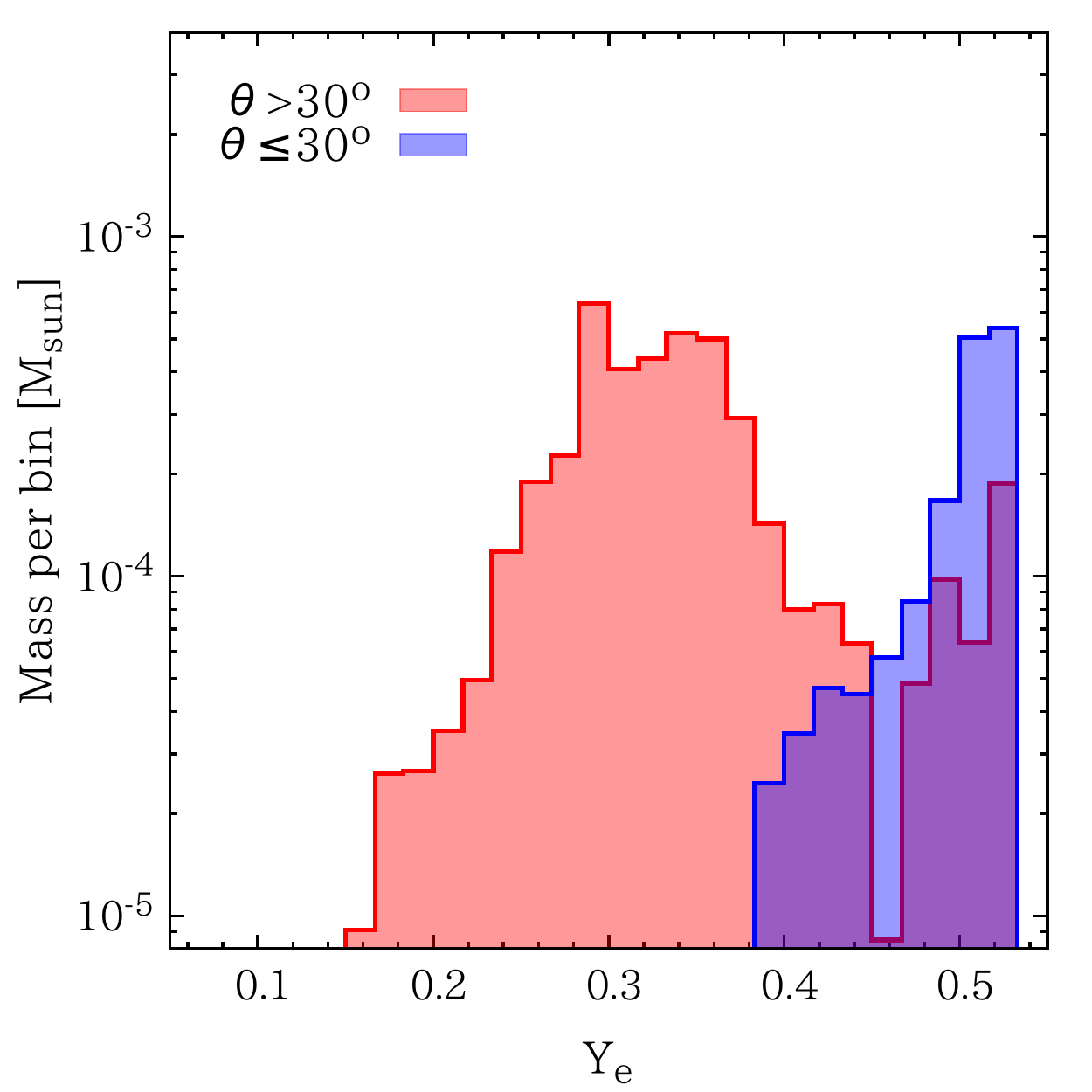}
\end{minipage}
\begin{minipage}{0.68\hsize}
\includegraphics[bb=0 0 1000 550,width=\hsize]{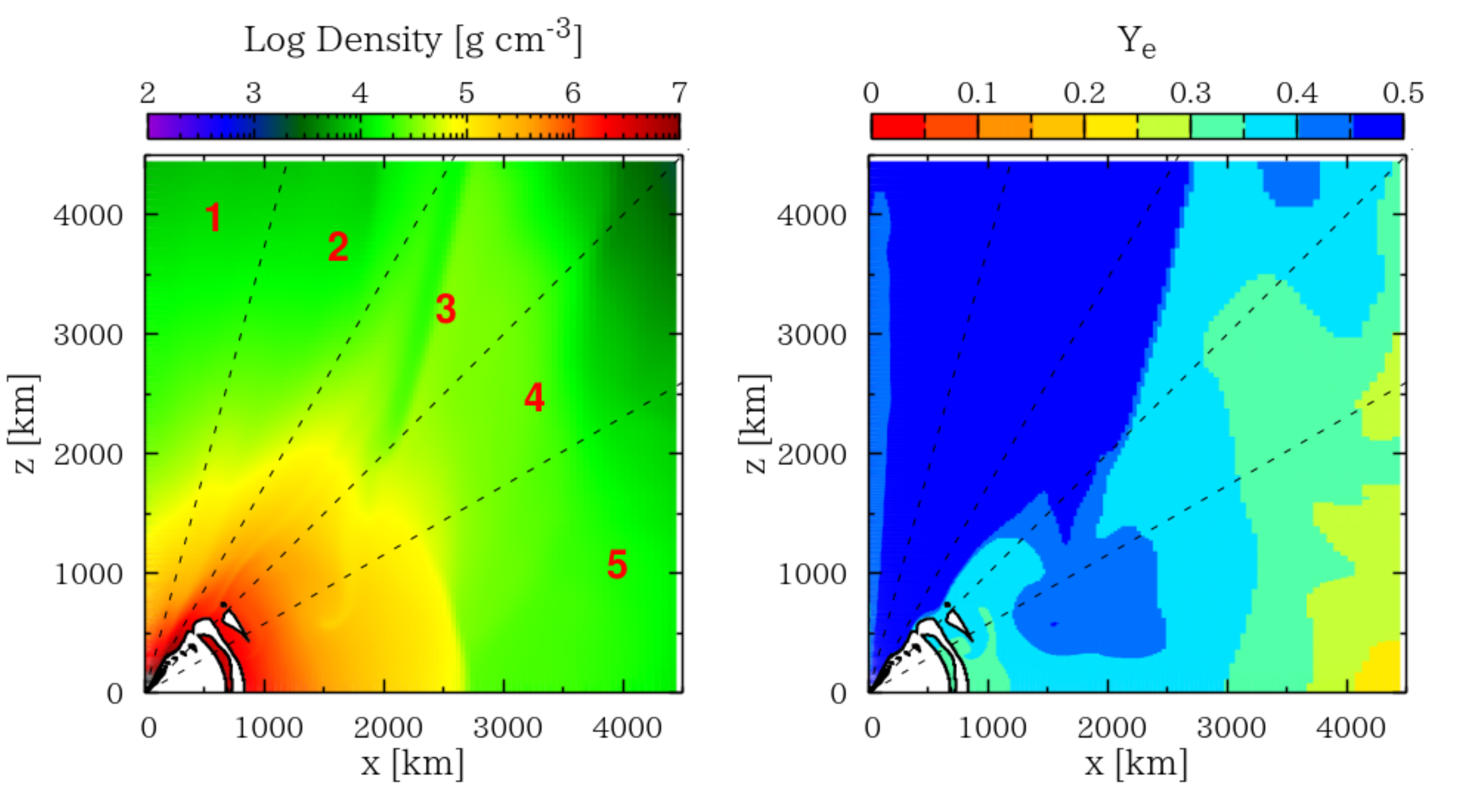}
\end{minipage}
\begin{minipage}{0.3\hsize}
\vspace{8mm}
\includegraphics[width=\hsize]{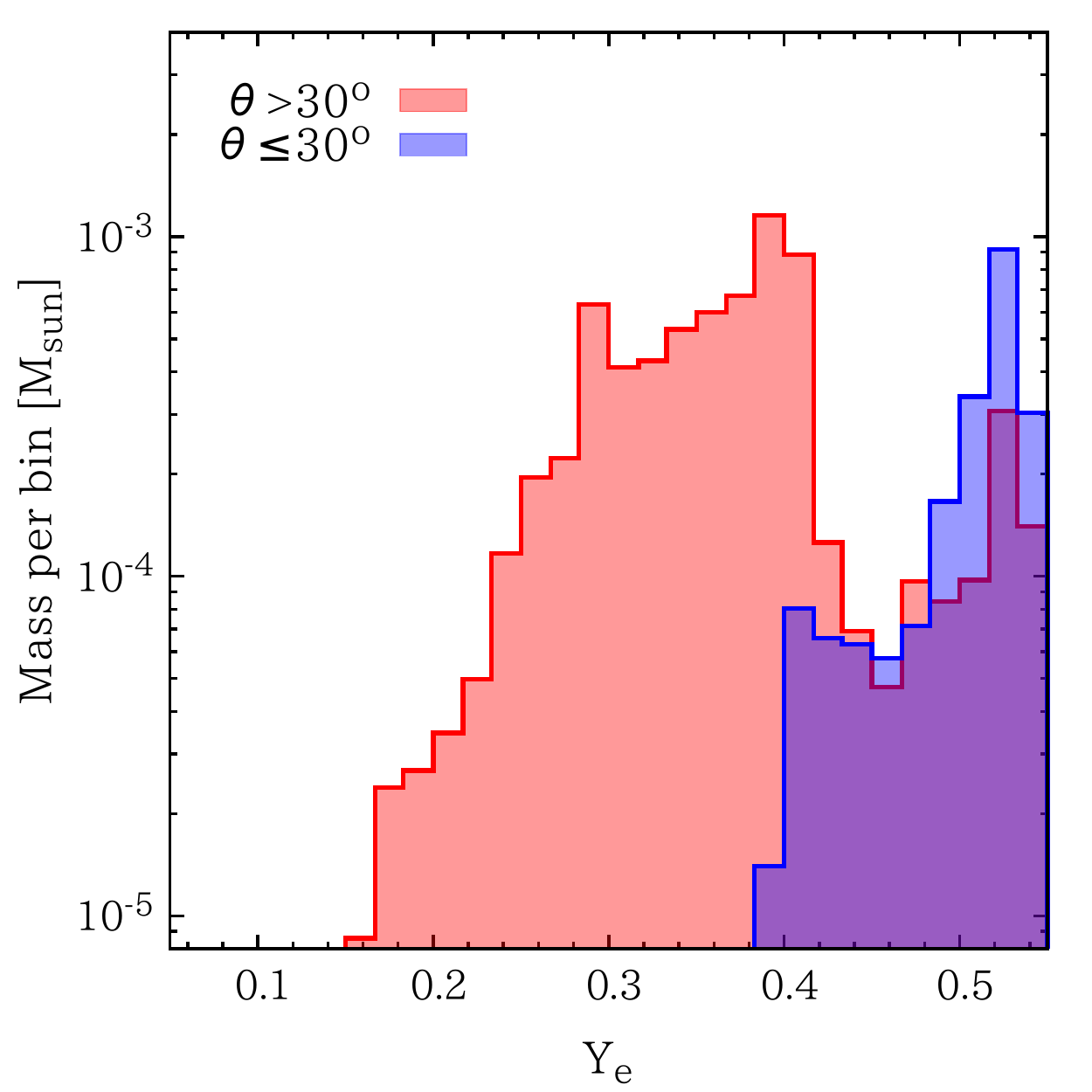}
\end{minipage}

\caption{Left panels: snapshots of the density profiles of
  the ejecta for the models DD2-135135-0.02-H (top) and
  DD2-135135-0.04-H (bottom) at $t=0.1$\,s.  Four black dashed lines
  and numbers label the five angular regions ($0^\circ\leq \theta <
  15^\circ$, $15^\circ\leq \theta < 30^\circ$, $30^\circ\leq \theta <
  45^\circ$, $45^\circ\leq \theta < 60^\circ$, and $60^\circ\leq
  \theta \leq 90^\circ$ with polar angle $\theta$).
Middle panels: snapshots of the electron fraction profiles for the same models.
Right panels: mass histogram of the ejecta as a function of $Y_e$  for $\theta \leq 30^\circ$ and $30^\circ < \theta \leq 90^\circ$.
Note that the mass histogram is generated for all of the ejecta components calculated by Eq.~\eqref{eq:mej}.
}
\label{fig:ejprof100}
\end{center}
\end{figure*}

The left two panels of Fig.~\ref{fig:ejprof100} show the density and electron fraction profiles of the ejecta at $t=0.1$\,s for $\alpha_{\rm vis}=0.02$ and 0.04.
For both models, the low-electron fraction ejecta exists near the equatorial plane (region 5 in the left panels of Fig.~\ref{fig:ejprof100}).
This is composed partly of the dynamical ejecta and mainly of the early viscosity-driven ejecta.
Since the power for the early viscosity-driven mass ejection is higher for $\alpha_{\rm vis}=0.04$, the low-electron fraction ejecta is located at a more distant zone for this model.
We find that the electron fraction of the early viscosity-driven ejecta is in the range 0.2--0.5 with the peak at $Y_e\sim0.3$--0.4 (see the right panels of Fig.~\ref{fig:ejprof100}).

\begin{figure*}[t]
\begin{center}
\begin{minipage}{0.68\hsize}
\includegraphics[bb=0 0 1000 550,width=\hsize]{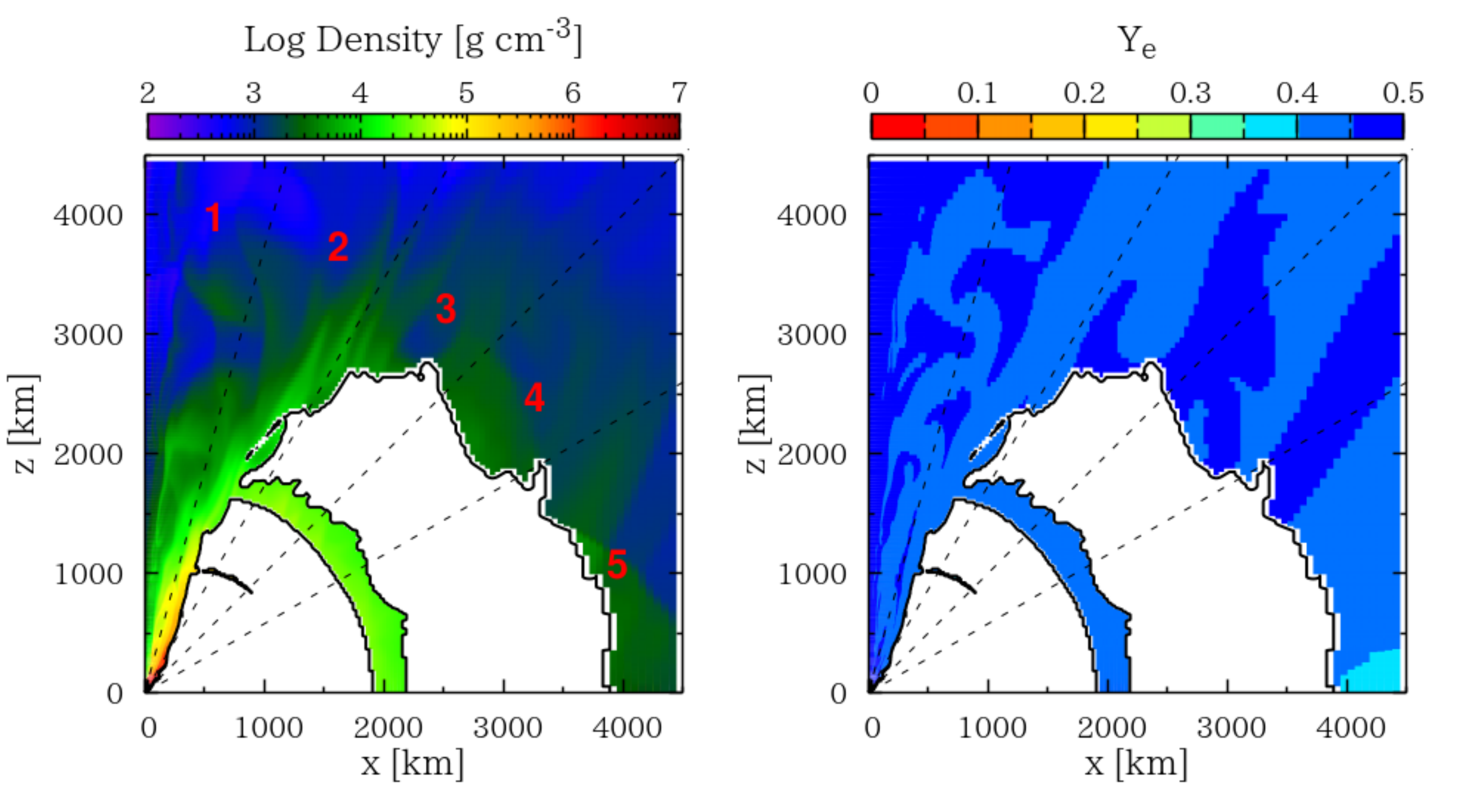}
\end{minipage}
\begin{minipage}{0.3\hsize}
\vspace{8mm}
\includegraphics[width=\hsize]{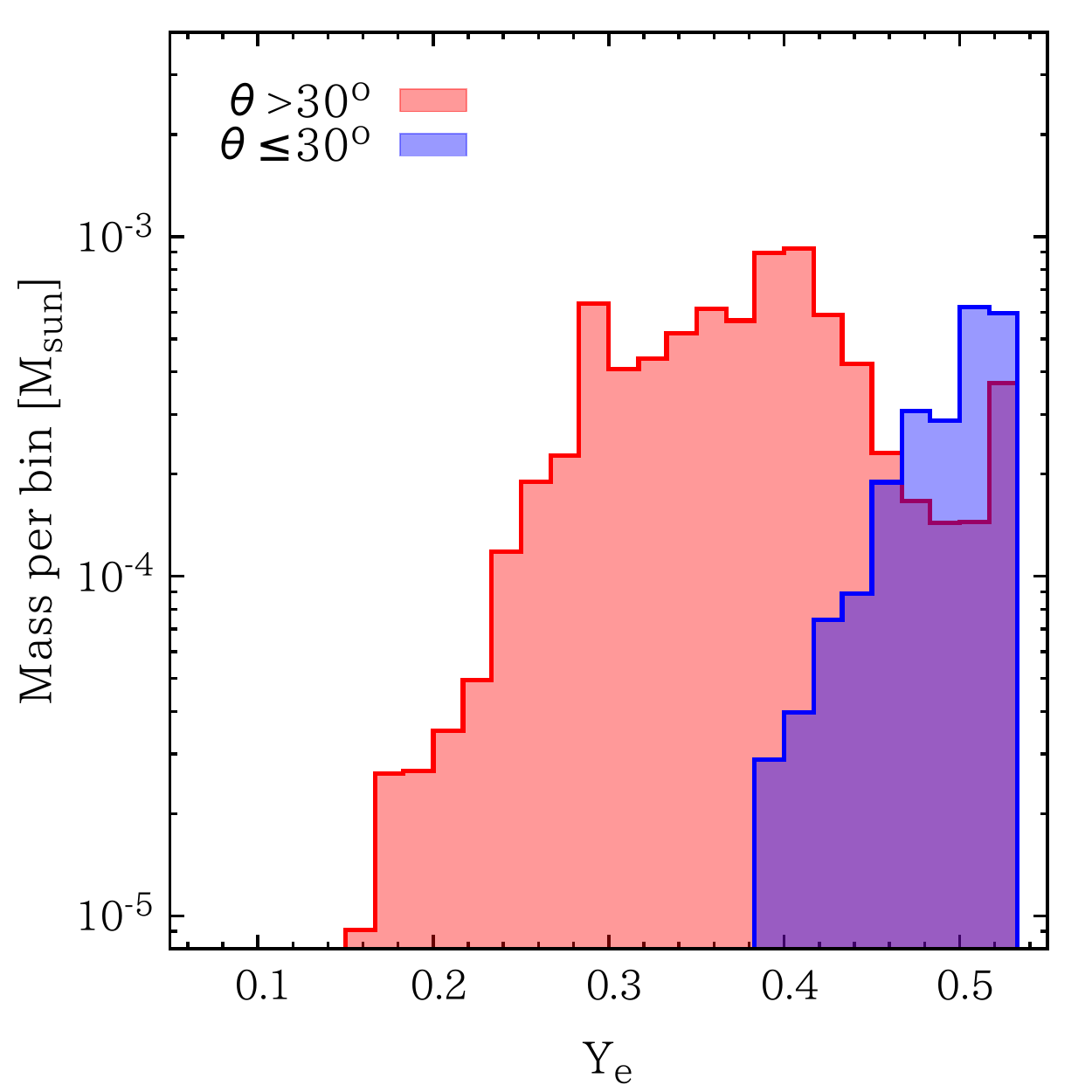}
\end{minipage}
\begin{minipage}{0.68\hsize}
\includegraphics[bb=0 0 1000 550,width=\hsize]{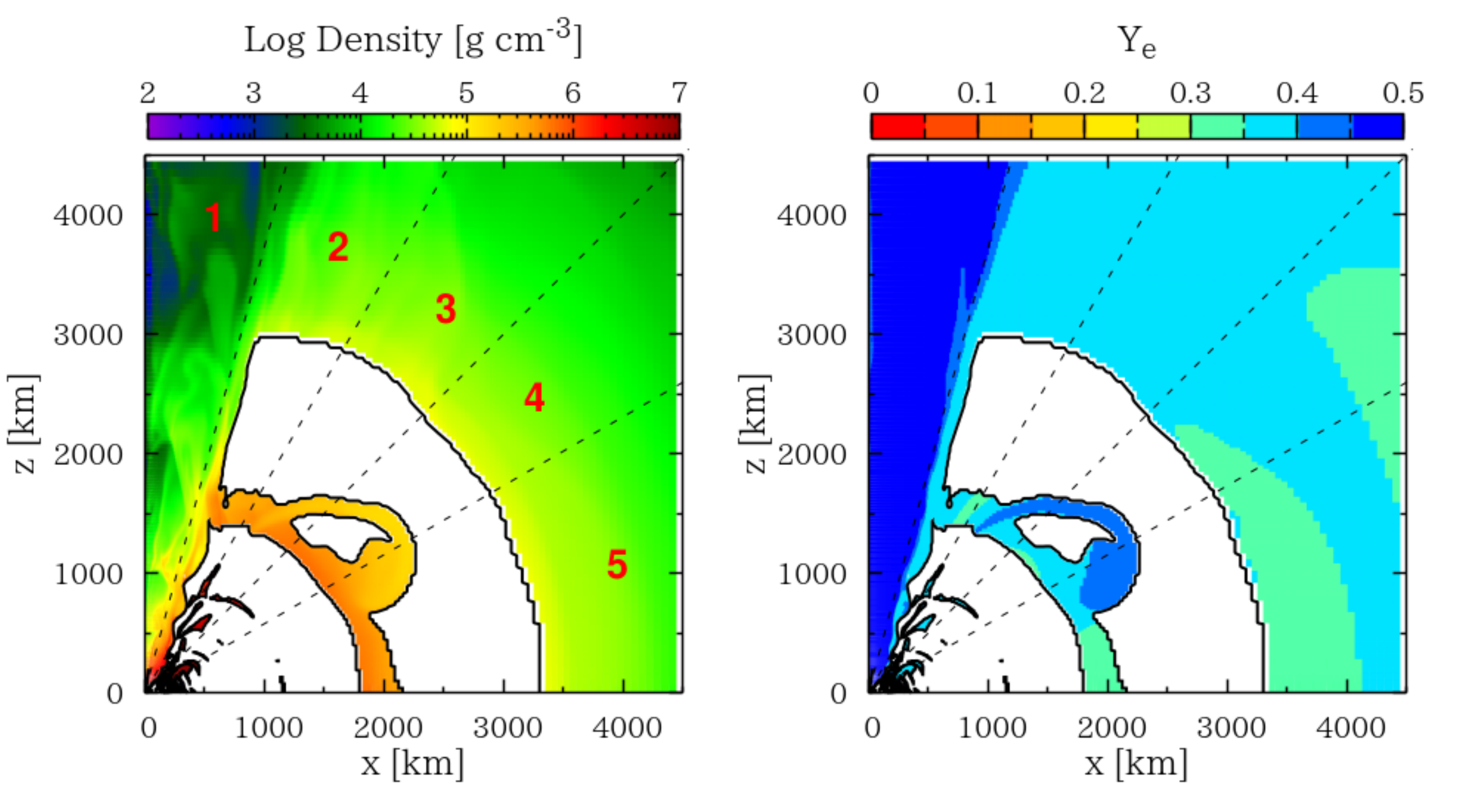}
\end{minipage}
\begin{minipage}{0.3\hsize}
\vspace{8mm}
\includegraphics[width=\hsize]{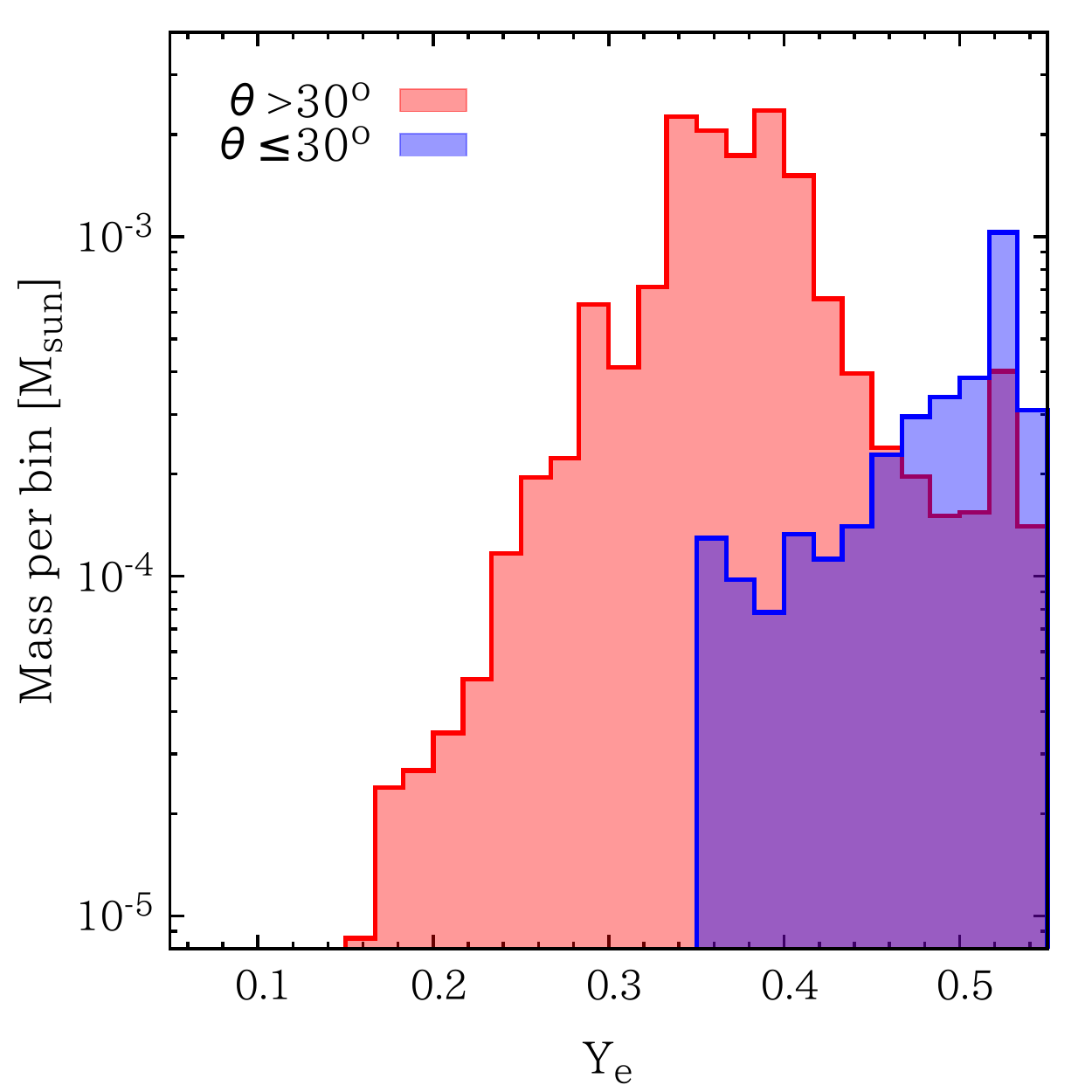}
\end{minipage}
\caption{
Same as Figure~\ref{fig:ejprof100} but at $t=1$\,s.
For $\alpha_{\rm vis}=0.02$ (top panels), the ejecta shown in these panels is composed mainly of the viscosity-driven component toward the polar direction, while for $\alpha_{\rm vis}=0.04$ (bottom panels), it is composed primarily of the viscosity-driven component toward the equatorial direction.
In the white region, matter is gravitationally bound ($|hu_t|<1$).
}
\label{fig:ejprof500}
\end{center}
\end{figure*}

In the presence of a long-lived MNS, the late-time viscosity-driven mass ejection occurs toward the polar direction for $t\gtrsim 0.2$\,s.
The top left and middle panels of Fig.~\ref{fig:ejprof500} show the density and electron fraction profiles of this polar ejecta for the fiducial model at $t=1$\,s.
The ejecta for $\theta \alt 30^\circ$ is launched from the region near the central MNS in assistance with the neutrino irradiation toward the polar direction (regions 1 and 2 in the left panels of Fig.~\ref{fig:ejprof500}); hence, the velocity of the ejecta is $\sim 0.15\,c$.
Due to the strong neutrino irradiation, the electron fraction of the ejecta is increased to be quite high as 0.4--0.5.
The properties of the polar ejecta for the model with $\alpha_{\rm vis}=0.04$ are similar to those of the ejecta for the fiducial model.

The late-time viscosity-driven mass ejection from the expanded torus occurs as we found for the model with $\alpha_{\rm vis}=0.04$ (see bottom left and middle panels of Fig.~\ref{fig:ejprof500}).
Even for the lower viscosity parameter models, this type of the ejecta is likely to be induced for a later phase, as discussed in Sec.~3.4.
This mass ejection is likely to occur irrespective of the presence of the long-lived MNS, and the velocity of
  this ejecta is $\sim 0.03-0.05\,c$~\citep{2014MNRAS.441.3444M}.
  This velocity is appreciably smaller than
  that for the polar ejecta component because it is launched from the outer
  region of the torus for which the typical velocity scale is $<0.1\,c$.
In the presence of the long-lived MNS, the effect of the neutrino irradiation is significant enough to increase the electron fraction above $\sim 0.3$ for this ejecta.
The bottom middle panel of Fig.~\ref{fig:ejprof500} shows that the late-time viscosity-driven ejecta is moderately neutron-rich as $Y_e=0.3$--0.4 for $\theta \agt 30^\circ$.
In the absence of the long-lived MNS, the neutrino irradiation is quite minor; hence, the electron faction of the ejecta would be relatively low, as shown by~\cite{2014MNRAS.441.3444M} and \cite{2015MNRAS.448..541J}.
Note that for a shorter lifetime of the MNS with $\lesssim 1$\,s, the amount of the neutron-rich matter with $Y_e<0.3$ would be larger than what we found.

  Because the mass accretion onto the MNS is significantly
  suppressed for the late time $t \agt 1$\,s, an appreciable fraction
  of the torus material is likely to be ejected as 
  the late-time viscosity-driven ejecta.
The torus mass for the late time ($t\agt
  1$\,s) is $\sim 0.05M_\odot$ (see the top panel of
  Fig.~\ref{fig:accretion1}); hence, the mass of this late-time
  ejecta could be $\sim 0.05M_\odot$. We note that this mass can be
  larger if the initial torus mass is larger.

In Table~\ref{tab:ejecta}, we summarize the mass, typical
  velocity, electron fraction, direction of the ejection, and duration
  for each ejecta component. This is likely to show a universal
  picture of the mass ejection process from the merger remnant for
  the case that a long-lived MNS is formed (Case I in Fig.~\ref{fig:ponchi}).
  On the other hand, if a long-lived MNS is not formed after the merger (Case II in Fig.~\ref{fig:ponchi}), early viscosity-driven
  ejecta and late-time polar viscosity-driven ejecta would be minor, and the late-time equatorial viscosity-driven ejecta would have a much lower value of $Y_e$.
We also note that if the lifetime of the MNS is shorter than $\alt 1$\,s, the value of $Y_e$ for the late-time viscosity-driven ejecta could be smaller than 0.3~\citep{2014MNRAS.441.3444M,2017MNRAS.472..904L}.

\begin{table*}[t]
\caption{ Properties of the ejecta for the DD2 EOS.
Table shows the
  component of ejecta, ejecta mass, average velocity, electron
  fraction, direction of the mass ejection, and ejection duration.
  $t_\nu$ and $t-t_{\rm merge}$ denote the duration of the neutrino
  emission and the time after the onset of merger, respectively.
}
\begin{center}

\begin{threeparttable}

\begin{tabular}{llllll}
\hline \hline
Type of Ejecta & Mass $(M_\odot)$ & $V_{\rm ej}/c$ & $Y_e$
& Direction & Duration\\ \hline 
Dynamical ejecta & $O(10^{-3})$ &
$\sim 0.2$ & 0.05--0.5 & $\theta\gtrsim45^\circ$ & $t-t_{\rm merge}\lesssim
10$\,ms \\ 
Early viscosity-driven ejecta & $\sim 10^{-2} (\alpha_{\rm
  vis}/0.02)$ & $\sim 0.15-0.2$ & 0.2--0.5 & $\theta \agt 30^\circ$& $t-t_{\rm
  merge}\lesssim 0.1$\,s \\ 
Late-time viscosity-driven ejecta (polar)
& $\sim 10^{-3} (t_\nu/{\rm s}) $ & $\sim 0.15$ & 0.4--0.5\tnote{a}
& $\theta\lesssim30^\circ$ & $t-t_{\rm merge}\sim t_\nu\sim 10$\,s
\\ 
Late-time viscosity-driven ejecta (equatorial)
& $\gtrsim 10^{-2} $ & $\sim 0.05$ &
0.3--0.4\tnote{a} & $\theta \gtrsim 30^\circ$ & $t-t_{\rm merge}\sim$ 1--10\,s
\\ \hline
\end{tabular}

\label{tab:ejecta}
\begin{tablenotes}\footnotesize 
\item[a] We note that for the EOS in which the value of $M_{\rm max}$ is not as high as that for the DD2 EOS, the MNS could collapse to a black hole within $\sim1$\,s. Even for such EOS, the late-time viscosity-driven mass ejection should continue after the black hole formation, but because of shorter neutrino irradiation time, the value of $Y_e$ is likely to be smaller than 0.3.
\end{tablenotes}
\end{threeparttable}
\end{center}
\end{table*}

At the end of this subsection, we emphasize that the viscous hydrodynamics we employ in this work is an effective approach to take into account the angular momentum transport and heating due to the MHD turbulence; hence, the values of the viscosity parameter we assumed in this work should be justified by performing very high-resolution MHD simulations that resolve a variety of MHD instabilities in the future.

\subsection{Elemental Abundance in the Ejecta and Implications for the Electromagnetic Signals}\label{sec4.2}
Here we discuss the elemental abundance in the post-merger ejecta.
We note that we do not consider the dynamical ejecta in this subsection.

\begin{figure*}[t]
\begin{minipage}{0.48\hsize}
\includegraphics[width=\hsize]{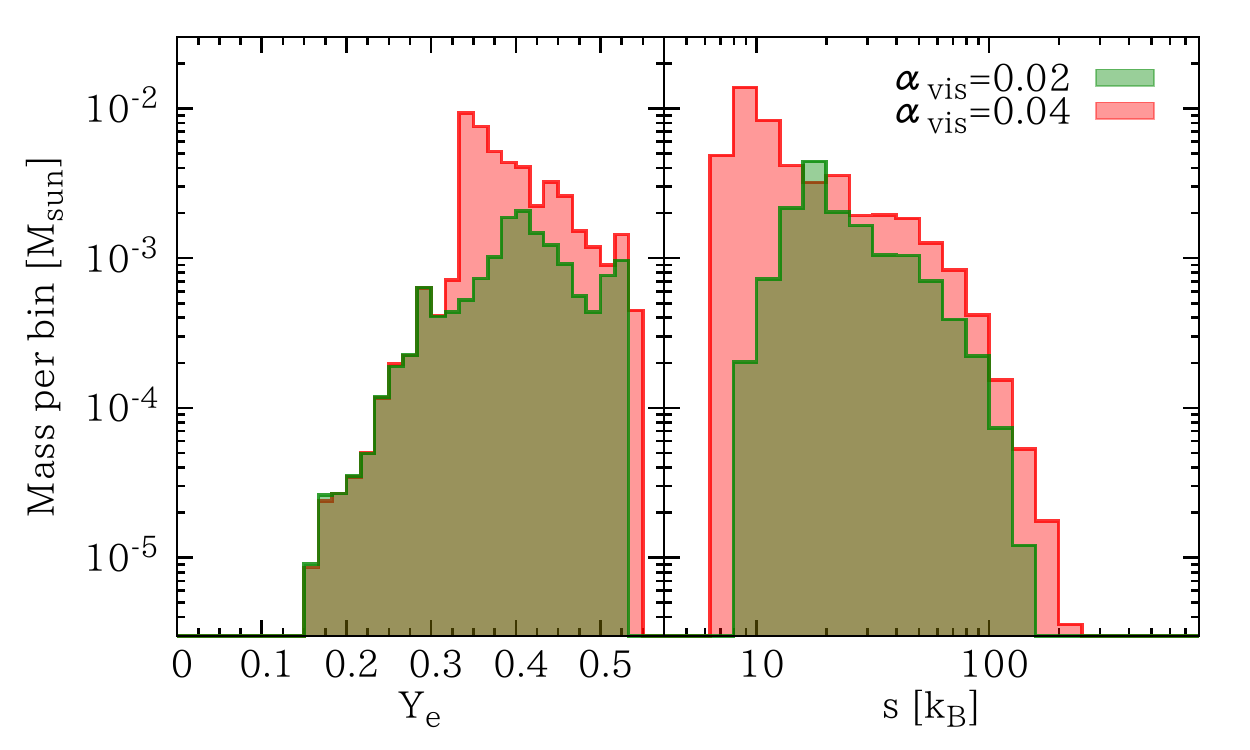}
\caption{ Mass histogram of the ejecta as a function of $Y_e$ (left)
  and $s/k_B$ (right) at $t=2.8$\,s.  The distributions of the
  electron fraction (left panel) and the specific entropy (right
  panel) are plotted.  The red and green curves denote the
  results for the models DD2-135135-0.04-H and DD2-135135-0.02-H,
  respectively.  }
\label{fig:hist}
\end{minipage}
\hspace{5mm}
\begin{minipage}{0.48\hsize}
\includegraphics[width=\hsize]{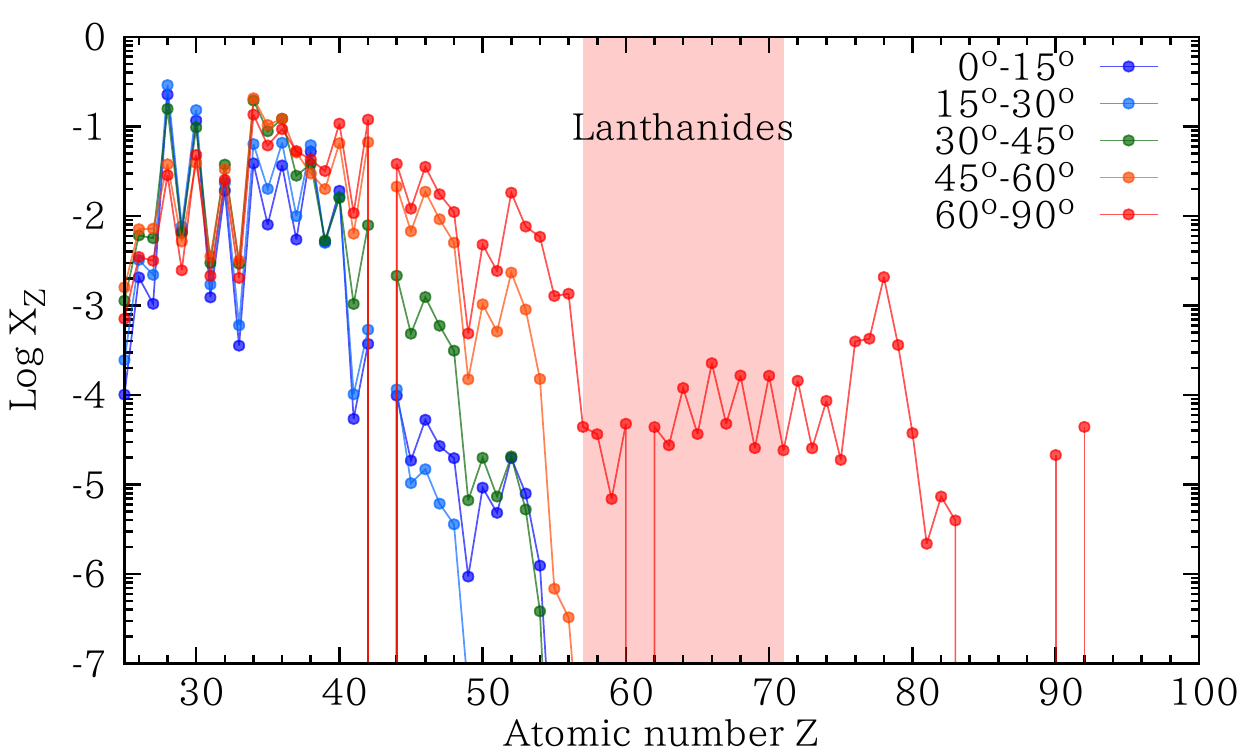}
\caption{ Mass fraction of the nuclei as a function of atomic number
  for the model DD2-135135-0.04-H at $t=2.8$\,s.  The different color curves
  correspond to the results in different angular regions.  The shaded
  region indicates the range of the lanthanide elements ($Z=$57--71).
 }
\label{fig:xz}
\end{minipage}
\end{figure*}

\begin{table}[t]
\caption{
%
Mass Fraction of Lanthanides $(Z=57-71)$ and Actinides $(Z=83-103)$, Baryon Mass of the Material Ejected for the Angular Regions Shown in Fig.\,\ref{fig:ejden}.
}
\begin{center}
\begin{tabular}{cllll}
\hline \hline
& \multicolumn{2}{l}{$\alpha=0.04$\, ($t=2.8$\,s)} & \multicolumn{2}{l}{$\alpha=0.02$\, ($t=3.3$\,s)} \\
\hline
Region & $X_{\rm lan+ac}$ & $M_{\rm ej}/M_\odot$ & $X_{\rm lan+ac}$ & $M_{\rm ej}/M_\odot$ \\
\hline
1 & $2.1\times10^{-10}$ & 0.0031 &$8.3\times10^{-12}$ & 0.0015\\
2 & $2.0\times10^{-12}$ & 0.0088 & $9.2\times10^{-13}$ & 0.0041\\
3 & $1.2\times10^{-11}$ & 0.0095 & $2.3\times10^{-11}$ & 0.0038\\
4 & $7.1\times10^{-8}$ & 0.0089 & $6.0\times10^{-8}$ & 0.0036\\
5 & $1.1\times10^{-3}$ & 0.0162 & $1.2\times10^{-3}$ & 0.0048\\
\hline
Total & $3.8\times10^{-4}$ & 0.046 & $3.1\times10^{-4}$ & 0.018\\
\hline
\end{tabular}
{\footnotesize
Notes. The material ejected by $t=2.8$ and 3.3\,s is analyzed for the $\alpha=0.04$ and 0.02 models, respectively.
The mass fraction of actinide elements is minor compared to that of lanthanides for all angular regions.
The actinide mass fraction is less than 10\% of that of lanthanide elements even for region 5, where the actinide mass fraction is highest.
}
\end{center}
\label{tab:lanthanide}
\end{table}

Figure~\ref{fig:hist} shows the mass histogram of the ejecta as a function of
the electron fraction and specific entropy at $t=1$\,s for
$\alpha_{\rm vis}=0.02$ and 0.04.
As found in this figure, the electron fraction of the ejecta is widely
distributed, but the mass of the ejecta component with $Y_e \lesssim 0.25$ is minor.
In this ejecta, it is expected that the $r$-process elements heavier than the second peak, including lanthanide elements, are not significantly synthesized~\citep[e.g.,][]{2014ApJ...789L..39W,2015ApJ...813....2M, 2018ApJ...852..109T}.

Note that the mass histogram of the ejecta for $\alpha_{\rm vis}=0.04$ exhibits remarkable excess in $Y_e \approx 0.3-0.4$ and $s/k_{\rm B} \lesssim 10$, compared to those for $\alpha_{\rm vis}=0.02$.
The reason for this is that for $\alpha_{\rm vis}=0.02$, the ejecta at $t=1$\,s is composed mainly of the early viscosity-driven and the late-time polar viscosity-driven ejecta, while for $\alpha_{\rm vis}=0.04$, it is additionally composed of the late-time equatorial viscous-driven ejecta.
Thus, the excess found for $\alpha_{\rm vis}=0.04$ indicates that the late-time equatorial viscosity-driven ejecta has $Y_e=0.3$--0.4 and $s/k_B \alt 10$. 

We performed a nucleosynthesis calculation as a post-process to the
ejecta as done in~\cite{2014ApJ...789L..39W}.
In our scheme, the temporal evolution of temperature, density, and $Y_{\rm e}$ are
obtained by using a tracer particle method~\citep[see ][for details
  and methodology]{2015ApJ...810..109N}. In this work, we employed a
nuclear reaction network by \cite{2016PhLB..756..273N}, of which the
base theoretical nuclear mass formula is the Finite-Range Droplet Model (FRDM)
\citep{1995ADNDT..59..185M}.

Figure~\ref{fig:xz} shows the mass fraction of the nuclei as a function of atomic number $Z$ for five angle bins shown in Figs.~\ref{fig:ejprof100} and \ref{fig:ejprof500} for the model with $\alpha_{\rm vis}=0.04$, in which the late-time equatorial viscosity-driven ejecta is found clearly.
As expected from the $Y_e$ histogram, the $r$-process nucleosynthesis does not proceed sufficiently, and, remarkably, the mass fraction of lanthanide elements is very small.  Especially, the material ejected to angular regions 1--4 in Fig.~\ref{fig:ejprof100} (i.e., $\theta < 60^\circ$) is approximately lanthanide-free.
In Table\,\ref{tab:lanthanide}, the mass fraction of lanthanide and actinide elements in the individual angular regions is shown.
We found that the mass fraction is $\lesssim10^{-7}$ for angular regions 1--4.
If the ejecta is contaminated only minorly by the lanthanide elements, the opacity of the ejecta is $\ll 10\,{\rm cm^2\, g^{-1}}$ and would be $\sim 0.1-1\,{\rm cm^2\, g^{-1}}$ \citep{2018ApJ...852..109T}.
According to the standard macronova/kilonova model~\citep{1998ApJ...507L..59L, 2010MNRAS.406.2650M}, if we observe the post-merger ejecta directly, the time to reach the peak emission, peak luminosity, and its effective temperature are estimated to give
\begin{widetext}
\begin{align}
t_{\rm peak} &\approx  \sqrt{\frac{\kappa M_{\rm ej}}{4\pi c V_{\rm ej}}\xi}
 \approx 1.5\,{\rm days}\ \biggl(\frac{V_{\rm ej}}{0.1\,c}\biggr)^{-1/2} 
\biggl(\frac{M_{\rm ej}}{0.03\ M_\odot}\biggr)^{1/2}
\biggl(\frac{\kappa}{0.3\,{\rm cm^2\,g^{-1}}}\biggr)^{1/2} \xi^{1/2},\\
L_{\rm peak} &\approx \frac{fM_{\rm ej}c^2}{t_{\rm peak}} 
\approx 4.3\times 10^{41} {\rm erg\,s^{-1}} \ 
\biggl(\frac{f}{10^{-6}}\biggr) 
\biggl(\frac{V_{\rm ej}}{0.1\,c}\biggr)^{1/2} 
\biggl(\frac{M_{\rm ej}}{0.03\ M_\odot}\biggr)^{1/2}
\biggl(\frac{\kappa}{0.3\,{\rm cm^2\,g^{-1}}}\biggr)^{-1/2} \xi^{-1/2},\\
T_{\rm eff,peak} &\approx \biggl[\frac{L_{\rm peak}}
{4\pi (V_{\rm ej}t_{\rm peak})^2 \sigma_{\rm SB}}\biggr]^{1/4} 
\approx 8 \times 10^3\,{\rm K}\ \biggl(\frac{f}{10^{-6}}\biggr)^{1/4} 
\biggl(\frac{V_{\rm ej}}{0.1\,c}\biggr)^{1/8} 
\biggl(\frac{M_{\rm ej}}{0.03\ M_\odot}\biggr)^{-1/8}
\biggl(\frac{\kappa}{0.3\,{\rm cm^2\,g^{-1}}}\biggr)^{-3/8}\xi^{-3/8},
\end{align}
\end{widetext}
where $f$ is the radioactive energy deposition factor,\footnote[2]{We note that $f$ is time-varying and proportional to $\approx t^{-1.3}$~\citep[e.g.,][]{2010MNRAS.406.2650M,2017MNRAS.468...91H}.}
$\kappa$ is the opacity of the material, $\xi (\leq 1)$ is a geometric
factor, and $\sigma_{\rm SB}$ is the Stefan--Boltzmann constant.
Here we suppose that the average velocity is $0.1\,c$.
These estimates show that 
if we observe this post-merger ejecta directly (i.e., from a low opening angle $\theta \lesssim 45^\circ$), the electromagnetic signal would be of a short-timescale, high-luminosity, and blue transient.

We note that in the ejecta component for which $Y_e\gtrsim 0.4$, the nucleosynthesis products are likely to have a smaller heating rate (or smaller value of $f$).
\cite{2014ApJ...789L..39W} showed that the specific heating rate for the ejecta with $Y_e\gtrsim 0.35$ is much smaller than that of more neutron-rich ejecta.
Thus, the high-electron-fraction ejecta may play a minor role as the energy source of the electromagnetic signal (i.e., $f$ could be much smaller than $10^{-6}$) even if the ejecta mass is much larger than that of the dynamical ejecta.
We should also note that \cite{2014ApJ...789L..39W} considered only dynamical ejecta.
The late-time ejecta irradiated by neutrinos has higher entropy than the dynamical ejecta.
Even in the high-electron-fraction material, heavy elements can be synthesized if the material has sufficiently high entropy and expansion velocity \citep[see, e.g.,][]{1997ApJ...482..951H}.

As we found in this paper, the early viscosity-driven ejecta and late-time equatorial viscosity-driven ejecta could have large mass $\agt 0.01M_\odot$ and moderately small values of $Y_e$ (0.2--0.5 and 0.3--0.4, respectively).
Such ejecta is likely to be the major heating source and contribute to electromagnetic counterparts as the energy source.
Note that the mass of the late-time polar viscosity-driven ejecta is likely to be much smaller, and $Y_e$ is large as $\gtrsim 0.4$; hence, their contribution would be minor.

We note that the maximum mass for cold spherical neutron stars for the DD2 EOS is $M_{\rm max}\approx 2.4M_\odot$. If the value of $M_{\rm max}$ is not as high as this value for the neutron star EOS in nature, the MNS could collapse into a black hole in a few seconds after the merger.
In this case, the electron fraction of the late-time equatorial viscosity-driven ejecta becomes lower than that in the presence of the MNS~\citep{2014MNRAS.441.3444M,2015MNRAS.448..541J,2017MNRAS.472..904L}; hence, the viscosity-driven ejecta would be more lanthanide-rich.
We plan to perform simulations for such an EOS in the future work.

\subsection{Effects of the Assumptions and Approximations Made in Our Simulations}

\begin{figure}[t]
\includegraphics[width=\hsize]{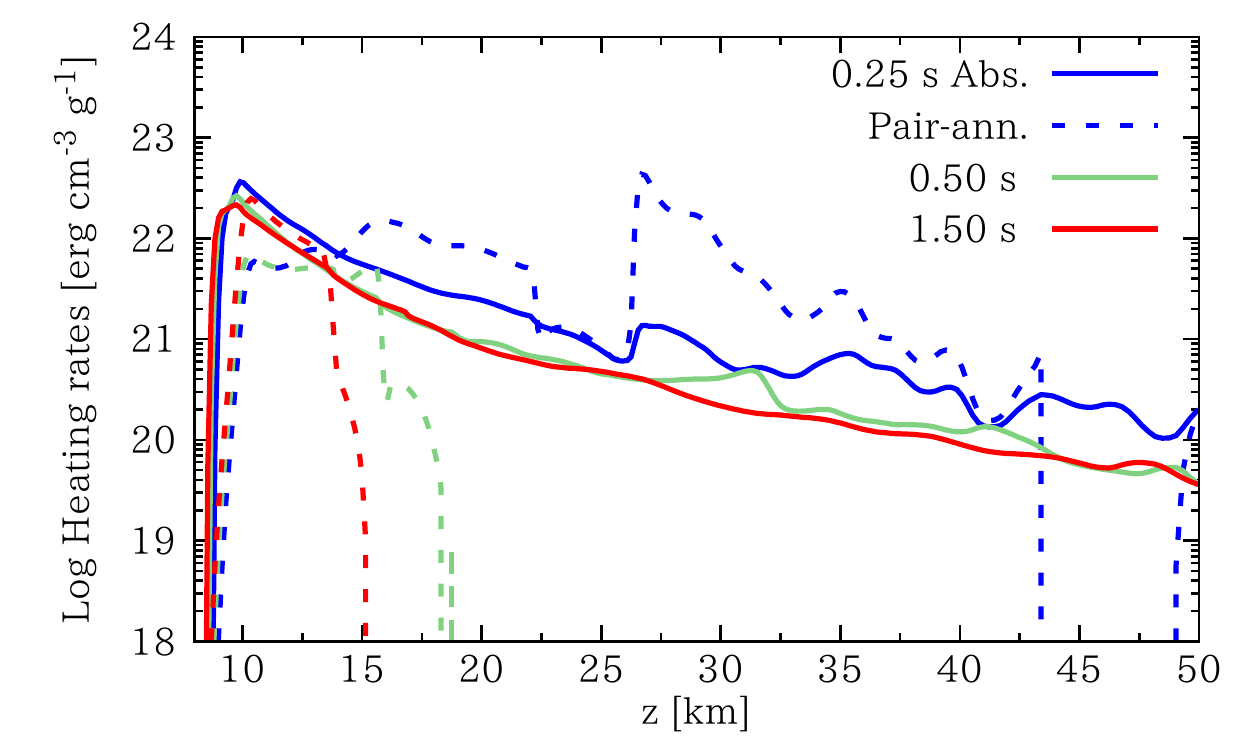}
\caption{
Specific heating rates for neutrino absorption (solid) and pair-annihilation (dashed) processes at $t=0.25$, 0.5, and 1.5\,s.
}
\label{fig:heating}
\end{figure}
\begin{figure}[t]
\includegraphics[width=\hsize]{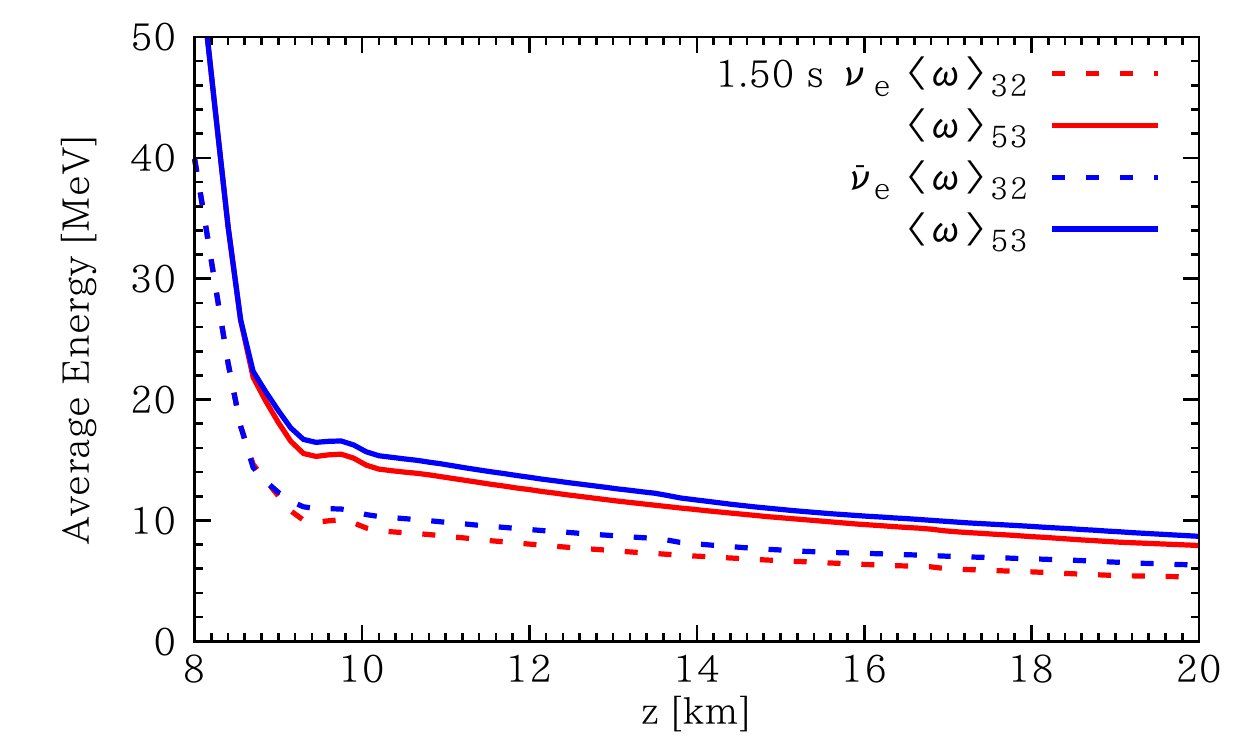}
\caption{
Average energy of electron neutrinos (red) and electron antineutrinos (blue) estimated by Equations \,(40) and (41) in \cite{2017ApJ...846..114F} (dashed and solid curves, respectively) along the rotational axis at $t=1.5$\,s.
}
\label{fig:avee}
\end{figure}


In our simulation, we make several approximations and assumptions that could affect the results.
In \cite{2015MNRAS.448..541J}, the neutrino energy density and flux obtained by the M1 scheme are compared to those calculated by a ray-tracing method for a black hole-torus system.
Their result suggests that the neutrino energy density and flux are overestimated by a factor of $\approx 2$ in the polar direction (with the polar angle $\theta \lesssim 30^\circ$), and underestimated in the more equatorial direction.
In the phase where the neutrino emission from the torus is stronger than that of the MNS, the profile of the neutrino emissivity is nonspherical, so that the neutrino energy density would be overestimated in the polar region, which leads to the overestimation of the neutrino heating rate.
Thus, in the earlier phase of the evolution with $t\lesssim 0.5$\,s during which the torus dominates over the whole neutrino emission (see Fig.\,\ref{fig:accretion2}), the polar ejecta may be affected by the behavior of the M1 scheme.

However, the outflow launched toward the polar direction is initially slow \citep[see Fig.\,7 in][]{2017ApJ...846..114F}, so that the electron fraction of the ejecta in the polar region is settled into an equilibrium value by the neutrino absorption \citep[][]{1996ApJ...471..331Q},
\begin{align}
Y_{e,{\rm eq}} \approx \biggl(1+\frac{L_{\bar{\nu}_e}}{L_{\nu_e}}\frac{\epsilon_{\bar{\nu}_e} -2\Delta}{\epsilon_{\nu_e} +2\Delta}\biggr)^{-1}, \label{eq:yeeq}
\end{align}
where $L_{\nu_e}$ and $\epsilon_{\nu_e}$ are the luminosity and average energy of electron neutrinos, $L_{\bar{\nu}_e}$ and $\epsilon_{\bar{\nu}_e}$ are those of electron antineutrinos, and $\Delta \approx 1.293$\,MeV is the mass difference between neutron and proton.
The equilibrium value is unchanged if the ratio of the fluxes of electron neutrinos and antineutrinos and average energy of neutrinos are the same.
Thus, the effects of the M1 scheme on the electron fraction of the polar ejecta would be minor if the overestimation of the energy density arises for electron neutrinos and antineutrinos at the same level.
On the other hand, in the later phase ($t\gtrsim 0.5$\,s), the MNS dominates over the whole neutrino emission of the system; hence the neutrino emissivity profile becomes more spherical.
Thus, the artificial effects due to the M1 scheme would be minor in this phase.

We should note that for the neutrino pair annihilation, the heating rate also depends on the angular distribution of the neutrinos; hence, its heating rate would be affected artificially by the M1 scheme.
Figure\,\ref{fig:heating} compares the heating rates due to the neutrino absorption and pair-annihilation processes.
While the pair-annihilation heating dominates over the whole heating rate for $t=0.25$\,s, it becomes comparable to the neutrino absorption heating at $z\lesssim 20$\,km for later times due to the decrease of the neutrino emission rate.
Thus, at least for the early phase ($t\lesssim 0.3$\,s), the outflow is primarily powered by the pair-annihilation heating.
Pair-annihilation heating has not been considered in most of the recent works \citep[e.g.,][]{2017MNRAS.472..904L, 2015ApJ...813....2M}.
However, the effect could be large, as found here; hence, we need to consider the pair-annihilation heating appropriately to obtain the properties of the ejecta quantitatively.

Since we approximately determine the average energy of neutrinos for calculating the neutrino absorption rates, the uncertainty in the estimated energy would affect the equilibrium value of the electron fraction of the ejecta.
If the average energy of neutrinos is higher than that estimated in our simulation, the equilibrium value of the electron fraction becomes lower because the mass difference between neutron and proton becomes unimportant for the higher average energy of neutrinos.
Figure\,\ref{fig:avee} shows the average energy of electron-type neutrinos along the rotational axis.
This figure shows that the difference between the average energy estimated by Eqs.\,(40) and (41) in \cite{2017ApJ...846..114F} is $\lesssim 50$\,\%.
Using Eq.\,\ref{eq:yeeq}, the 50\,\% difference in the average energy around 15\,MeV leads to the difference of $\approx 0.06$ in the equilibrium value of the electron fraction if the flux of the electron-type neutrinos is the same.

We approximate viscous effects due to MHD turbulence by solving viscous hydrodynamics equations.
However, if sufficiently strong magnetic fields are globally formed, the Lorentz force would accelerate the ejecta.
This would happen in the polar region because of the low density.
The left panel of Fig.\,\ref{fig:ejden} shows that the density at $z=20$\,km on the rotational axis is $\approx10^7\,{\rm g\,cm^{-3}}$, which suggests that the Lorentz force would be important in the region in which the magnetic field strength exceeds $5\times10^{14}$\,G.
Global magnetic fields could be formed when the outflow is driven because the field line is stretched in the outflow.
Such an outflow is driven possibly by the viscous heating and/or neutrino heating.
Thus, another ejecta component could be generated as the magnetically accelerated wind from the strongly magnetized MNS.
\cite{2018ApJ...856..101M} suggested that a significant mass ejection ($\gtrsim0.01M_\odot$) is possible if the MNS is sufficiently magnetized and rapidly rotating.
In addition, the density profiles shown in Fig.\,\ref{fig:ejden} would be modified in the existence of the global magnetic fields (e.g., \cite{2007ApJ...659..561M} for winds launched from magnetized neutron stars).
We should check whether global magnetic fields are formed in the relevant timescale after the merger and whether they affect the properties of the ejecta by performing neutrino radiation MHD simulations.

The remnants of binary neutron star mergers have been considered to power relativistic jets that would drive short-duration gamma-ray bursts \citep[][]{1989Natur.340..126E, 2007PhR...442..166N}.
If a relativistic jet is launched from the merger remnant, the jet may inject a part of its kinetic energy into the surrounding ejecta and modify its expansion velocity \citep[cocoon;][]{2014ApJ...784L..28N, 2014ApJ...788L...8M}.
Thus, the timescale of the macronova/kilonova emission would be modified to be shorter.
The thermal energy in the cocoon injected from the jet would power another component of electromagnetic transient \citep[cocoon emission;][]{2017ApJ...834...28N}.
Since the post-merger ejecta is lanthanide-poor in our simulation, the timescale of the cocoon emission is likely to be short; hence, this could contribute to the electromagnetic signal in the early phase ($\lesssim$ day).

\section{Summary}

We performed a general relativistic neutrino-radiation-viscous hydrodynamics simulation for a remnant of the binary neutron star merger.
Starting from data for a merger remnant obtained from a fully general relativistic merger simulation, we evolved the remnant MNS and torus together.
This is the first work in which such a remnant system is evolved in a self-consistent manner taking into account the effect of angular momentum transport.

We found that there would be two viscous effects on the evolution of
the merger remnant.  One is the viscous effect on the
differentially rotating MNS, which results in the transition of the
rotational profile of the remnant MNS from a differentially rotating
one to a rigidly rotating one in $\sim 10$\,ms.  The other plays an
important role in the long-term viscous evolution of the torus. 

These viscous effects introduce the mass ejection mechanisms,
  which do not exist in the inviscid case.  As a result of the
  transition of the MNS density profile due to the redistribution of
  its angular momentum, a sound wave that becomes a shock wave eventually is formed in the central region,
  and then the material in the outer region of the torus
  ($r\sim100-1000$ km) is ejected by the shock wave for the duration of
  $\lesssim 0.1$\,s.  After this early viscosity-driven mass ejection
  ceases, the late-time viscosity-driven mass ejection takes place from the torus.  The
  mass ejection with neutrino irradiation is activated toward the polar direction first.  After the
  neutrino cooling becomes inefficient in the torus,
  the viscosity-driven mass ejection from the torus toward the equatorial direction is activated.

Table~\ref{tab:ejecta} summarizes the properties of the ejecta by
  various mass ejection processes. As found from this table, the
  electron fraction of the post-merger ejecta is distributed between
  0.2 and 0.5.  In particular, for the polar direction ($\theta < 45^\circ$), the ejecta has
  higher values of the electron fraction with $Y_e \gtrsim 0.3$.  In such
  ejecta, lanthanide elements are not efficiently synthesized.
The dynamical ejecta of the low-electron fraction, which would contain
  lanthanide elements, is ejected mainly near the equatorial plane.
  Therefore, if we observe the system from the viewing angle less than
  $45^\circ$, the radioactive emission from the viscosity-driven
  ejecta does not suffer from the ``lanthanide curtain"
  \citep{2015MNRAS.450.1777K} of the dynamical ejecta, and we will
  observe a rapid, bright, and blue electromagnetic transient.

This indicates that the electromagnetic emission from the viscosity-driven ejecta could approximately reproduce the electromagnetic signals in the optical--infrared bands associated with GW170817. Our interpretation of these electromagnetic counterparts and further discussion are described in an accompanying paper~\citep{2017PhRvD..96l3012S}.

\acknowledgements
We thank Rodrigo Fern{\'a}ndez, Kunihito Ioka, Brian Metzger, Yudai Suwa, and Takahiro Tanaka for fruitful
discussions.  This work was supported by JSPS Grants-in-Aid for
Scientific Research (17H06363, 17H06361, 16H02183, 15H00836, 15K05077, 15H00783, 16K17706,
16H06341, 15H00782, 26400237), by the HPCI Strategic Program of Japanese MEXT,
and by the MEXT program ``Priority Issue 9 to be tackled by using Post K
computer" (project numbers hp160211, hp170230, hp170313).
Numerical simulations were carried out on the Cray XC40 at the Yukawa Institute for Theoretical Physics, Kyoto University; Cray XC30 at the Center for Computational Astrophysics, National Astronomical Observatory of Japan; and the FX10 at the Information Technology Center, the University of Tokyo.

\appendix

\section{A. Dependency of the Results on the Grid Resolution}

\begin{figure}[t]
\begin{center}
\includegraphics[width=0.5\hsize]{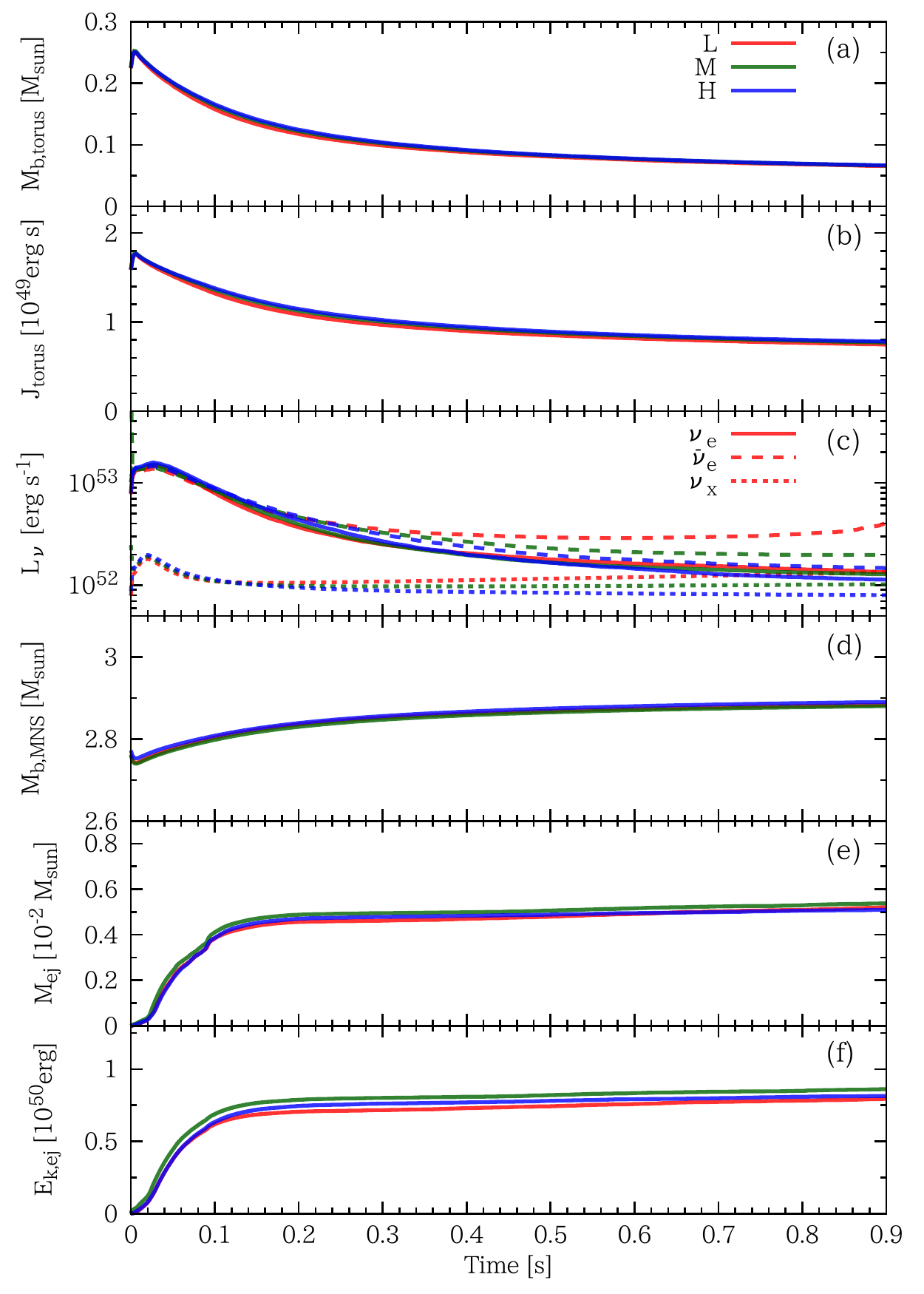}
\caption{
Time evolution of the torus mass (panel (a)), torus angular momentum (panel (b)), neutrino emission rate (panel (c)), baryon mass of the MNS (panel (d)), ejecta mass (panel (e)), and ejecta kinetic energy (panel (f)) for three different resolution models: DD2-135135-0.01-H, DD2-135135-0.01-M, and DD2-135135-0.01-L.
For the panel (c), the solid, dashed, and dotted curves denote the emission rates of electron neutrinos, electron antineutrinos, and the other neutrino species, respectively.
In panel (c), the emission rates of electron neutrinos, electron antineutrinos, and the other neutrino species are shown in the solid, dashed, and dotted curves, respectively.
}
\label{fig:conv}
\end{center}
\end{figure}

In Fig.~\ref{fig:conv}, we plot the results for the lower-resolution models DD2-135135-0.01-M and DD2-135135-0.01-L, together with those for DD2-135135-0.01-H for $\alpha_{\rm vis}=0.01$.  
Only for the time evolutions of the emission rates of electron antineutrinos and heavy lepton neutrinos, the agreement with different resolution models becomes poor for the late time, while the agreement with different resolution models is well achieved for the other quantities.
This trend is the same as that in the inviscid case, as described in \cite{2017ApJ...846..114F}.
A possible reason for this poor behavior is that the density gradient at the surface of the MNS, around which neutrinos are most significantly emitted, becomes steeper for that time; hence the diffusion process of neutrinos is not accurately resolved with the low resolution.
For the $\alpha_{\rm vis}=0.01$ model, we may conclude that $L_{\bar{\nu}_e}\lesssim 2\times10^{52}\,{\rm erg\,s^{-1}}$, $L_{\nu_x}\lesssim 8\times 10^{51}\,{\rm erg\,s^{-1}}$ at $t=0.8$\,s.
We found, from the middle panel of Fig.~\ref{fig:accretion2}, that the emission rate of electron antineutrinos increases for $t\gtrsim 1.2$\,s.
This behavior also would be a numerical artifact due to the steep density gradient at the MNS surface for the late phase; hence, the electron fraction of the ejecta would be underestimated because the difference of the emission rates of electron neutrinos and electron antineutrinos would be smaller than that found in our simulation.
The differences among the different resolution models at $t=0.8$\,s are within 3 \%, 4 \%, 0.3 \%, 6 \%, and 8 \% for the torus mass, torus angular momentum, baryon mass of the MNS, ejecta mass, and ejecta kinetic energy, respectively.

\vspace{1cm}

\bibliographystyle{apj}
\bibliography{apj-jour,reference}

\end{document}